\documentclass[10pt]{article}

\usepackage[T1]{fontenc}
\usepackage{lmodern}
\usepackage[margin=2.5cm]{geometry}
\usepackage{graphicx}
\usepackage{booktabs}
\usepackage{amsmath}
\usepackage{setspace}

\usepackage{microtype}
\usepackage{xcolor}
\usepackage[labelfont=bf,labelsep=period,font=small,justification=justified,singlelinecheck=false]{caption}
\usepackage[superscript,sort,compress]{cite}
\usepackage{longtable,array,calc,needspace,fvextra,newunicodechar,etoolbox}
\newunicodechar{−}{\ensuremath{-}}
\usepackage[hidelinks]{hyperref}
\hypersetup{pdftitle={Groups of AI agents underuse resources despite shared evidence of increased capacity},pdfauthor={Takahiro Ezaki, Naoto Imura and Katsuhiro Nishinari}}

\newcommand{\real}[1]{#1}
\AtBeginEnvironment{longtable}{\footnotesize\setlength{\tabcolsep}{3pt}}

\makeatletter
\renewcommand\@biblabel[1]{#1.}
\makeatother

\newcommand{\sectionhead}[1]{\section*{#1}}
\newcommand{\subhead}[1]{\subsection*{#1}}

\begin{document}

\begin{center}
{\LARGE\bfseries Groups of AI agents underuse resources despite shared evidence of increased capacity\par}
\vspace{1.0em}
{\large Takahiro Ezaki$^{1\ast}$, Naoto Imura$^{1}$ and Katsuhiro Nishinari$^{1,2}$\par}
\vspace{0.6em}
{\small $^{1}$Research Center for Advanced Science and Technology, The University of Tokyo,\\
4-6-1 Komaba, Meguro-ku, Tokyo 153-8904, Japan\par
$^{2}$Department of Aeronautics and Astronautics, School of Engineering, The University of Tokyo,\\
7-3-1 Hongo, Bunkyo-ku, Tokyo 113-8656, Japan\par
$^\ast$Corresponding author: tkezaki@g.ecc.u-tokyo.ac.jp (ORCID 0000-0003-4175-3028)\par}
\end{center}

\vspace{0.6em}

\begin{abstract}
\noindent When AI agents share limited resources, such as computing services or public infrastructure, their individual choices affect how much work they complete as a group. Can sharing evidence of what is available let a group make full use of its resources after capacity changes? We tested this with language-model agents choosing between two simulated servers. One server recovered while its status report still listed it as unavailable. Each server had limited capacity per round, so rejected jobs required another attempt before the deadline. Telling agents that recovery was possible reduced missed deadlines. In controlled tests with Anthropic's Claude Haiku 5.5, no agent tried the recovered server after a rule reminder, yet all chose it when shown a successful request record. After sharing, all five models tested left capacity unused. Keeping the success record fixed, we gave agents numbers without instructions, using each value equally often but giving group members either identical or distinct numbers. With OpenAI's GPT-6.1 Sol or Google's Gemini 3.5 Flash-Lite, distinct numbers spread simultaneous requests across servers and increased job completions even though overall choice frequencies barely changed. Telling AI agents which resources are available leaves a second design challenge: helping them spread demand across those resources.
\end{abstract}

\onehalfspacing

\sectionhead{Introduction}

AI agents already operate in production across diverse domains\cite{pan2026}. When many act for their own users on shared infrastructure, for example by submitting jobs to compute services or reserving limited capacity, the natural goal is an efficient population, one that completes as many jobs as the available capacity allows. Making each agent well informed does not reach that goal. Instead, the population's performance depends on how its members' decisions combine\cite{kleinberg2021}. Collective efficiency has three requirements that can pull in different directions: agents must discover that conditions have changed, act on evidence of the change, and spread their actions across the available capacity. A shared signal can help each agent recognize that a resource is available while simultaneously drawing more demand than the resource can serve\cite{acemoglu2018}. The problem becomes acute when observations arrive late, because a service-status page or traffic report can remain accurate about the past after conditions have changed. Waiting wastes an opportunity, but responding together can waste capacity elsewhere\cite{mitzenmacher2000}. Here we ask what lets a population discover a change before its shared report registers it, and whether agents that each act on a shared discovery also exploit it as a population.

Research on social learning explains both why access to others' discoveries is valuable and why it is not enough\cite{march1991,laland2004}. Copying saves the cost of individual learning, but a population of copiers has nothing new to copy\cite{boyd1995}. Moreover, when one agent's experiment informs everyone, each agent has a reason to wait for others to go first\cite{bolton1999}. How a group learns therefore shapes its performance: depending on the strategies in use, fast diffusion can either speed learning or pull a group toward a local optimum\cite{rendell2010,lazer2007,mason2012,barkoczi2016,toyokawa2019,bernstein2018}. Limited resources add a further requirement, because useful actions must be spread across options\cite{couzin2005,shaffer2013}. When a choice pays less the more others make it, as in the El Farol bar problem and in selfish routing\cite{arthur1994,roughgarden2002}, better information need not improve collective welfare\cite{morris2002,acemoglu2018}, and acting on stale load information produces poor balancing and oscillations\cite{mitzenmacher2000,pender2017}. Service recovery under a delayed report combines these elements: a change that someone must discover, a public signal that may crowd the response, and a short window before routine reporting catches up.

Language-model agents let us manipulate both what individuals are told about possible change and how a common observation is presented. Populations of such agents develop conventions and biases\cite{ashery2025}, show group-size effects\cite{flint2026}, struggle to sustain shared resources\cite{piatti2024}, and spread news among themselves\cite{park2023}. Furthermore, cooperation benchmarks show that models adapt to partners but plan poorly together\cite{zhang2024,agashe2025}, and multi-agent workflows fail through flaws in design, coordination, and verification\cite{cemri2025}. Because many agents may share one decision procedure, algorithmic monoculture is a further concern\cite{kleinberg2021}. Our earlier experiments showed that shared state descriptions shape synchronization in such populations\cite{ezaki2026a}, that a shared warning about other drivers aligned agents' anticipatory route choices and sustained congestion\cite{ezaki2026b}, and that, in a group task with a collective loss and a target participation rate, reasoning agents mapped a shared identifier to the same choice through threshold rules and became unanimous, whereas private random numbers produced group losses close to an independent-choice reference\cite{ezaki2026c}. Here, the shared input is correct evidence of recovery that each agent should use, each agent pursues only its own job, no target split is given, and the need to spread demand arises from a recovered service with limited capacity. We ask whether such evidence can still leave capacity unused, at which stage the loss arises, and whether changing only how the same inputs are assigned within groups changes the work completed.

\begin{figure}[!tbp]
\centering
\makebox[\textwidth][c]{\includegraphics[width=183mm]{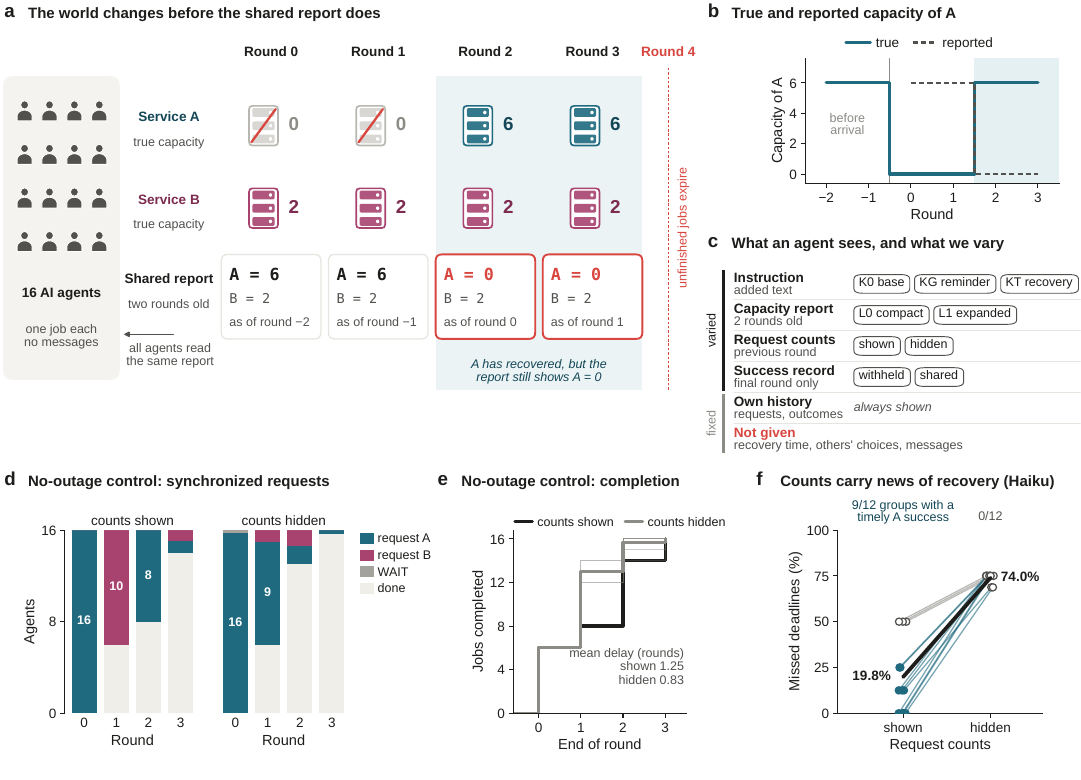}}
\caption{\textbf{The world changes before the shared report does.}
(\textbf{a}) Sixteen agents, each with one job, choose service A, service B, or WAIT in rounds 0--3. Unfinished jobs expire at round 4. A is down in rounds 0--1 and recovers to capacity 6 at round 2; B has capacity 2 throughout. Every agent reads the same capacity report, which describes the state two rounds earlier. At rounds 2 and 3, it still shows A = 0 (red frames).
(\textbf{b}) True and reported capacity of A. Shading marks the rounds in which A has recovered but is reported as zero.
(\textbf{c}) What each agent is given at a decision and the factors we vary: the instruction (K0, KG, KT), the layout of the shared capacity report (L0, L1), whether request counts are shown, and, in constructed histories, whether a success record is shared. Every agent also sees its own request history and the registered population of 16; no agent sees the recovery time, other agents' current choices, or any message.
(\textbf{d}) No-outage control: A has capacity 6 and B has capacity 2 throughout. Shown are the states of the 16 Luna agents (K0, L0) in each round: already done, requesting A, requesting B, or waiting. Areas show means over 12 groups. With request counts shown, all 12 groups followed the same sequence: 16 requests to A, then 10 to B, and then 8 to A. WAIT here denotes valid choices (three of 352 decisions with counts hidden, none of 432 with counts shown); there were no response failures.
(\textbf{e}) Jobs completed by the end of each round in the same no-outage control. Thin lines show individual groups, and thick lines show means. Every job was completed in both conditions.
(\textbf{f}) Haiku (K0, L1) during recovery, with request counts shown or hidden. Lines connect the same environment seed. Filled symbols and colored lines mark groups in which an A success occurred early enough to inform the final decision. The black line connects means.}
\label{fig:setup}
\end{figure}

\looseness=-1
We used service recovery to separate these events (Fig.~\ref{fig:setup}). Agents submitted jobs before a deadline to two simulated servers, which we call services A and B. Each could process a limited number of jobs per round. A failed service recovered while the shared capacity report still described it as unavailable, creating capacity that stayed unused unless agents discovered it. We varied what agents knew about possible recovery, how the report was presented, and whether recent traffic was visible. In addition, we used histories constructed by code to test discovery, the response to a shared success, and the resulting allocation from fixed starting points. Knowing what could change sharply reduced missed deadlines, and sharing a success led agents to use the recovered service, yet the population left capacity unused in all five models we tested. We distinguish individual use of shared evidence, measured by each agent's choice, from collective use, measured by the work the population completes. Agents given the same input respond with the same probabilities, so individual use does not ensure collective use. We separate the discovery of change, the response to evidence of success, and the allocation of work across limited capacity, and show that improvements in the first two did not guarantee efficient allocation. Inputs that differed between agents reduced this loss in two models without any instruction on their use, and capacity, demand, and remaining time determined how much completed work the remaining concentration cost.

\sectionhead{Results}

\subhead{Agents face a recovered service still reported as unavailable}

Each of the 16 agents controlled one job and aimed to have it processed before its deadline. At each decision round (a slot in the simulator), an agent chose service A, service B, or WAIT. Decisions took place at rounds 0--3, and jobs expired before actions at round 4. Each service admitted requests up to its capacity, using a uniform lottery when oversubscribed, and agents exited once their jobs completed. A had capacity zero for the first two rounds and six from round 2 onward. B could complete two jobs per round throughout (Fig.~\ref{fig:setup}a). Each agent was instructed to complete its own job; no agent was given a population objective.

The shared capacity report described the state two rounds earlier, and its observation time advanced at every decision. Accordingly, the first two reports showed the pre-failure capacity of six. At both post-recovery decisions (rounds 2 and 3), the report still showed A at zero (Fig.~\ref{fig:setup}b). We experimentally varied whether agents also received the previous round's request, success, and rejection counts for each service (hereafter, request counts). In the counts-shown condition, these totals appeared in a shared traffic report; in the counts-hidden condition, that report was omitted. Both conditions displayed the number of active agents and each agent's own recent outcomes (Methods).

The factorial experiment used Luna (\texttt{gpt-6-luna}, OpenAI) and Haiku (\texttt{claude-haiku-5-5}, Anthropic), two models run with little or no extended reasoning. We crossed the base instruction (K0), a reminder of rules already in it (KG), and a 69-word description of how A's capacity could change (the recovery description, KT) with two report layouts and request counts shown or hidden (Fig.~\ref{fig:setup}c), giving 288 episodes with 12 matched environment seeds per condition. Both layouts reported the same delayed capacities. The compact report (L0) gave one shared observation time. The expanded report (L1) gave each service its own observation time, labeled the values as delayed, repeated them in a raw field, and included an empty list of confirmation records. Later experiments added Sol (\texttt{gpt-6.1-sol}, OpenAI) and Opus (\texttt{claude-opus-5-5}, Anthropic), both run with medium reasoning effort, and Gemini (\texttt{gemini-3.5-flash-lite}, Google). Invalid or refused outputs were executed as WAIT. Such outputs were rare in the original stages, at most 3.0\% in the cells entering their primary comparisons (Methods, Supplementary Note~7, and Supplementary Table~8); Opus refusals in the five-model comparison are reported with those results. We first examine discovery, then the use of a shared success, and finally how the resulting demand is distributed, what changes that distribution, and when its concentration costs work.

\subhead{Shared request counts synchronize demand and carry evidence of recovery}

Before introducing the failure, we examined the effect of the shared traffic report on its own. Without a service failure (A serving six jobs and B two in every round; Luna, K0, L0), shared request counts synchronized demand: in all 12 groups, all 16 agents first chose A, all ten rejected agents then switched to the idle B, and all eight remaining agents returned to A. When request counts were hidden, rejected agents mostly retried A, and jobs completed sooner (Fig.~\ref{fig:setup}d,e).

During recovery, request counts carried a different kind of information: a request had succeeded at A even though the capacity report still showed zero. Specifically, a success at round 2 appeared in the traffic report at round 3, beside a capacity report describing A's failure at round 1, and showed the remaining agents that A had resumed processing jobs before the capacity report caught up. In Haiku under K0 and L1, the missed-deadline rate was 19.8\% with request counts shown and 74.0\% with them hidden (Fig.~\ref{fig:setup}f). With request counts, an A success occurred at round 2 in nine of the 12 groups, early enough to inform the final decision. Without them, no group attempted A while it was recovered but reported as zero. The same shared feedback could thus either direct agents away from recent crowding or reveal that a supposedly unavailable service was actually working. However, request counts could reveal recovery only after an agent had succeeded at A. We next asked what leads agents to test A in the first place.

\subhead{A short recovery description sharply improves completion}

\begin{figure}[!tbp]
\centering
\makebox[\textwidth][c]{\includegraphics[width=183mm]{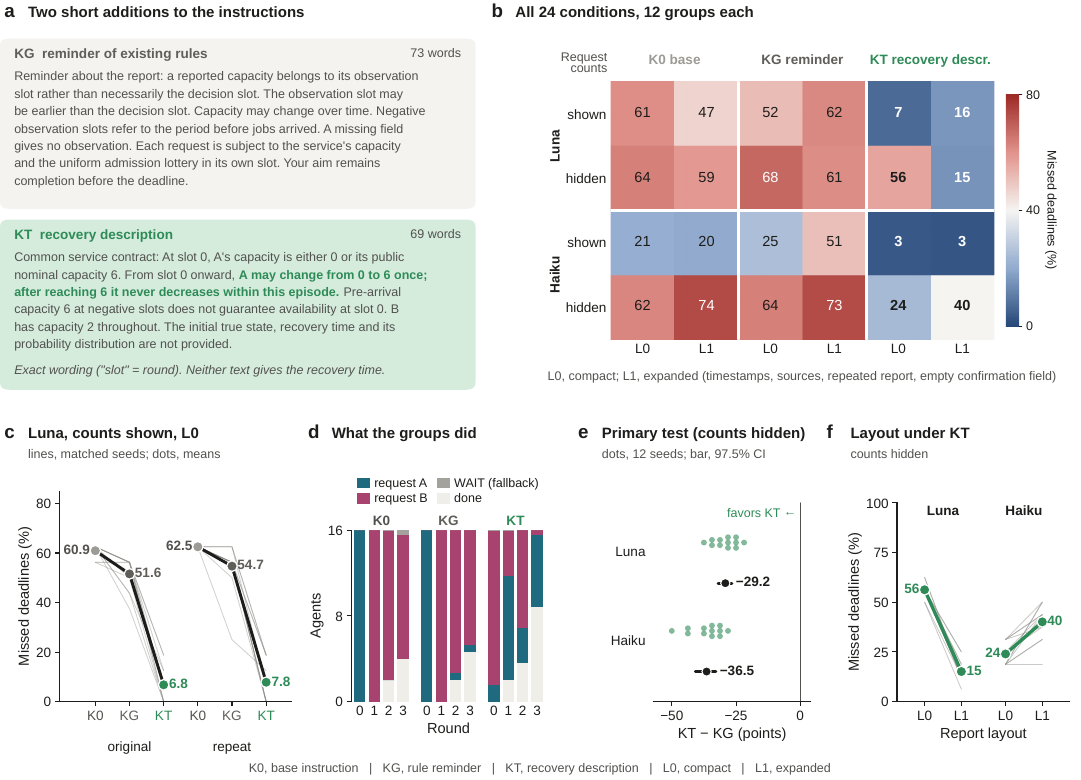}}
\caption{\textbf{Describing possible recovery improves completed work.}
(\textbf{a}) Exact text of the two additions to the base instruction (``slot'' is a decision round). Highlighting marks the transition rule; neither text gives the recovery time.
(\textbf{b}) Missed deadlines (rounded to whole percentages) in all 24 factorial conditions (two models $\times$ three instructions $\times$ two layouts $\times$ request counts shown or hidden), with 12 groups each.
(\textbf{c}) Luna with L0 and request counts shown, in the original factorial and in a separate repeat with fresh seeds. Gray lines follow one environment seed across K0, KG, and KT; dots and black lines show means. Both acquisitions used the 256-token Luna output limit (the acquisition controls are documented in Supplementary Note~6).
(\textbf{d}) State of the 16 agents in each round of the repeat acquisition (means over 12 groups). Gray fallback actions are WAIT actions executed for nonvalid outputs (Supplementary Note~7); no valid WAIT choices occurred. A has recovered from round 2 onward.
(\textbf{e}) Primary test, KT $-$ KG with request counts hidden, averaged over layouts within seed. Green dots show the 12 seed-level differences; black symbols show means with 97.5\% paired bootstrap intervals. All 12 differences favored KT in each model.
(\textbf{f}) Effect of layout under KT with request counts hidden. Thin lines connect the same seed, and thick lines connect means.}
\label{fig:knowledge}
\end{figure}

Discovery requires trying a service that the report still marks as unavailable. We tested whether describing how A could change would lead agents to do so. KT added 69 words specifying that A could start unavailable, recover once to capacity six, and then remain available, while B remained at capacity two (Fig.~\ref{fig:knowledge}a). It supplied neither the actual recovery time nor an instruction to explore. By contrast, KG added a reminder of similar length (73 words) about report timestamps and admission rules already present in the base instruction. The key distinction concerns what an old zero means: KT makes recovery possible after that observation, so retrying A can discover usable capacity before the report changes, whereas KG reiterates when the observation was made but describes no such transition.

With the compact report (L0) and request counts displayed, Luna missed 60.9\% of deadlines under the base instruction (K0). The recovery description (KT) reduced this rate to 6.8\%, whereas the reminder (KG) reduced it to 51.6\% (Fig.~\ref{fig:knowledge}c). A repeat with new environment seeds reproduced this pattern (62.5\%, 54.7\%, and 7.8\% under K0, KG, and KT), and groups under K0 and KG almost never requested A after it recovered (Fig.~\ref{fig:knowledge}d).

The primary comparisons against KG confirmed the benefit even with request counts hidden. Averaged within each seed across the two layouts, KT reduced missed deadlines by 29.2 percentage points in Luna and 36.5 points in Haiku. All 12 matched differences favored KT in each model, with prespecified 97.5\% intervals of $-32.0$ to $-26.6$ and $-40.9$ to $-32.8$ points, respectively (Fig.~\ref{fig:knowledge}e and Supplementary Table~5).

Under KT with request counts hidden, the effect of layout differed between the two models (Fig.~\ref{fig:knowledge}b,f), and response failures do not explain the difference (Supplementary Note~7 and Supplementary Fig.~1). We next fixed a B-only history to test whether agents would explore A without prior failures there.

\subhead{Knowledge and report layout jointly enable exploration}

\begin{figure}[!tbp]
\centering
\makebox[\textwidth][c]{\includegraphics[width=183mm]{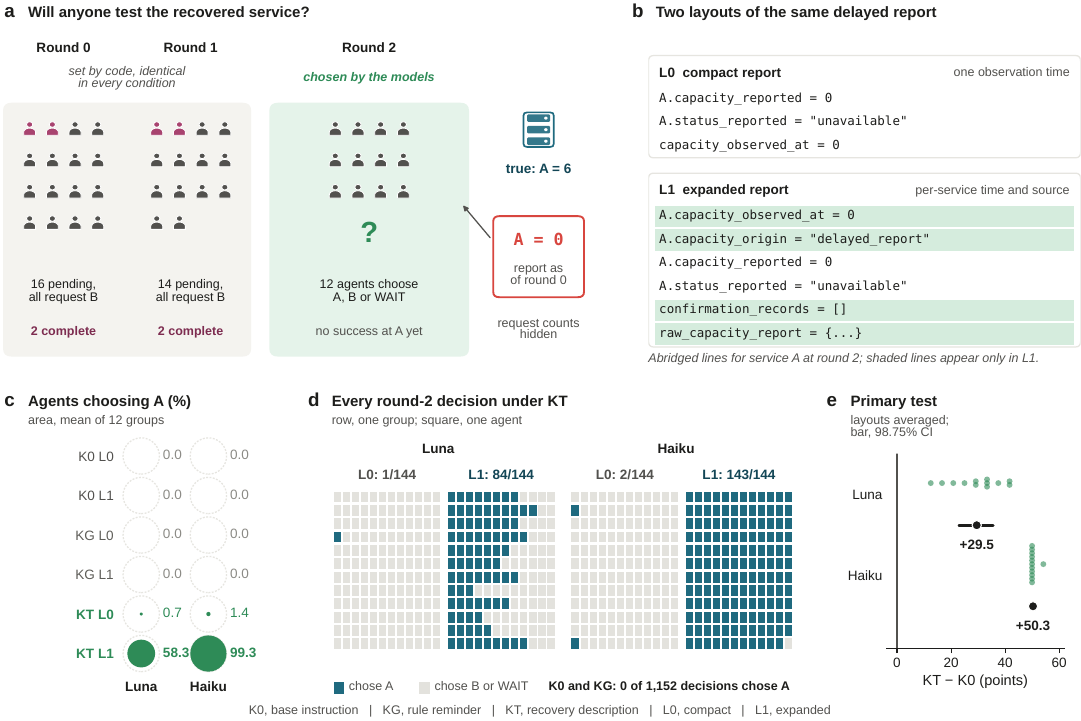}}
\caption{\textbf{Exploration depends on knowledge and its presentation.}
(\textbf{a}) Constructed history. Code sends every unfinished agent to B in rounds 0 and 1 (two jobs complete per round). At round 2, the 12 remaining agents choose freely; A has recovered, the report shows A = 0, and no agent has succeeded at A. Request counts are hidden, and the active-client count is 12.
(\textbf{b}) Abridged lines for service A in the two report layouts at round 2. Both carry the same delayed values; shaded lines appear only in the expanded layout (L1).
(\textbf{c}) Percentage of agents choosing A in all 12 conditions. Disk area is proportional to the mean over 12 groups, and dashed circles mark 100\%.
(\textbf{d}) Every round-2 decision under KT. Each row is one group and each square is one agent. Under K0 and KG, none of the 1,152 decisions (576 per model) chose A.
(\textbf{e}) Primary test, KT $-$ K0, averaged over layouts within seed. Green dots show seed-level differences; black symbols show means with 98.75\% paired bootstrap intervals.}
\label{fig:explore}
\end{figure}

Free trajectories mix discovery with retries after earlier failures and with each group's own history. To isolate discovery, the first requirement, we used a history constructed by code, so that every compared group started from the same events. Specifically, code assigned every unfinished agent to B at rounds 0 and 1, completing four jobs. The remaining 12 agents then chose freely at round 2, when A had recovered, its report still showed zero, no agent had succeeded there, and request counts were hidden (Fig.~\ref{fig:explore}a,b). Thus, these agents had experienced only rejections at B.

Under K0 and KG, neither primary model attempted A in any of these controlled batches: none of the 1,152 decisions (576 per model) chose A. By contrast, KT elicited few attempts with L0 but substantial exploration with L1. The A-choice rate of Luna rose from 0.7\% to 58.3\%, and that of Haiku from 1.4\% to 99.3\% (Table~\ref{tab:explore} and Fig.~\ref{fig:explore}c,d). The primary KT $-$ K0 differences, averaged across layouts, were positive in every seed for both models (Fig.~\ref{fig:explore}e). However, exploration also crowded the recovered service. Under KT with the expanded report, Haiku sent no request to B in 11 of the 12 batches and completed 6.1 of the eight attainable round-2 jobs on average, whereas Luna, with 58\% choosing A, used both services in every batch and completed 7.5 jobs (post hoc; Supplementary Note~3).

\begin{table}[!tbp]
\centering
\caption{\textbf{Exploration from a constructed history.} Percentage of the 12 remaining agents choosing A at round 2. Each cell contains 12 generation batches indexed by simulator seed.}
\label{tab:explore}
\begin{tabular}{llrrr}
\toprule
Model & Layout & Base K0 (\%) & Reminder KG (\%) & Recovery description KT (\%)\\
\midrule
Luna  & L0 & 0.00 & 0.00 & 0.69\\
Luna  & L1 & 0.00 & 0.00 & 58.33\\
Haiku & L0 & 0.00 & 0.00 & 1.39\\
Haiku & L1 & 0.00 & 0.00 & 99.31\\
\bottomrule
\end{tabular}
\end{table}

In a separate acquisition that decomposed the expanded report, service-specific timestamps alone raised A choice relative to L0 by 91.7 percentage points in Luna and 70.8 in Haiku (Supplementary Note~8, Supplementary Fig.~2, and Supplementary Tables~13 and 14).

Recovery knowledge enabled exploration, but its frequency depended strongly on report layout, and under K0 and KG no agent explored at all. When only some agents discover a change, the population depends on whether the others use their success, which we tested next.

\subhead{A shared success restores use of A but can leave B idle}

\begin{figure}[!tbp]
\centering
\makebox[\textwidth][c]{\includegraphics[width=167mm]{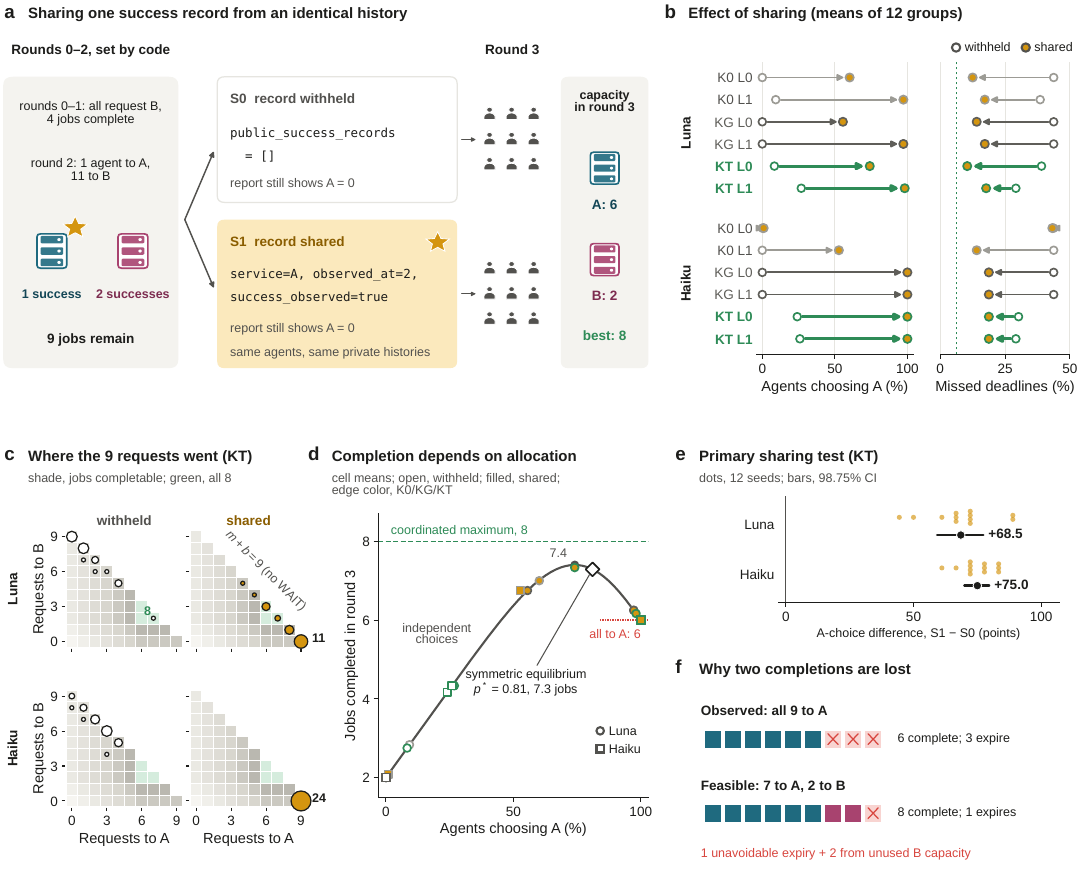}}
\captionsetup{skip=9pt}
\caption{\textbf{A shared success improves completion while leaving capacity idle.}
(\textbf{a}) Design. After a coded prefix (rounds 0--2), nine agents remain. In one branch, the success record is withheld (S0); in the matched branch, the record of the round-2 success at A is shared (S1). The capacity report still shows A = 0 in both. At round 3, A and B can complete six and two jobs, respectively.
(\textbf{b}) Change from withheld (open) to shared (filled) in A choice (left) and missed deadlines (right) for each model, instruction, and layout (means of 12 matched groups). Denominators are the nine pending agents and the original 16 jobs, respectively. The dashed line marks the attainable minimum loss, $1/16=6.25\%$.
(\textbf{c}) Round-3 allocations under KT, pooled over layouts (24 branches per panel). Each cell represents one combination of requests to A and to B. Its shade indicates the number of jobs it can complete, and green cells complete all eight. Branches on the diagonal $m+b=9$ contain no WAIT. Circle area is proportional to the number of branches; numbers mark cells with at least 10 branches.
(\textbf{d}) Jobs completed in round 3 plotted against the percentage of agents choosing A. The curve is a mathematical reference for nine independent A/B choices with no WAIT; its horizontal coordinate is $100p$. The dashed line is the coordinated maximum. The open diamond marks the symmetric mixed-strategy equilibrium ($p^{*}=0.81$, 7.3 expected completions), at which an agent's expected success is the same for either request given the other eight (Supplementary Note~5). Symbols show observed condition means (open, withheld; filled, shared; edge color: gray, K0; dark gray, KG; green, KT). The independent-choice maximum, 7.40 expected completions, is at $p\approx0.741$, close to Luna's empirical KT/L0 shared-record rate of $80/108$.
(\textbf{e}) Prespecified sharing contrasts under KT, averaged across layouts within seed. Dots show the 12 seed-level differences, and black symbols show means with 98.75\% paired bootstrap intervals.
(\textbf{f}) Capacity accounting for the observed all-A allocation and a feasible seven-A/two-B allocation. Filled squares show completed jobs, and crosses show missed deadlines. The feasible allocation is an illustrative benchmark, not a new intervention.}
\label{fig:share}
\end{figure}

To test the second requirement, acting on evidence of a change, we supplied a successful demonstration while holding the preceding events fixed (Fig.~\ref{fig:share}a). After the same two B-only rounds, code sent one agent to A and 11 to B at round 2; one job completed at A and two at B, leaving nine agents for the final decision. From that state, we generated decisions in two branches, with the success record either withheld (S0) or shared (S1). Both branches hid request and active-agent counts and contained identical private histories. The shared record stated that A had served a request at round 2, while the capacity report still described zero capacity at round 1. Under KT, which states that recovery persists, that success established that A was available.

Under KT, sharing increased A choice by 68.5 percentage points in Luna and 75.0 in Haiku, averaged across layouts (98.75\% intervals, 59.3 to 77.3 and 69.9 to 79.6; Fig.~\ref{fig:share}e). Correspondingly, the missed-deadline rate fell from 34.1\% to 14.1\% in Luna and from 29.7\% to 18.8\% in Haiku (Table~\ref{tab:share} and Fig.~\ref{fig:share}b). Even under KG, which produced no exploration, Haiku shifted from no A requests to all A requests, using a demonstration that its population had failed to produce. Luna also responded under K0 and KG, so the secondary prediction that KT would enlarge its sharing response was not supported (Supplementary Table~6).

\begin{table}[!tbp]
\centering
\caption{\textbf{Sharing a success record under the recovery description.} A choice uses the nine remaining agents as its denominator; missed deadlines use the original 16 jobs. The record is withheld in S0 and shared in S1. Each comparison contains 12 matched generation batches indexed by simulator seed.}
\label{tab:share}
\begin{tabular}{llrrrr}
\toprule
 & & \multicolumn{2}{c}{A choice (\%)} & \multicolumn{2}{c}{Missed deadlines (\%)}\\
\cmidrule(lr){3-4}\cmidrule(lr){5-6}
Model & Layout & S0 & S1 & S0 & S1\\
\midrule
Luna  & L0 & 8.33  & 74.07  & 39.06 & 10.42\\
Luna  & L1 & 26.85 & 98.15  & 29.17 & 17.71\\
Haiku & L0 & 24.07 & 100.00 & 30.21 & 18.75\\
Haiku & L1 & 25.93 & 100.00 & 29.17 & 18.75\\
\bottomrule
\end{tabular}
\end{table}

Sharing resolved whether A was usable, but it did not assign different services to different agents. Every remaining agent could act on the same success simply by choosing A. At the final decision, nine jobs remained; A could complete six and B two.

\begin{samepage}
Let $C(m,b)$ denote the total number of jobs completed in that round when $m$ agents request A and $b$ request B, with $w$ agents waiting. Each service completes the smaller of its request count and its capacity, giving
\begin{equation}
C(m,b)=\min(m,6)+\min(b,2),\qquad m+b+w=9.
\end{equation}
\end{samepage}
An allocation using both capacities could complete eight jobs, for example seven requests to A and two to B (Fig.~\ref{fig:share}f). By contrast, sending all nine requests to A completed only six jobs and left B idle, as happened in all 24 Haiku KT branches with the shared record (Fig.~\ref{fig:share}c). Given eight others choosing A, an agent would succeed with probability $6/9$ by also choosing A but with certainty by choosing B.

Eight completions require agents to act differently, but the remaining agents in each branch received identical prompts and could not see one another's choices, so each choice was a draw from the same response distribution. If each of the nine agents chose A independently with probability $p$, expected completion would peak at 7.40 jobs at $p \approx 0.74$ (Fig.~\ref{fig:share}d). This optimized independent-choice benchmark, a mathematical reference rather than a fitted model, is the relevant reference for agents given the same input. The all-A response lost 1.40 jobs relative to it, and two relative to a coordinated assignment. Where every agent chose A, each agent used the shared evidence, but the population did not: it completed fewer jobs than this optimized independent-choice benchmark. Luna came close to the benchmark with L0, where 80 of 108 choices went to A and 7.33 jobs were completed on average, but with L1 its A choice rose to 98.1\% and completions fell to 6.17 (Table~\ref{tab:share}).

The all-A allocation is not an equilibrium. Had the agents known the capacities and that nine agents remained, the symmetric mixed equilibrium would have each agent choose A with probability $p^{*}=0.81$, giving 7.29 expected completions (Supplementary Note~5). Agents were not told that nine remained; we test this information gap below by revealing the count.

\subhead{The value of a success record depends on timing and model}

\begin{figure}[!tbp]
\centering
\makebox[\textwidth][c]{\includegraphics[width=183mm]{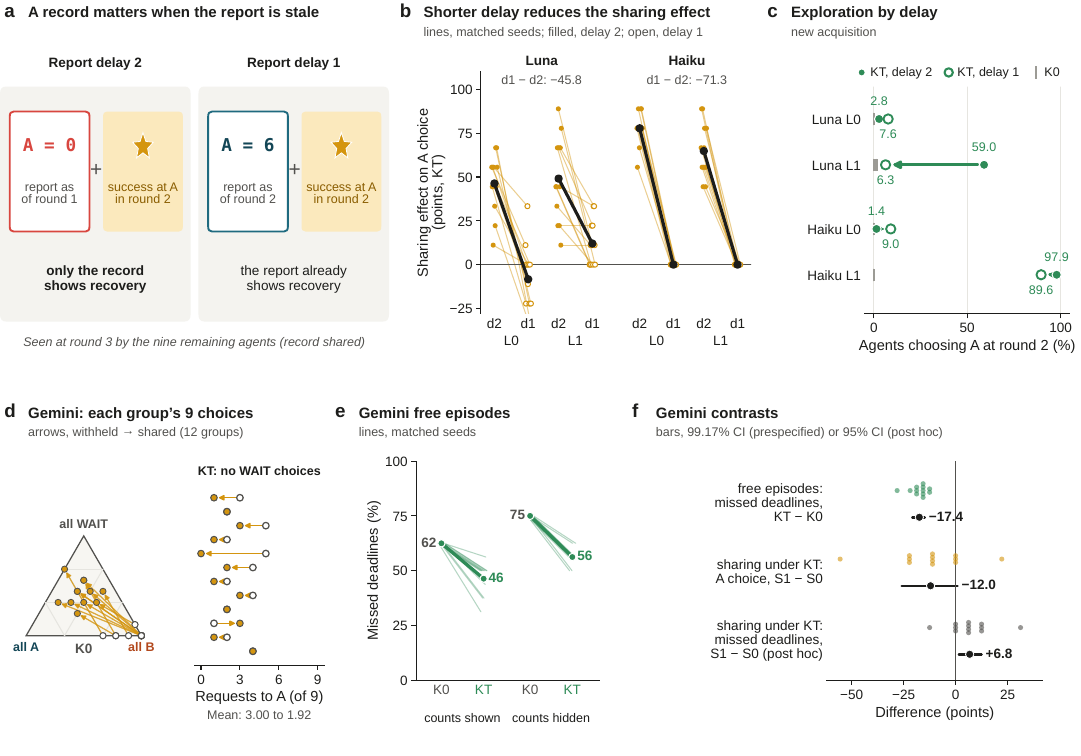}}
\caption{\textbf{Report timing and model change the effect of sharing.}
(\textbf{a}) What the nine remaining agents see at round 3 when the record is shared. Under delay 2, the capacity report still shows A = 0; under delay 1, it already shows the recovery.
(\textbf{b}) Effect of sharing on A choice (S1 $-$ S0) under KT. Thin lines link the same seed under delay 2 (filled) and delay 1 (open); black lines link means. The primary contrast (delay 1 $-$ delay 2, averaged over layouts) is printed above each model.
(\textbf{c}) Exploration from the constructed round-2 history under KT, from delay 2 (filled) to delay 1 (open). Gray ticks show K0.
(\textbf{d}) Third model, Gemini (L1, delay 2). Left: the nine choices of each K0 group in the simplex of A, B, and WAIT shares. Right: A requests in each KT group, with one horizontal row per matched seed; all other requests were to B. Arrows link withheld (open) and shared (filled) branches. Row positions distinguish groups and do not encode WAIT.
(\textbf{e}) Missed deadlines of Gemini in free episodes (L1). Thin lines link matched seeds, and thick lines link means.
(\textbf{f}) Gemini contrasts. Dots show seed-level differences; black symbols show means with 99.17\% paired bootstrap intervals for the two prespecified contrasts and a 95\% interval for the post hoc deadline comparison.}
\label{fig:timing}
\end{figure}

If the success record helps because it is newer than the capacity report, it should matter less when the report is less stale. We tested this prediction with fresh seeds and contemporaneous delay-1 and delay-2 conditions. At the final decision, the delay-1 capacity report already described the recovery of A at round 2, so agents could learn that A was available from the ordinary report even when the success record was withheld. Under delay 2, by contrast, the ordinary report still showed zero, and the success record supplied the newer observation needed to recognize recovery (Fig.~\ref{fig:timing}a).

As predicted, shortening the delay reduced the additional A-choice effect of sharing under KT by 45.8 percentage points in Luna and 71.3 in Haiku, averaged across layouts (Fig.~\ref{fig:timing}b). Exploration at round 2 was more model-dependent: with delay 1 and L1, the KT $-$ K0 difference was 6.3 points in Luna and 89.6 in Haiku (Fig.~\ref{fig:timing}c). The success record thus mattered most when the capacity report still described failure.

To check whether these responses depend on the model, we tested a third model. Gemini, run with minimal thinking, also benefited from the recovery description, which reduced missed deadlines by 17.4 percentage points relative to K0 (Fig.~\ref{fig:timing}e,f), but sharing did not increase its A choice. Even without the record, it already sent two-thirds of its requests to B. Under KT, sharing reduced A requests from 36/108 to 23/108 (difference $-12.0$ points, prespecified interval $-25.9$ to 0.9; Fig.~\ref{fig:timing}d), and missed deadlines rose from 25.0\% to 31.8\% (post hoc 95\% interval for the difference, 1.6 to 12.5 points). Haiku and Gemini thus lost work through opposite allocations.

\begin{figure}[!tbp]
\centering
\makebox[\textwidth][c]{\includegraphics[width=176mm]{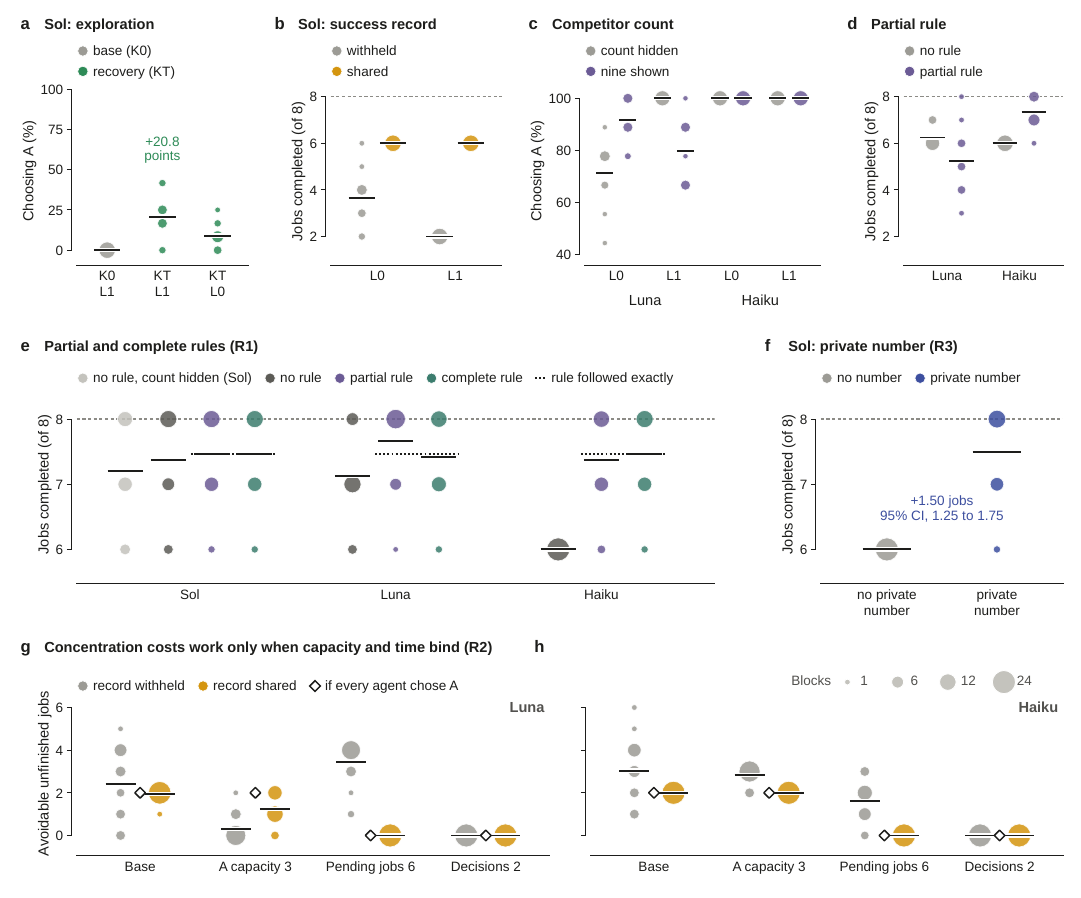}}
\captionsetup{skip=4pt,font=footnotesize}
\caption{\textbf{What lets identical agents respond differently, and when concentration costs work.}
(\textbf{a}) Sol exploration under the base instruction (K0) and the recovery description (KT), with medium reasoning effort. The main knowledge contrast compares K0 and KT at L1; KT/L0 is also shown.
(\textbf{b}) Sol final-round completions with the success record withheld (S0) or shared (S1) under KT. All 24 shared branches chose A unanimously, completing six jobs.
(\textbf{c}) Effect of revealing nine active clients in Luna and Haiku, with KT and the success record supplied in both branches. Luna's opposite layout-specific shifts cancel out in the primary layout-averaged comparison.
(\textbf{d}) A private-draw rule (F3) under KT/L1, with the success record supplied in both branches. Both branches include a private number; each agent's number is fixed across branches, and the treatment directs draws below 0.22 to B and leaves other choices free.
(\textbf{e}) R1 completions with KT/L1, a shared success, and private numbers in every branch. All branches show nine pending agents except Sol's count-hidden branch. The partial rule assigns numbers below 0.22 to B and leaves other choices free; the complete rule assigns those other choices to A. Dotted segments over the rule conditions mark the 7.46 mean completions from complete adherence to these realized draws; the number of B assignments varies across blocks.
(\textbf{f}) R3 Sol completions with and without a private number. Both branches use KT/L1 and the shared success, with no competitor count or distribution rule. Supplementary Fig.~4 shows choices by number decile.
(\textbf{g} and \textbf{h}) R2 avoidable unfinished jobs for Luna and Haiku. Open diamonds show the capacity reference conditional on always choosing A. The base setting has nine pending jobs, capacities 6/2, and one remaining decision. Variants change A capacity to three, pending jobs to six (from 13 registered), or remaining decisions to two with observations updating normally. Loss excludes jobs beyond the feasible maximum. No private numbers or rules are supplied.
Each circle represents the acquisition blocks that share a value, with area proportional to their number (12 blocks per condition in \textbf{a} to \textbf{d}, 24 in \textbf{e} to \textbf{h}); black lines show means. Dashed lines mark eight attainable completions. Supplementary Tables 11, 12, 14, 15, 18 to 20, and 21 report the conditions and comparisons.}
\label{fig:followup}
\end{figure}

Because concentration might reflect limited reasoning, we repeated the constructed-history tests with Sol, run with medium reasoning effort (864 decisions). Under L1, recovery knowledge increased A exploration from 0 to 20.8\% (98.33\% interval for the difference, 12.5 to 29.2 points; Fig.~\ref{fig:followup}a). Sharing increased A choice by 90.7 points, averaged across layouts (86.1 to 94.9). In all 24 shared branches every remaining agent chose A, and mean final-round completions rose from 2.83 to 6.00, which is 1.40 below the independent-choice benchmark (Fig.~\ref{fig:followup}b and Supplementary Table~11). These single-model checks motivated the five-model comparison below.

\subhead{Response probabilities for identical inputs account for mean allocation}

Why did agents that each acted on correct evidence concentrate their requests? Within each constructed branch, agents chose before admission and could not observe one another's choices, so following others' observed choices was excluded by design. Moreover, complete-prompt hashes were identical across all agents and all 12 batches within each of 115 conditions without private draws (Supplementary Note~10 and Supplementary Table~16). Identical inputs could therefore concentrate choices simply through shared response probabilities.

To test how far these probabilities explain completed work, we estimated A, B, and WAIT probabilities for each of 72 final-round conditions from 11 batches and predicted completions in the held-out batch assuming independent choices. In 38 conditions the choices did not vary, and the prediction followed directly. In the other 34 conditions, the mean absolute discrepancy between observed and predicted condition means was 0.056 jobs (residuals $-0.291$ to 0.286; Supplementary Fig.~3 and Supplementary Table~17). Concentrated response probabilities for a shared input thus account for the mean allocation loss, although this test does not address the full joint distribution of choices.

\subhead{Private information and rules change the distribution of demand}

We next tested whether revealing competitor counts, providing allocation rules, or adding private numbers would change the distribution of demand. Showing Luna and Haiku the total of nine active agents, which closes the information gap noted above and makes the excess demand explicit, did not remove concentration (with KT and the shared record in both branches). Haiku chose A in every decision across all 48 branches, with or without the count. In Luna, the layout-averaged effect was zero (97.5\% interval, $-4.6$ to 5.1 points) because the count moved A choice in opposite directions in the two layouts, from 71.3 to 91.7\% under L0 and from 100 to 79.6\% under L1 (Fig.~\ref{fig:followup}c and Supplementary Tables~12 and 14).

A partial distribution rule (F3) gave each agent a private uniform draw in $[0,1)$ and instructed those with draws below 0.22 to request B, leaving the rest to choose freely; the matched no-rule branch showed the same draws. Under KT/L1 with the shared record, the rule raised Haiku's completions from 6.00 to 7.33 (difference 1.33, 97.5\% interval 0.92 to 1.75) but lowered Luna's from 6.25 to 5.25 ($-1.00$, $-1.75$ to $-0.17$; Fig.~\ref{fig:followup}d), because Luna's freely choosing agents also sent 47 of 87 requests to B and left A capacity unused (Supplementary Table~12). A complete rule (R1), which also instructed undesignated agents to request A, was followed exactly by Sol and Haiku and yielded 7.46 completions, the value implied by the realized draws; it falls short of eight because the number of agents drawing below the threshold varies across groups (Fig.~\ref{fig:followup}e, Supplementary Note~11, and Supplementary Table~18). For Luna, contrary to prediction, the complete rule did not improve on the partial rule (Supplementary Table~20). Rules thus improved allocation only where they fixed every agent's choice, and even then some capacity went unused.

Sol distributed demand even without a rule. Its no-rule R1 branch with the count hidden completed 7.21 jobs, compared with six in an earlier branch whose prompt differed only in lacking the number. A contemporaneous test (R3), with KT/L1, the shared record, and no count or rule, confirmed that adding only a private number raised Sol's completions from 6.00 to 7.50 (prespecified difference 1.50, 95\% interval 1.25 to 1.75; Fig.~\ref{fig:followup}f and Supplementary Table~21). All 24 no-number branches chose A unanimously, compared with two of 24 with the number, and B was chosen in none of the 150 decisions with numbers below 0.7 but in 44 of the 66 at or above 0.7 (Supplementary Fig.~4). This agrees with our observation that Sol maps identifiers and private random numbers to choices through thresholds\cite{ezaki2026c}; here, however, the number came with no instruction, no target split, and no collective objective. Haiku and Luna had received the same field in F3 without comparable dispersion. We next tested whether this effect of a private number extends beyond Sol and whether it depends on numbers differing between agents.

\subhead{Capacity remains unused after sharing in five models}

\begin{figure}[!tbp]
\centering
\makebox[\textwidth][c]{\includegraphics[width=183mm]{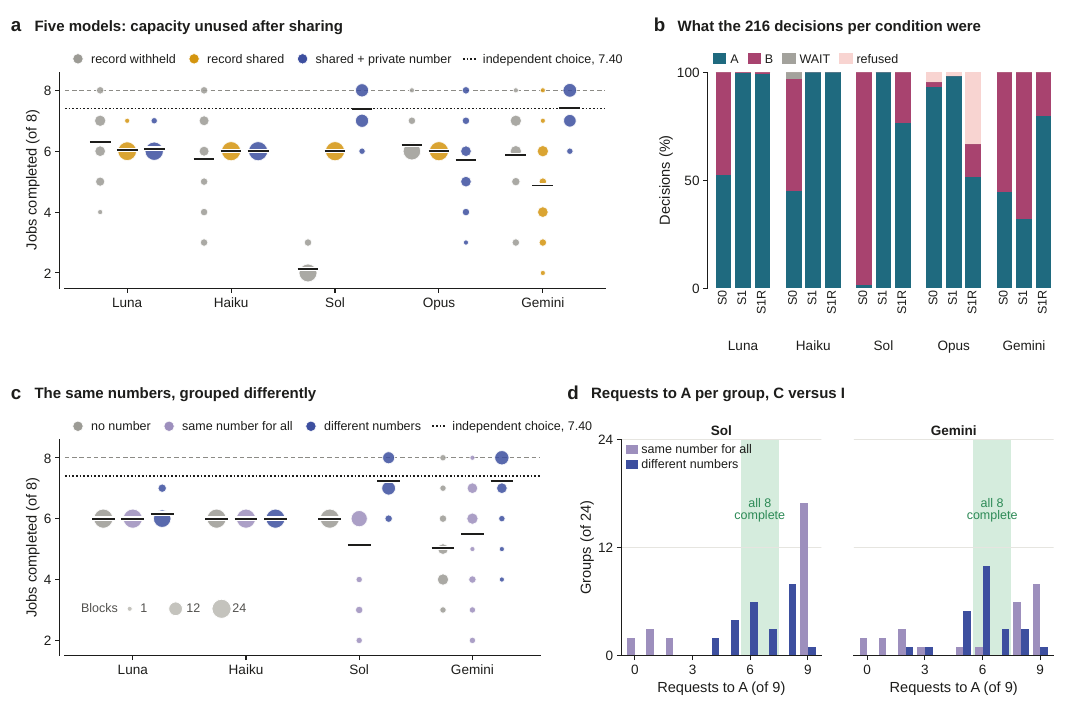}}
\captionsetup{skip=4pt,font=footnotesize}
\caption{\textbf{The allocation loss recurs in five models, and different inputs within a group reduce it.}
(\textbf{a}) Final-round completions in five models under KT/L1 with the success record withheld (S0), shared (S1), or shared with a private random number (S1R). Each circle represents the blocks that share a value, with area proportional to their number (24 blocks per condition); black lines show means. The dashed line marks eight attainable completions, and the dotted line marks 7.40, the maximum expected completion when all nine agents choose independently with the same optimized probability. In S0 and S1, all blocks and models received identical prompts; in S1R, blocks differed only in the private numbers, which were the same across models.
(\textbf{b}) Composition of the 216 decisions per condition: requests to A, requests to B, valid WAIT choices, and refused responses executed as WAIT.
(\textbf{c}) Completions with the success record shared and no number (N), the same number for all nine agents in a group (common, C), or different numbers for the nine agents (individual, I), in four models; lines and circles as in \textbf{a}. C and I use the same multiset of numbers and complete prompts; only the grouping differs. Luna was acquired in a subsequent stage with the same inputs.
(\textbf{d}) Number of groups (of 24) with each count of requests to A under C and I in Sol and Gemini. Groups that send at least six requests to A and at least two to B complete all eight jobs (shaded); because agents may also WAIT, the count of requests to A alone does not determine completions. Supplementary Note~15 and Supplementary Tables~22 to 25 report every condition and comparison.}
\label{fig:fivemodel}
\end{figure}

To test whether the allocation loss and the effect of a private number depend on the model, we extended the single-model checks above and repeated the final-round test with inputs matched across five models (Luna, Haiku, Sol, Opus, and Gemini), with 24 blocks per condition and three conditions: record withheld (S0), record shared (S1), and record shared with a private random number (S1R) (Fig.~\ref{fig:fivemodel}a,b, and Supplementary Tables~22 and 23). After sharing, no model reached the optimized independent-choice benchmark of 7.40 jobs. Luna, Haiku, Sol, and Opus sent all or nearly all valid requests to A (215, 216, 216, and 212 of 216; Opus refused the other four) and completed 6.04, 6.00, 6.00, and 6.00 jobs. Gemini lost work differently. Its agents did not choose alike, as 23 of 24 groups used both services, but they leaned toward B, which received 147 of 216 requests, and completed 4.88 jobs. Without the record, completions were also below 7.40 in every model (2.12 to 6.29), so sharing did not create the loss; rather, it changed where demand went, moving nearly all requests to A in four models and more of them toward B in Gemini.

The individual response to the record replicated: sharing raised A choice from 52\% to nearly 100\% in Luna and from 45\% to 100\% in Haiku. However, its benefit to completion depended on the withheld-record baseline, which varied across acquisitions of the same KT/L1 condition (27 to 52\% of requests to A in Luna and 26 to 45\% in Haiku). In this acquisition many agents already chose A without the record, so sharing changed completions by only $-0.25$ and $+0.25$, with intervals including zero (Supplementary Note~15). By contrast, sharing raised Sol's completions from 2.12 to 6.00 (difference 3.88, 95\% interval 3.75 to 4.00; Holm-adjusted $P = 1.2 \times 10^{-6}$ across ten planned comparisons). The decreases in Gemini ($-1.00$) and Opus ($-0.21$) had intervals excluding zero but were not significant under the planned test (adjusted $P = 0.31$ and 0.75). Adding a private number raised completions in both Sol (from 6.00 to 7.38; difference 1.38, 1.12 to 1.62, adjusted $P = 4.3 \times 10^{-6}$) and Gemini (from 4.88 to 7.42; difference 2.54, 1.96 to 3.17, adjusted $P = 1.1 \times 10^{-5}$). Meanwhile, Luna and Haiku remained almost unanimous in choosing A. Opus refused 72 of 216 requests when the number was present and completed 5.71 jobs; we report this result as executed and do not evaluate its allocation with refusals removed.

\subhead{Distributing different numbers within groups improves allocation}

A private number could help because each agent receives a different input or merely because the prompt contains a number. To separate these explanations, we compared three conditions with the same shared record in Luna, Haiku, Sol, and Gemini: no number (N), the same number for all nine agents in a group (common, C), and different numbers for the nine agents (individual, I) (Fig.~\ref{fig:fivemodel}c,d, and Supplementary Tables~24 and 25). C and I used the same 24 numbers, each shown to nine agents, and therefore the same set of complete prompts; only the assignment of numbers to groups differed.

In Sol, distinct numbers raised mean completions to 7.25, up from 6.00 without a number (difference 1.25, 95\% interval 1.00 to 1.50) and from 5.12 with a common number (difference 2.12, 1.50 to 2.79). In Gemini, the corresponding increases were from 5.04 and 5.50 to 7.25 (differences 2.21, 1.46 to 2.88, and 1.75, 1.08 to 2.46). All four comparisons remained significant after Holm correction. By contrast, Haiku chose A in every condition. In a later Luna acquisition with the same inputs, N and C again produced unanimous A choice, whereas I produced four B choices in 216 decisions (6.17 completions), a change that was not significant.

The overall share of requests to A was nearly the same under C and I (160 and 158 of 216 in Sol; 142 and 144 in Gemini), yet nine Sol groups and 13 Gemini groups completed all eight jobs under I, compared with none and one under C. A common number was worse than none in Sol (5.12 compared with 6.00) because whole groups responded to it in the same way: 17 of 24 groups sent every request to A and completed six jobs, just as without a number, whereas the other seven sent most requests to B and completed only two to four. In both Sol and Gemini, the same numbers and nearly the same rate of choosing each service thus produced markedly different collective outcomes, depending solely on whether agents within a group shared their input.

\subhead{Capacity, demand, and time determine when concentration loses work}

The remedies above concern how demand is divided, but whether concentration costs work at all depends on the environment. To test this, we varied three features of the constructed sharing task in Luna and Haiku (R2), with 24 matched blocks per environment, KT/L1, hidden counts, and no numbers or rules: a recovered A capacity of three rather than six, six rather than nine pending jobs, and two remaining decisions rather than one (Supplementary Note~12). If every pending agent requests A, the avoidable completion gap is
\[
G_A=\min\{n,h(a+b)\}-\min\{n,ha\},
\]
where $n$ is pending demand, $a,b$ are capacities, and $h$ is the number of remaining decisions; the first term is what both services could finish before expiry and the second what A alone could finish. The predicted gap is two jobs in the base and smaller-capacity settings and zero with fewer pending jobs or an extra decision.

Haiku's shared-record branches followed this prediction exactly (Fig.~\ref{fig:followup}h): every choice was A in all four environments, the gap stayed at two with smaller A capacity, and it vanished in every block with fewer pending jobs or an extra decision. Luna nearly always chose A after sharing in the base setting (gap 1.96). With smaller A capacity, its A choice fell to 90.7\% and the gap to 1.25 (primary contrast $-0.71$ jobs, 99.17\% interval $-1.12$ to $-0.33$). Both other variants eliminated Luna's gap in every block, as predicted (Fig.~\ref{fig:followup}g and Supplementary Table~20).

Sharing could also lower completions: with smaller A capacity, it raised Luna's A choice from 36.1 to 90.7\% and lowered completions from 4.71 to 3.75 (secondary difference $-0.96$, descriptive 95\% interval $-1.33$ to $-0.54$; Supplementary Tables~19 and 20). The same concentrated response need not waste work when demand fits within the chosen service's remaining capacity.

\sectionhead{Discussion}

Knowing what could change helped AI populations adapt before their shared reports caught up with the environment, and a shared success led agents to use the recovered service. However, these useful individual responses did not produce an efficient allocation. After sharing, four of the five models sent nearly every request to the recovered service, and the fifth spread its requests but favored the smaller service; none reached the optimized independent-choice benchmark. Our contribution is an experimental separation of three requirements for an efficient population: discovering change, acting on evidence of it, and distributing those actions across limited resources. Holding histories fixed identifies the stage at which an intervention acts, and measuring completed work shows whether the response realizes the available capacity. Whether sharing raised completions depended on the model and on how the population behaved without the record, and it varied between acquisitions; what recurred was the allocation after sharing.

Agents that receive the same input respond with the same probabilities. When agents choose independently, such a population can at best reach the optimized independent-choice benchmark, and concentrated probabilities do much worse. Indeed, response probabilities estimated from repeated inputs accounted for the mean loss. The intervention that reduced the loss was a difference in input. A number with no stated purpose raised completions in Sol and Gemini to between 7.21 and 7.50 of eight. Regrouping the same prompts, a comparison of the same kind as in our earlier study\cite{ezaki2026c}, showed that the benefit required agents within a group to receive different numbers: at nearly the same aggregate choice rates, distinct numbers completed more jobs and a common number completed fewer. These results sit one layer beneath our earlier findings. Those studies concern a later layer of control, in which the information given to a population comes with something meant to manage the collective response: a message that made every agent consider what the others would do, or a target for the group to implement. Here agents received only correct evidence about their environment, with no collective objective and no target split, yet the loss still arose. A natural response is to make agents take one another into account. In the present experiments, however, showing the number of competing agents did not remove concentration. Our route-choice experiments tested a stronger version, a warning that others might follow the same advice. Because every agent received the same warning and applied similar decision rules, their anticipatory choices aligned and congestion persisted\cite{ezaki2026b}. Similarly, in the group task, which gave the population a collective loss and a target participation rate, agents sharing an identifier became unanimous, and private random numbers produced group losses close to an independent-choice reference\cite{ezaki2026c}. Across these layers, additions common to all agents kept choices aligned, and the additions that helped differed between agents. Even so, independent numbers leave the count sent to each service random, so under a fixed response rule the best one-round expectation is 7.40 (Fig.~\ref{fig:share}d), and the higher means in R1 and R3 reflect finite sets of draws. Reaching eight completions reliably requires coordinated roles, negatively dependent assignments, or another decision with feedback.

The allocation result differs from herding and from the costs of communication in collective search. Informational cascades and social influence can sustain mistaken choices or reduce the diversity needed for accurate aggregate judgments\cite{bikhchandani1992,lorenz2011}, and rapid communication can reduce the exploration that generates good solutions\cite{lazer2007,mason2012,barkoczi2016,bernstein2018}. Here, the shared success was correct, and the obstacle appeared after the valuable option had been found, when simultaneous users competed for its capacity. Public-information and congestion theory establish that individual responses to information can impose collective costs\cite{morris2002,acemoglu2018}; our experiment measures that cost directly as unused capacity after a controlled disclosure. Conversely, Haiku under the rule reminder generated no exploration yet chose A unanimously once the record was supplied. A population able to exploit a discovery can thus still fail to generate it, which is why strategic experimentation treats the supply of information as a collective problem\cite{bolton1999}.

The delay and boundary manipulations show when these effects matter. A success record changed choices most when the ordinary report still described failure. Information freshness is a distinct systems objective\cite{kaul2012}, and the consequences of delayed observations depend on the policy that acts on them\cite{mitzenmacher2000,pender2017}. Furthermore, the capacity and horizon variants showed that unanimity alone does not indicate a system failure: Haiku's unanimous A requests lost nothing when demand fit within A's capacity or when another decision allowed the remaining jobs to finish. Rather, the relevant boundary relates the demand sent to a service to what that service can process before expiry.

For designers pursuing an efficient population of agents, improving each agent's information and judging each agent's decision in isolation are not enough. Recovery information should be paired with a mechanism that gives agents different inputs or roles\cite{hong2004,kleinberg2021}, and the resulting demand should be tested against capacity and deadlines rather than against unanimity, much as engineering practice spreads retries with jitter and limits probes during recovery\cite{microsoft_cb,brooker2019}. Reasoning did not remove the problem, since Sol and Opus both concentrated on A after sharing. Which inputs produce different choices is an empirical property of each model: Sol and Gemini showed substantial allocation gains from private numbers without instruction, Luna's improvement was small and not significant, and Haiku continued to choose A unanimously. Meanwhile, Opus refused many requests that contained a number. In these shared-record allocation tests, Haiku distributed requests only under an explicit rule. Even when Sol and Haiku followed the assignment rule exactly, the number of jobs directed to each service varied across groups, and some capacity went unused.

This suggests a role for a coordinating AI that can see pending jobs across the group. Using current capacity estimates, it could assign each job to a service and update the capacity still available before assigning the next job. A direct test would give the coordinator and a simple capacity-based scheduler the same observations and compare how many jobs each completes by the deadline under delayed reporting.

The experiments cover two services and one persistent recovery, with local changes to capacity, demand, and remaining time. In the controlled branches, agents could not observe one another's current choices. Moreover, agents were instructed to complete their own jobs, and we did not test an explicitly collective objective. Cross-model comparisons describe the tested systems, whose reasoning modes and output limits differed. Response rates also varied across acquisitions with identical prompts, so a response rate is not a fixed property of a model, and our intervals describe repeated generation within an acquisition, not variation across dates. Whether models adjust their use of numerical inputs to changing capacity ratios, and why Gemini shifted toward the smaller service, remain to be tested. Within this scope, the experiments identify where an intervention changes adaptation and quantify the work lost when useful individual responses leave collective capacity unused.

\sectionhead{Methods}

\subhead{Simulator and observations}

In the base environment, all 16 jobs arrived at round 0, required one successful request, and expired at the beginning of round 4. Pending agents (clients in the simulator) submitted one action per round in \{A, B, WAIT\}. Actions were collected before service admission, so the order in which API calls completed did not determine admission. Excess requests were rejected by a uniform lottery. Rejected jobs remained pending, completed clients exited, and there was no server queue or duplicate execution.

In the main recovery condition, A had capacities of 0, 0, 6, and 6 in rounds 0--3, and B had capacity two throughout. Pre-arrival observations at negative rounds showed A at six. Boundary experiments set A to six throughout or zero throughout. Capacity reports at round $t$ described round $t-d$, with delay $d=2$ unless specified otherwise. Optional request, success, and rejection counts described round $t-1$. Each decision used a fresh English prompt that combined the current shared report with the agent's serialized private state, including recent request outcomes, and requested a JSON action. Previous public reports and earlier model responses were not carried over as a conversation history, and agents could not communicate directly. The stated objective was for each client to complete its own job before its deadline.

\subhead{Interventions and experimental stages}

All conditions shared the basic service rules and a clarification that capacity is fixed within a round. KG added a 73-word reminder of existing rules, and KT added 69 words defining possible initial states, one possible persistent recovery of A, and constant capacity at B. The actual initial state, the recovery time, and its distribution were withheld. L0 and L1 encoded the same delayed observations using the layouts described in Results. Exact instructions appear in Supplementary Note~2.

Each environment seed set the admission lotteries and the internal agent identities of one simulated group, and conditions run with the same seed formed a matched block. Each stage below used a new set of seeds unless stated.

The factorial stage crossed two models, three instructions, two layouts, and request counts shown or hidden, with 12 seeds per condition (288 trajectories). When request counts were hidden in free trajectories, active-client counts and private outcomes remained visible.

Experiments with constructed histories used 12 seeds. Exploration batches stopped after the free round-2 decision by the 12 pending clients. For sharing, code sent the pending client with the lowest identifier to A and the other 11 to B at round 2, and then branched the nine-client state at round 3. S0 contained an empty success-record list, whereas S1 contained one record with \texttt{service=A}, \texttt{observed\_at=2}, and \texttt{success\_observed=true}; both explained that an empty list meant the record was withheld. Both branches hid request and active-client counts so that agents could not infer recovery from population attrition; the registered population of 16 remained visible. Within each branch, all nine pending agents received the same serialized private history and report. Coded actions passed through the same simulator step as free actions, so each agent's history listed them as its own requests with their actual outcomes, without marking them as assigned (Supplementary Note~2 reproduces complete round-3 prompts). The same seeds were also used for free trajectories that repeated the recovery episodes or kept A always available or never available (Supplementary Note~1 and Supplementary Table~3). The B-only prefix isolates first exploration and evidence use conditional on that history, whereas free trajectories also measure retries after failures at A.

The transfer experiments crossed delays 1 and 2, K0 and KT, L0 and L1, and Luna and Haiku. Gemini was tested under L1 and delay 2, in free trajectories crossing K0 and KT with request counts shown or hidden and in matched sharing pairs.

Four subsequent experiments (F1, Sol; F2, active-agent counts; F3, partial distribution rule; F4, report components) reused the constructed histories and seeds of the constructed-history stage with newly acquired responses, delay 2, and request counts hidden (Supplementary Notes~8 and 9). The competitor-count experiment varied only whether \texttt{active\_clients=9} appeared. The distribution-rule experiment held an agent-specific private draw fixed across branches; the added instruction assigned B when the draw was below 0.22 and otherwise left the decision free. The report-component experiment compared L0 and L1 with four intermediate layouts under KT (Supplementary Tables~13 and 14). No response from an earlier acquisition was used as a contemporary control.

R1 and R2 used 24 seeds under KT/L1 and delay 2 (Supplementary Notes~11 to 13). R1 compared the count-shown condition across no rule, a partial rule, and a complete rule in Sol, Luna, and Haiku; Sol also included a count-hidden no-rule branch. Every branch shared the success record and displayed a private draw. The partial-rule instruction retained the wording from F3, and the complete-rule instruction changed only the free-choice clause to request A. R2 crossed Luna and Haiku with four environments and withheld or shared success records. The base environment retained capacities of 6 and 2, nine pending jobs from 16 registered jobs, and one decision before expiry. The variants changed A's recovered capacity to three, started with 13 registered jobs (leaving six pending), or allowed a second decision before expiry. Request counts and active-client counts were hidden throughout R2, and no private draws or distribution rules were supplied. For a pending population $n$, capacities $a,b$, and $h$ remaining decisions, the feasible maximum was $C^{*}=\min\{n,h(a+b)\}$. Avoidable unfinished jobs were defined as $C^{*}$ minus actual completions by expiry.

The private-number test, R3, used 24 seeds with Sol under KT/L1, with the success record shared and both competitor and request counts hidden. Neither branch included a distribution rule; the sole difference in input was the numeric \texttt{private\_draw} field (Supplementary Note~14). The five-model comparison used 24 seeds under the same conditions and crossed Luna, Haiku, Sol, Opus, and Gemini with S0, S1, and S1R. In S0 and S1 the visible prompt was identical across blocks and models; S1R added only the \texttt{private\_draw} field. The grouping test used 24 seeds with the same environment and shared record, in models chosen from the five-model results: Sol and Gemini, whose completions had improved with a private number, and Haiku, which had not, followed by Luna in a separate stage. Twenty-four uniform random numbers were generated once. Condition C gave each group one number for all nine agents; condition I arranged the same numbers so that each group received nine distinct numbers and each number appeared in nine groups. The multisets of numbers and complete prompts were therefore identical between C and I, and I is a balanced arrangement of a finite set rather than an independent draw. In no condition were agents told whether others' numbers matched theirs (Supplementary Note~15).

\subhead{Models, acquisition, and response failures}

Luna used temperature 1, no reasoning, and a 256-token output limit. Haiku used adaptive reasoning at low effort, with a 1,024-token limit in the factorial stage and 2,048 tokens in later stages. Gemini used temperature 1, minimal thinking, and a 1,024-token limit. Sol used medium reasoning effort and an 8,192-token limit, and Opus used adaptive thinking at medium effort with the same limit. Haiku's later limit and Sol's limit were selected by diagnostics run before behavioral acquisition (Supplementary Notes~6 and 8). These provider-specific settings were operational configurations rather than matched reasoning effort (Supplementary Table~7). Responses were acquired on 14, 15, 18, and 25 October 2026 (UTC); provider identifiers do not guarantee fixed model weights.

Strict parsing used the final answer. Invalid and refused responses were executed as WAIT. Communication failures were handled by documented reacquisition rules that did not depend on the content of responses, and excluded attempts were stored separately (Supplementary Notes~6 and 13 and Supplementary Table~10). In the two primary models, nonvalid outputs occurred in none of the 1,728 controlled exploration decisions and in one of the 2,592 matched sharing decisions. In the factorial stage, Luna's highest nonvalid rate was 6.6\%, with the expanded report and counts shown. A control with a 1,024-token limit for Luna under KT/L1 lowered nonvalid rates and missed deadlines slightly (Supplementary Note~7 and Supplementary Table~6), so that factorial cell combines behavioral responses with output-limit effects. Finally, Opus returned explicit refusals in a diagnostic before acquisition, and refusals were recorded as such and executed as WAIT (Supplementary Note~15).

Across all behavioral datasets, 64,302 decisions were accepted from 66,460 requests, and 2,158 were excluded; diagnostic calls are counted separately. Reproducibility checks replayed the saved observations, prompts, parsed actions, admission lotteries, and expiry accounting. Supplementary Note~6 lists the seeds, sample sizes, model settings, diagnostics, replacements, and exclusions of every stage.

\subhead{Outcomes and statistical analysis}

In the original 16-job setting, the missed-deadline rate was the number of unfinished jobs at expiry divided by 16. Restricted delay was the mean of completion round minus arrival round, with unfinished jobs assigned a value of four rounds. Exploration was the fraction of the 12 pending clients that chose A in the round-2 control. The sharing effect was the S1 $-$ S0 difference in the fraction of the nine remaining clients that chose A at round 3. Unused final-round capacity was eight minus the number of completions in that round. Complete condition summaries and secondary process measures appear in Supplementary Tables~1 to 4.

Analyses followed protocols frozen internally before each acquisition stage; none was publicly preregistered. The inferential unit was the matched acquisition block (12 or 24 per contrast) rather than the individual agent. Conditions with identical prompts across blocks measure repeated model generations from fixed inputs. Earlier stages used paired bootstrap intervals, with nominal confidence levels adjusted for families of primary contrasts. The five-model and grouping stages used exact sign-flip permutation tests with Holm correction and descriptive bootstrap intervals. Stage-specific comparisons, resampling procedures, multiplicity adjustments, post hoc checks, and allocation benchmarks are detailed in Supplementary Notes~4 and 5, with post hoc sign tests in Supplementary Table~9.

\noindent\textbf{Use of generative AI.} We used generative AI tools to help edit the manuscript and write code to run the experiments and generate the figures.

\subhead{Data availability}
The article and Supplementary Information report aggregate data.

\begin{singlespace}
\small
\raggedright

\end{singlespace}

\begin{singlespace}
\small
\sectionhead{Acknowledgements}
This work was partially supported by JSPS KAKENHI Grant Number JP25H01000.

\sectionhead{Author contributions}
T.E. designed research, performed research, analysed data and wrote the paper. N.I. and K.N. supervised the research. T.E., N.I. and K.N. reviewed and edited the paper.

\sectionhead{Competing interests}
The authors declare no competing interests.

\noindent\hyperlink{si-start}{\textbf{Supplementary Information:}} Supplementary Notes~1 to 15, Supplementary Tables 1 to 25, and Supplementary Figs.~1 to 4.

\end{singlespace}
\clearpage
\newgeometry{margin=2cm}
\singlespacing
\renewcommand{\rmdefault}{ptm}
\renewcommand{\ttdefault}{pcr}
\normalfont\normalsize
\setlength{\parindent}{0pt}
\setlength{\parskip}{4pt}
\setlength{\LTleft}{0pt}
\setlength{\LTright}{0pt plus 1fill}
\setcounter{secnumdepth}{0}
\makeatletter
\def\toclevel@subsection{1}
\def\toclevel@subsubsection{2}
\makeatother
\setcounter{table}{0}
\setcounter{figure}{0}
\renewcommand{\theHtable}{supp.\arabic{table}}
\renewcommand{\theHfigure}{supp.\arabic{figure}}
\captionsetup[figure]{name=Supplementary Fig.}
\pdfbookmark[0]{Supplementary Information}{supplementary-information}
\hypertarget{si-start}{}
\sloppy
\begin{center}
{\large Supplementary Information for\par}
\vspace{0.5em}
{\LARGE\bfseries Groups of AI agents underuse resources despite shared evidence of increased capacity\par}
\vspace{0.8em}
Takahiro Ezaki$^{\ast}$, Naoto Imura and Katsuhiro Nishinari\par

\vspace{0.3em}
{\small $^\ast$Corresponding author. Email: tkezaki@g.ecc.u-tokyo.ac.jp\par}
\end{center}
\vspace{1em}
\noindent\textbf{The Supplementary Information includes:}\\
Supplementary Notes 1 to 15\\
Supplementary Tables 1 to 25\\
Supplementary Figures 1 to 4
\par\Needspace{32mm}

These materials provide exact instructions, complete condition summaries,
prespecified contrasts, acquisition settings, and response-failure checks.
Tables use the completed analysis files and display percentages to two
decimal places.

\noindent\textbf{Guide to these materials.} Notes 1--7 cover the basic design and original results; Notes 8--9 cover additional controls for reasoning, competitor counts, allocation rules, and report components; Notes 10--14 cover the response-probability analysis, allocation interventions, and environmental boundaries; and Note 15 covers the five-model comparison and the grouping of numerical inputs. Note 4 collects the statistical methods for all stages. The dataset inventory in Note 1 and the tables in Notes 3 and 4 cover the original three stages.

\hypertarget{terminology-and-retained-analysis-labels}{%
\subsection{Terminology and retained analysis
labels}\label{terminology-and-retained-analysis-labels}}

\begin{longtable}[]{@{}
  >{\raggedright\arraybackslash}p{(\columnwidth - 2\tabcolsep) * \real{0.5000}}
  >{\raggedright\arraybackslash}p{(\columnwidth - 2\tabcolsep) * \real{0.5000}}@{}}
\toprule\noalign{}
\begin{minipage}[b]{\linewidth}\raggedright
Manuscript term
\end{minipage} & \begin{minipage}[b]{\linewidth}\raggedright
Saved label or definition
\end{minipage} \\
\midrule\noalign{}
\endhead
\bottomrule\noalign{}
\endlastfoot
Decision round & \texttt{slot} in the simulator; decisions at rounds
0--3 \\
Request counts & Previous-round request, success, and rejection counts
for each service; called \texttt{aggregate\ outcomes} in earlier
records \\
Recovery description & \texttt{KT}, previously called
\texttt{transition\ knowledge}; 69 words describing possible capacity
changes \\
Reminder & \texttt{KG}; 73 words reminding agents of existing report
rules \\
Original / expanded report & \texttt{L0} / \texttt{L1}; L1 combines
per-service observation times, source labels, repeated observations, and
a confirmation field \\
Identical constructed histories & \texttt{common-state} design in the
analysis records \\
Success record withheld / shared & \texttt{S0} / \texttt{S1}; the
original \texttt{disclosure} comparison \\
Decision branch & One final-round group decision with the success record
withheld or shared; \texttt{endpoint} in the original allocation analysis \\
Matched pair & One shared-record and one withheld-record decision branch
from the same constructed history \\
Request counts shown / hidden & \texttt{full} / \texttt{capacity\_\allowbreak{}only};
private outcomes remain available \\
\end{longtable}

The exact instruction blocks, field names, and source-file paths retain their acquisition wording; table rows use the manuscript terms (for example, counts shown or hidden for \texttt{full} or \texttt{capacity\_only}).

\Needspace{40mm}
\hypertarget{s1.-dataset-inventory-and-prospective-specifications}{%
\subsection{Supplementary Note 1. Dataset inventory and prospective
specifications}\label{s1.-dataset-inventory-and-prospective-specifications}}

\begin{longtable}[]{@{}
  >{\raggedright\arraybackslash}p{(\columnwidth - 14\tabcolsep) * \real{0.1250}}
  >{\raggedright\arraybackslash}p{(\columnwidth - 14\tabcolsep) * \real{0.1250}}
  >{\raggedright\arraybackslash}p{(\columnwidth - 14\tabcolsep) * \real{0.1250}}
  >{\raggedright\arraybackslash}p{(\columnwidth - 14\tabcolsep) * \real{0.1250}}
  >{\raggedright\arraybackslash}p{(\columnwidth - 14\tabcolsep) * \real{0.1250}}
  >{\raggedright\arraybackslash}p{(\columnwidth - 14\tabcolsep) * \real{0.1250}}
  >{\raggedright\arraybackslash}p{(\columnwidth - 14\tabcolsep) * \real{0.1250}}
  >{\raggedright\arraybackslash}p{(\columnwidth - 14\tabcolsep) * \real{0.1250}}@{}}
\toprule\noalign{}
\begin{minipage}[b]{\linewidth}\raggedright
Stage
\end{minipage} & \begin{minipage}[b]{\linewidth}\raggedright
Free trajectories
\end{minipage} & \begin{minipage}[b]{\linewidth}\raggedright
Exploration batches
\end{minipage} & \begin{minipage}[b]{\linewidth}\raggedright
Success-record pairs
\end{minipage} & \begin{minipage}[b]{\linewidth}\raggedright
Decision branches
\end{minipage} & \begin{minipage}[b]{\linewidth}\raggedright
Analyzed API decisions
\end{minipage} & \begin{minipage}[b]{\linewidth}\raggedright
Acquired requests
\end{minipage} & \begin{minipage}[b]{\linewidth}\raggedright
Excluded requests
\end{minipage} \\
\midrule\noalign{}
\endhead
\bottomrule\noalign{}
\endlastfoot
Factorial and auxiliary & 384 & 0 & 0 & 0 & 20759 & 21113 & 354 \\
Constructed-history and boundary & 312 & 144 & 144 & 288 & 18877 & 19900 &
1023 \\
Timing and third model & 48 & 192 & 216 & 432 & 8918 & 9049 & 131 \\
Total & 744 & 336 & 360 & 720 & 48554 & 50062 & 1508 \\
\end{longtable}

Every condition in the original factorial, constructed-history, and delay/model-transfer stages summarized here contains 12 environment seeds. The 192
output-limit diagnostic responses are not included above. The original
diagnostic, earlier exploratory experiments, and incomplete acquisition
attempts remain separate. Each stage had a written protocol that fixed
its design, analysis, and contrasts before acquisition. These were internal stage-specific freezes, not public preregistrations.

\hypertarget{s2.-exact-instructions}{%
\subsection{Supplementary Note 2. Exact instructions}\label{s2.-exact-instructions}}

The following instruction text is copied from the original factorial
freeze. The report and private state were appended to these instructions
at each decision; their values depended on the current simulator state.

\hypertarget{base-instruction-original-layout-k0l0}{%
\subsubsection{Base instruction, original layout
(K0/L0)}\label{base-instruction-original-layout-k0l0}}

\begin{Verbatim}[breaklines=true,fontsize=\scriptsize,frame=single,framerule=0.3pt]
You operate one client's jobs in a simulated processing service.
Your objective is to maximize the fraction of your client's jobs completed before their deadlines.
At each integer slot you may send the oldest pending job to service A or B, or WAIT.
Each service handles at most its true capacity of requests in that slot. Excess requests are
rejected using a fresh uniform lottery. No queue or duplicate execution remains on the server.
Each job needs one successful request. A rejected job remains pending until its exclusive
expiry slot; at that slot it is expired before actions. You can make at most one request per slot.
Reported capacities and status describe the capacity_observed_at slot, not necessarily now.
Capacity can change over time. Negative observation slots refer to the period before jobs arrived.
Any previous traffic or diagnostic evidence describes traffic_observed_at, not a forecast.
Diagnostic labels mean: success_observed = at least one request succeeded;
requests_without_success = requests occurred but none succeeded; unobserved = no requests occurred.
Missing fields provide no observation. All active clients receive the same report.
Others' current choices are unknown. No direct communication.
Choose using the report and your client's information. Output only JSON: {"action":"A"},
{"action":"B"}, or {"action":"WAIT"}.
True capacity remains fixed throughout each integer slot; a capacity report stamped with the current slot reports that slot's true capacity.
\end{Verbatim}

\hypertarget{base-instruction-expanded-layout-k0l1}{%
\subsubsection{Base instruction, expanded layout
(K0/L1)}\label{base-instruction-expanded-layout-k0l1}}

In K0/L1, replace the following K0/L0 sentence:
\begin{Verbatim}[breaklines=true,fontsize=\scriptsize,frame=single,framerule=0.3pt]
Reported capacities and status describe the capacity_observed_at slot, not necessarily now.
\end{Verbatim}
with:
\begin{Verbatim}[breaklines=true,fontsize=\scriptsize,frame=single,framerule=0.3pt]
Each service's reported capacity and status describe its own capacity_observed_at slot, not necessarily now.
\end{Verbatim}

\hypertarget{additional-instruction-kg}{%
\subsubsection{Additional instruction
KG}\label{additional-instruction-kg}}

\begin{Verbatim}[breaklines=true,fontsize=\scriptsize,frame=single,framerule=0.3pt]
Reminder about the report: a reported capacity belongs to its observation slot rather than necessarily the decision slot.
The observation slot may be earlier than the decision slot. Capacity may change over time.
Negative observation slots refer to the period before jobs arrived. A missing field gives no observation.
Each request is subject to the service's capacity and the uniform admission lottery in its own slot.
Your aim remains completion before the deadline.
\end{Verbatim}

\hypertarget{additional-instruction-kt}{%
\subsubsection{Additional instruction
KT}\label{additional-instruction-kt}}

\begin{Verbatim}[breaklines=true,fontsize=\scriptsize,frame=single,framerule=0.3pt]
Common service contract: At slot 0, A's capacity is either 0 or its public nominal capacity 6.
From slot 0 onward, A may change from 0 to 6 once; after reaching 6 it never decreases within this episode.
Pre-arrival capacity 6 at negative slots does not guarantee availability at slot 0.
B has capacity 2 throughout. The initial true state, recovery time and its probability distribution are not provided.
\end{Verbatim}

\hypertarget{additional-instruction-kb}{%
\subsubsection{Additional instruction
KB}\label{additional-instruction-kb}}

\begin{Verbatim}[breaklines=true,fontsize=\scriptsize,frame=single,framerule=0.3pt]
Common service contract: At slot 0, A's capacity is either 0 or its public nominal capacity 6.
From slot 0 onward, A may change from 0 to 6 once; after reaching 6 it never decreases within this episode.
Pre-arrival capacity 6 at negative slots does not guarantee availability at slot 0.
B has capacity 2 throughout. The initial true state, recovery time and its probability distribution are not provided.
The platform may replace one active client's request with a confirmation request to A.
Confirmation requests enter the same uniform admission lottery as all other requests and have no priority.
Their public records specify the requester, observation slot and actual outcome, and become visible only at the next slot.
A rejection does not prove zero capacity. A successful confirmation at nonnegative slot s establishes capacity 6 at s
under the binary-capacity contract, and persistence thereafter under the nondecreasing contract.
Each service's capacity_reported and status_reported describe its own capacity_observed_at slot.
raw_capacity_report preserves the original delayed report. capacity_origin identifies the source of the primary report.
The platform may derive a primary capacity report from these same records and this common contract.
No confirmation records means no observation; it does not mean that a confirmation failed.
\end{Verbatim}

\hypertarget{shared-sharing-legend}{%
\subsubsection{Shared sharing legend}\label{shared-sharing-legend}}

\begin{Verbatim}[breaklines=true,fontsize=\scriptsize,frame=single,framerule=0.3pt]
Public success records refer only to successful processing in their stated previous slot. An empty record list means no disclosure, not an observation of zero successes.
\end{Verbatim}

S0 uses \texttt{public\_\allowbreak{}success\_\allowbreak{}records={[}{]}}. S1 uses one record
with \texttt{service=A}, \texttt{observed\_\allowbreak{}at=2}, and
\texttt{success\_\allowbreak{}observed=true}. The same record legend appears in both
branches. Request counts and active-client counts are absent from both
branches.

\hypertarget{complete-round-3-prompts}{%
\subsubsection{Complete round-3 prompts in the sharing experiment}\label{complete-round-3-prompts}}

The following prompt was regenerated from the frozen simulator for seed 3106
(KT, expanded layout L1, success record shared, agent 4). The coded prefix
(rounds 0--2) appears in the agent's private history as its own rejected
requests to B; nothing marks these actions as assigned. The acquisition
replay checks that every saved prompt equals the regenerated prompt
for its state.

\begin{Verbatim}[breaklines=true,fontsize=\scriptsize,frame=single,framerule=0.3pt]
You operate one client's jobs in a simulated processing service.
Your objective is to maximize the fraction of your client's jobs completed before their deadlines.
At each integer slot you may send the oldest pending job to service A or B, or WAIT.
Each service handles at most its true capacity of requests in that slot. Excess requests are
rejected using a fresh uniform lottery. No queue or duplicate execution remains on the server.
Each job needs one successful request. A rejected job remains pending until its exclusive
expiry slot; at that slot it is expired before actions. You can make at most one request per slot.
Each service's reported capacity and status describe its own capacity_observed_at slot, not necessarily now.
Capacity can change over time. Negative observation slots refer to the period before jobs arrived.
Any previous traffic or diagnostic evidence describes traffic_observed_at, not a forecast.
Diagnostic labels mean: success_observed = at least one request succeeded;
requests_without_success = requests occurred but none succeeded; unobserved = no requests occurred.
Missing fields provide no observation. All active clients receive the same report.
Others' current choices are unknown. No direct communication.
Choose using the report and your client's information. Output only JSON: {"action":"A"},
{"action":"B"}, or {"action":"WAIT"}.
True capacity remains fixed throughout each integer slot; a capacity report stamped with the current slot reports that slot's true capacity.
Common service contract: At slot 0, A's capacity is either 0 or its public nominal capacity 6.
From slot 0 onward, A may change from 0 to 6 once; after reaching 6 it never decreases within this episode.
Pre-arrival capacity 6 at negative slots does not guarantee availability at slot 0.
B has capacity 2 throughout. The initial true state, recovery time and its probability distribution are not provided.
Public success records refer only to successful processing in their stated previous slot. An empty record list means no disclosure, not an observation of zero successes.

SHARED REPORT
A.capacity_observed_at = 1
A.capacity_origin = "delayed_report"
A.capacity_reported = 0
A.status_reported = "unavailable"
B.capacity_observed_at = 1
B.capacity_origin = "delayed_report"
B.capacity_reported = 2
B.status_reported = "operational"
confirmation_records = []
public_success_records = [{"service": "A", "observed_at": 2, "success_observed": true}]
raw_capacity_report.A = {"capacity_reported": 0, "status_reported": "unavailable", "capacity_observed_at": 1}
raw_capacity_report.B = {"capacity_reported": 2, "status_reported": "operational", "capacity_observed_at": 1}
registered_clients = 16
slot = 3
traffic_observed_at = 2

PRIVATE CLIENT STATE
{"attempts_this_job": 3, "pending_jobs": 1, "recent_history": [{"action": "B", "outcome": "rejected", "slot": 0}, {"action": "B", "outcome": "rejected", "slot": 1}, {"action": "B", "outcome": "rejected", "slot": 2}], "remaining_slots": 1}
\end{Verbatim}

Under the compact layout L0, the instruction block uses the K0/L0 base text
above, followed by the same KT and record-legend text, and the report and
private state read:

\begin{Verbatim}[breaklines=true,fontsize=\scriptsize,frame=single,framerule=0.3pt]
SHARED REPORT
A.capacity_reported = 0
A.status_reported = "unavailable"
B.capacity_reported = 2
B.status_reported = "operational"
capacity_observed_at = 1
public_success_records = [{"service": "A", "observed_at": 2, "success_observed": true}]
registered_clients = 16
slot = 3
traffic_observed_at = 2

PRIVATE CLIENT STATE
{"attempts_this_job": 3, "pending_jobs": 1, "recent_history": [{"action": "B", "outcome": "rejected", "slot": 0}, {"action": "B", "outcome": "rejected", "slot": 1}, {"action": "B", "outcome": "rejected", "slot": 2}], "remaining_slots": 1}
\end{Verbatim}

With the record withheld (S0), \texttt{public\_\allowbreak{}success\_\allowbreak{}records = []}
replaces the record line; all other input is identical between S0 and S1 within each layout.

\hypertarget{s3.-complete-condition-summaries}{%
\subsection{Supplementary Note 3. Complete condition
summaries}\label{s3.-complete-condition-summaries}}

\textbf{Supplementary Table 1. Factorial and auxiliary free trajectories.} Deadline
rates refer to all 16 jobs. Restricted delay is measured in rounds,
assigning four to an unfinished job. The \texttt{factorial},
\texttt{bridge}, and corrected-extraction rows identify distinct designs;
only the first is the main factorial experiment.

\begin{longtable}[]{@{}
  >{\raggedright\arraybackslash}p{(\columnwidth - 18\tabcolsep) * \real{0.1000}}
  >{\raggedright\arraybackslash}p{(\columnwidth - 18\tabcolsep) * \real{0.1000}}
  >{\raggedright\arraybackslash}p{(\columnwidth - 18\tabcolsep) * \real{0.1000}}
  >{\raggedright\arraybackslash}p{(\columnwidth - 18\tabcolsep) * \real{0.1000}}
  >{\raggedright\arraybackslash}p{(\columnwidth - 18\tabcolsep) * \real{0.1000}}
  >{\raggedright\arraybackslash}p{(\columnwidth - 18\tabcolsep) * \real{0.1000}}
  >{\raggedright\arraybackslash}p{(\columnwidth - 18\tabcolsep) * \real{0.1000}}
  >{\raggedright\arraybackslash}p{(\columnwidth - 18\tabcolsep) * \real{0.1000}}
  >{\raggedright\arraybackslash}p{(\columnwidth - 18\tabcolsep) * \real{0.1000}}
  >{\raggedright\arraybackslash}p{(\columnwidth - 18\tabcolsep) * \real{0.1000}}@{}}
\toprule\noalign{}
\begin{minipage}[b]{\linewidth}\raggedright
Design
\end{minipage} & \begin{minipage}[b]{\linewidth}\raggedright
Model
\end{minipage} & \begin{minipage}[b]{\linewidth}\raggedright
Knowledge
\end{minipage} & \begin{minipage}[b]{\linewidth}\raggedright
Layout
\end{minipage} & \begin{minipage}[b]{\linewidth}\raggedright
Request counts
\end{minipage} & \begin{minipage}[b]{\linewidth}\raggedright
Lag
\end{minipage} & \begin{minipage}[b]{\linewidth}\raggedright
n
\end{minipage} & \begin{minipage}[b]{\linewidth}\raggedright
Missed deadlines
\end{minipage} & \begin{minipage}[b]{\linewidth}\raggedright
Restricted delay
\end{minipage} & \begin{minipage}[b]{\linewidth}\raggedright
Timely A evidence
\end{minipage} \\
\midrule\noalign{}
\endhead
\bottomrule\noalign{}
\endlastfoot
bridge & Haiku & KB & L1 & counts hidden & 2 & 12 & 21.88\% & 2.2396 &
100.00\% \\
bridge & Haiku & KB & L1 & counts shown & 2 & 12 & 1.56\% & 2.0938 & 100.00\% \\
factorial & Haiku & K0 & L0 & counts hidden & 2 & 12 & 62.50\% & 3.2500
& 0.00\% \\
factorial & Haiku & K0 & L0 & counts shown & 2 & 12 & 20.83\% & 2.6510 &
100.00\% \\
factorial & Haiku & K0 & L1 & counts hidden & 2 & 12 & 73.96\% & 3.5938
& 0.00\% \\
factorial & Haiku & K0 & L1 & counts shown & 2 & 12 & 19.79\% & 2.5573 &
75.00\% \\
factorial & Haiku & KG & L0 & counts hidden & 2 & 12 & 63.54\% & 3.2812
& 0.00\% \\
factorial & Haiku & KG & L0 & counts shown & 2 & 12 & 25.00\% & 2.6510 &
83.33\% \\
factorial & Haiku & KG & L1 & counts hidden & 2 & 12 & 73.44\% & 3.5781
& 0.00\% \\
factorial & Haiku & KG & L1 & counts shown & 2 & 12 & 51.04\% & 3.1198 &
16.67\% \\
factorial & Haiku & KT & L0 & counts hidden & 2 & 12 & 23.96\% & 2.6146
& 58.33\% \\
factorial & Haiku & KT & L0 & counts shown & 2 & 12 & 3.12\% & 2.1927 &
100.00\% \\
factorial & Haiku & KT & L1 & counts hidden & 2 & 12 & 40.10\% & 2.6146
& 83.33\% \\
factorial & Haiku & KT & L1 & counts shown & 2 & 12 & 2.60\% & 2.0000 &
100.00\% \\
bridge & Luna & KB & L1 & counts hidden & 2 & 12 & 19.27\% & 2.2500 &
100.00\% \\
bridge & Luna & KB & L1 & counts shown & 2 & 12 & 25.52\% & 2.3906 & 66.67\% \\
factorial & Luna & K0 & L0 & counts hidden & 2 & 12 & 63.54\% & 3.2812
& 0.00\% \\
factorial & Luna & K0 & L0 & counts shown & 2 & 12 & 60.94\% & 3.2344 &
0.00\% \\
factorial & Luna & K0 & L1 & counts hidden & 2 & 12 & 58.85\% & 3.1042
& 0.00\% \\
factorial & Luna & K0 & L1 & counts shown & 2 & 12 & 46.88\% & 2.7708 &
33.33\% \\
factorial & Luna & KG & L0 & counts hidden & 2 & 12 & 68.23\% & 3.4219
& 0.00\% \\
factorial & Luna & KG & L0 & counts shown & 2 & 12 & 51.56\% & 3.1198 &
33.33\% \\
factorial & Luna & KG & L1 & counts hidden & 2 & 12 & 61.46\% & 3.2083
& 0.00\% \\
factorial & Luna & KG & L1 & counts shown & 2 & 12 & 61.98\% & 3.2292 &
0.00\% \\
factorial & Luna & KT & L0 & counts hidden & 2 & 12 & 56.25\% & 3.0208
& 8.33\% \\
factorial & Luna & KT & L0 & counts shown & 2 & 12 & 6.77\% & 2.1615 &
83.33\% \\
factorial & Luna & KT & L1 & counts hidden & 2 & 12 & 15.10\% & 2.0625
& 100.00\% \\
factorial & Luna & KT & L1 & counts shown & 2 & 12 & 15.62\% & 2.1458 &
91.67\% \\
corrected extraction & Luna & K0 & L0 & counts hidden & 0 & 12 & 0.00\% &
2.0000 & 100.00\% \\
corrected extraction & Luna & K0 & L0 & counts hidden & 2 & 12 & 64.58\% &
3.3125 & 0.00\% \\
corrected extraction & Luna & K0 & L0 & counts shown & 0 & 12 & 2.60\% & 1.9688 &
100.00\% \\
corrected extraction & Luna & K0 & L0 & counts shown & 2 & 12 & 60.94\% & 3.2344 &
0.00\% \\
\end{longtable}

Timely evidence is the cohort fraction with an A success that can be shared before expiry. With zero report delay, no decision occurs when A has recovered but is still
reported at zero; the recovered-but-reported-zero exploration rate is therefore undefined.

\textbf{Supplementary Table 2. All common-state exploration and sharing cells.}
Exploration percentages use 12 remaining clients; sharing A-choice
percentages use nine. Each cell has 12 environment seeds. A dash marks an outcome that does not apply, because exploration batches stop at round 2. 

\begin{longtable}[]{@{}
  >{\raggedright\arraybackslash}p{(\columnwidth - 14\tabcolsep) * \real{0.1250}}
  >{\raggedright\arraybackslash}p{(\columnwidth - 14\tabcolsep) * \real{0.1250}}
  >{\raggedright\arraybackslash}p{(\columnwidth - 14\tabcolsep) * \real{0.1250}}
  >{\raggedright\arraybackslash}p{(\columnwidth - 14\tabcolsep) * \real{0.1250}}
  >{\raggedright\arraybackslash}p{(\columnwidth - 14\tabcolsep) * \real{0.1250}}
  >{\raggedright\arraybackslash}p{(\columnwidth - 14\tabcolsep) * \real{0.1250}}
  >{\raggedright\arraybackslash}p{(\columnwidth - 14\tabcolsep) * \real{0.1250}}
  >{\raggedright\arraybackslash}p{(\columnwidth - 14\tabcolsep) * \real{0.1250}}@{}}
\toprule\noalign{}
\begin{minipage}[b]{\linewidth}\raggedright
Design
\end{minipage} & \begin{minipage}[b]{\linewidth}\raggedright
Model
\end{minipage} & \begin{minipage}[b]{\linewidth}\raggedright
Knowledge
\end{minipage} & \begin{minipage}[b]{\linewidth}\raggedright
Layout
\end{minipage} & \begin{minipage}[b]{\linewidth}\raggedright
Branch
\end{minipage} & \begin{minipage}[b]{\linewidth}\raggedright
n
\end{minipage} & \begin{minipage}[b]{\linewidth}\raggedright
A choice
\end{minipage} & \begin{minipage}[b]{\linewidth}\raggedright
Missed deadlines
\end{minipage} \\
\midrule\noalign{}
\endhead
\bottomrule\noalign{}
\endlastfoot
exploration & Luna & K0 & L0 & --- & 12 & 0.00\% & --- \\
exploration & Luna & KG & L0 & --- & 12 & 0.00\% & --- \\
exploration & Luna & KT & L0 & --- & 12 & 0.69\% & --- \\
exploration & Luna & K0 & L1 & --- & 12 & 0.00\% & --- \\
exploration & Luna & KG & L1 & --- & 12 & 0.00\% & --- \\
exploration & Luna & KT & L1 & --- & 12 & 58.33\% & --- \\
exploration & Haiku & K0 & L0 & --- & 12 & 0.00\% & --- \\
exploration & Haiku & KG & L0 & --- & 12 & 0.00\% & --- \\
exploration & Haiku & KT & L0 & --- & 12 & 1.39\% & --- \\
exploration & Haiku & K0 & L1 & --- & 12 & 0.00\% & --- \\
exploration & Haiku & KG & L1 & --- & 12 & 0.00\% & --- \\
exploration & Haiku & KT & L1 & --- & 12 & 99.31\% & --- \\
sharing & Luna & K0 & L0 & S0 & 12 & 0.00\% & 43.75\% \\
sharing & Luna & K0 & L0 & S1 & 12 & 60.19\% & 12.50\% \\
sharing & Luna & KG & L0 & S0 & 12 & 0.00\% & 43.75\% \\
sharing & Luna & KG & L0 & S1 & 12 & 55.56\% & 14.06\% \\
sharing & Luna & KT & L0 & S0 & 12 & 8.33\% & 39.06\% \\
sharing & Luna & KT & L0 & S1 & 12 & 74.07\% & 10.42\% \\
sharing & Luna & K0 & L1 & S0 & 12 & 9.26\% & 38.54\% \\
sharing & Luna & K0 & L1 & S1 & 12 & 97.22\% & 17.19\% \\
sharing & Luna & KG & L1 & S0 & 12 & 0.00\% & 43.75\% \\
sharing & Luna & KG & L1 & S1 & 12 & 97.22\% & 17.19\% \\
sharing & Luna & KT & L1 & S0 & 12 & 26.85\% & 29.17\% \\
sharing & Luna & KT & L1 & S1 & 12 & 98.15\% & 17.71\% \\
sharing & Haiku & K0 & L0 & S0 & 12 & 0.00\% & 43.75\% \\
sharing & Haiku & K0 & L0 & S1 & 12 & 0.93\% & 43.23\% \\
sharing & Haiku & KG & L0 & S0 & 12 & 0.00\% & 43.75\% \\
sharing & Haiku & KG & L0 & S1 & 12 & 100.00\% & 18.75\% \\
sharing & Haiku & KT & L0 & S0 & 12 & 24.07\% & 30.21\% \\
sharing & Haiku & KT & L0 & S1 & 12 & 100.00\% & 18.75\% \\
sharing & Haiku & K0 & L1 & S0 & 12 & 0.00\% & 43.75\% \\
sharing & Haiku & K0 & L1 & S1 & 12 & 52.78\% & 14.06\% \\
sharing & Haiku & KG & L1 & S0 & 12 & 0.00\% & 43.75\% \\
sharing & Haiku & KG & L1 & S1 & 12 & 100.00\% & 18.75\% \\
sharing & Haiku & KT & L1 & S0 & 12 & 25.93\% & 29.17\% \\
sharing & Haiku & KT & L1 & S1 & 12 & 100.00\% & 18.75\% \\
\end{longtable}

Round-2 completions in the exploration batches (A and B successes at the
free decision, from the same analysis records) show crowding of the
recovered service without any shared record. Under KT with L1, Haiku
completed 6.08 of the eight attainable jobs on average and sent no request
to B in 11 of 12 batches; Luna completed 7.50 and used both services in
every batch. All other exploration cells completed 2.00 to 2.17 jobs,
almost all at B. This description is post hoc.

\textbf{Supplementary Table 3. Free-trajectory repeats, output-limit controls, and
boundaries.} \texttt{repair} denotes newly acquired recovery
trajectories, and the output-limit rows use the additional Luna 1,024-token
limit. Boundary recovery rounds 0 and 4 denote normal service and no
recovery before expiry. The two response-failure columns report Luna decisions only; dashes for Haiku mean that this analysis was not run for Haiku, not that failures were zero. Response status in the recovery-description factorial cells appears in Supplementary Table 8, and the outcomes of the affected jobs are summarized in Supplementary Note 7.

\begin{longtable}[]{@{}
  >{\raggedright\arraybackslash}p{(\columnwidth - 20\tabcolsep) * \real{0.0909}}
  >{\raggedright\arraybackslash}p{(\columnwidth - 20\tabcolsep) * \real{0.0909}}
  >{\raggedright\arraybackslash}p{(\columnwidth - 20\tabcolsep) * \real{0.0909}}
  >{\raggedright\arraybackslash}p{(\columnwidth - 20\tabcolsep) * \real{0.0909}}
  >{\raggedright\arraybackslash}p{(\columnwidth - 20\tabcolsep) * \real{0.0909}}
  >{\raggedright\arraybackslash}p{(\columnwidth - 20\tabcolsep) * \real{0.0909}}
  >{\raggedright\arraybackslash}p{(\columnwidth - 20\tabcolsep) * \real{0.0909}}
  >{\raggedright\arraybackslash}p{(\columnwidth - 20\tabcolsep) * \real{0.0909}}
  >{\raggedright\arraybackslash}p{(\columnwidth - 20\tabcolsep) * \real{0.0909}}
  >{\raggedright\arraybackslash}p{(\columnwidth - 20\tabcolsep) * \real{0.0909}}
  >{\raggedright\arraybackslash}p{(\columnwidth - 20\tabcolsep) * \real{0.0909}}@{}}
\toprule\noalign{}
\begin{minipage}[b]{\linewidth}\raggedright
Design
\end{minipage} & \begin{minipage}[b]{\linewidth}\raggedright
Model
\end{minipage} & \begin{minipage}[b]{\linewidth}\raggedright
Knowledge
\end{minipage} & \begin{minipage}[b]{\linewidth}\raggedright
Layout
\end{minipage} & \begin{minipage}[b]{\linewidth}\raggedright
Request counts
\end{minipage} & \begin{minipage}[b]{\linewidth}\raggedright
Recovery round
\end{minipage} & \begin{minipage}[b]{\linewidth}\raggedright
n
\end{minipage} & \begin{minipage}[b]{\linewidth}\raggedright
Missed deadlines
\end{minipage} & \begin{minipage}[b]{\linewidth}\raggedright
Restricted delay
\end{minipage} & \begin{minipage}[b]{\linewidth}\raggedright
Luna invalid
\end{minipage} & \begin{minipage}[b]{\linewidth}\raggedright
Luna refusal
\end{minipage} \\
\midrule\noalign{}
\endhead
\bottomrule\noalign{}
\endlastfoot
repair & Luna & K0 & L0 & counts shown & 2 & 12 & 62.50\% & 3.2500 & 0.57\% &
0.43\% \\
repair & Luna & KG & L0 & counts shown & 2 & 12 & 54.69\% & 3.1302 & 0.00\% &
0.00\% \\
repair & Luna & KT & L0 & counts shown & 2 & 12 & 7.81\% & 2.1771 & 0.34\% &
0.00\% \\
repair & Luna & KT & L0 & counts hidden & 2 & 12 & 55.21\% & 2.9479 &
0.00\% & 0.00\% \\
repair & Luna & K0 & L1 & counts shown & 2 & 12 & 46.88\% & 2.7969 & 0.78\% &
0.63\% \\
repair & Luna & KG & L1 & counts shown & 2 & 12 & 61.46\% & 3.2083 & 0.29\% &
0.14\% \\
repair & Luna & KT & L1 & counts shown & 2 & 12 & 14.58\% & 2.1562 & 4.33\% &
6.40\% \\
repair & Luna & KT & L1 & counts hidden & 2 & 12 & 13.54\% & 2.0208 &
1.08\% & 0.72\% \\
boundary & Luna & K0 & L0 & counts shown & 0 & 12 & 0.00\% & 1.2500 & 0.00\% &
0.00\% \\
boundary & Luna & K0 & L0 & counts shown & 4 & 12 & 62.50\% & 3.2500 & 0.00\% &
0.14\% \\
boundary & Luna & K0 & L0 & counts hidden & 0 & 12 & 0.00\% & 0.8333 &
0.00\% & 0.00\% \\
boundary & Luna & K0 & L0 & counts hidden & 4 & 12 & 66.67\% & 3.3750 &
0.00\% & 0.00\% \\
boundary & Luna & KT & L0 & counts shown & 0 & 12 & 0.00\% & 1.1719 & 0.00\% &
0.00\% \\
boundary & Luna & KT & L0 & counts shown & 4 & 12 & 54.69\% & 2.8906 & 0.16\% &
0.16\% \\
boundary & Luna & KT & L0 & counts hidden & 0 & 12 & 0.00\% & 0.6198 &
0.00\% & 0.00\% \\
boundary & Luna & KT & L0 & counts hidden & 4 & 12 & 56.25\% & 2.9844 &
0.00\% & 0.00\% \\
boundary & Haiku & K0 & L0 & counts shown & 0 & 12 & 0.00\% & 1.0208 & --- &
--- \\
boundary & Haiku & K0 & L0 & counts shown & 4 & 12 & 62.50\% & 3.2500 & --- &
--- \\
boundary & Haiku & K0 & L0 & counts hidden & 0 & 12 & 0.00\% & 0.8958 &
--- & --- \\
boundary & Haiku & K0 & L0 & counts hidden & 4 & 12 & 62.50\% & 3.2500
& --- & --- \\
boundary & Haiku & KT & L0 & counts shown & 0 & 12 & 0.00\% & 0.7708 & --- &
--- \\
boundary & Haiku & KT & L0 & counts shown & 4 & 12 & 62.50\% & 3.0104 & --- &
--- \\
boundary & Haiku & KT & L0 & counts hidden & 0 & 12 & 0.00\% & 0.6562 &
--- & --- \\
boundary & Haiku & KT & L0 & counts hidden & 4 & 12 & 58.85\% & 3.0417
& --- & --- \\
\end{longtable}

\Needspace{47mm}
\textbf{Supplementary Table 4. All timing-transfer and third-model cells.} Delay-2 cells use new seeds 3301--3312 and are not pooled with the preceding acquisition.

\begin{longtable}[]{@{}
  >{\raggedright\arraybackslash}p{(\columnwidth - 18\tabcolsep) * \real{0.1000}}
  >{\raggedright\arraybackslash}p{(\columnwidth - 18\tabcolsep) * \real{0.1000}}
  >{\raggedright\arraybackslash}p{(\columnwidth - 18\tabcolsep) * \real{0.1000}}
  >{\raggedright\arraybackslash}p{(\columnwidth - 18\tabcolsep) * \real{0.1000}}
  >{\raggedright\arraybackslash}p{(\columnwidth - 18\tabcolsep) * \real{0.1000}}
  >{\raggedright\arraybackslash}p{(\columnwidth - 18\tabcolsep) * \real{0.1000}}
  >{\raggedright\arraybackslash}p{(\columnwidth - 18\tabcolsep) * \real{0.1000}}
  >{\raggedright\arraybackslash}p{(\columnwidth - 18\tabcolsep) * \real{0.1000}}
  >{\raggedright\arraybackslash}p{(\columnwidth - 18\tabcolsep) * \real{0.1000}}
  >{\raggedright\arraybackslash}p{(\columnwidth - 18\tabcolsep) * \real{0.1000}}@{}}
\toprule\noalign{}
\begin{minipage}[b]{\linewidth}\raggedright
Design
\end{minipage} & \begin{minipage}[b]{\linewidth}\raggedright
Model
\end{minipage} & \begin{minipage}[b]{\linewidth}\raggedright
Knowledge
\end{minipage} & \begin{minipage}[b]{\linewidth}\raggedright
Layout
\end{minipage} & \begin{minipage}[b]{\linewidth}\raggedright
Request counts
\end{minipage} & \begin{minipage}[b]{\linewidth}\raggedright
Lag
\end{minipage} & \begin{minipage}[b]{\linewidth}\raggedright
Branch
\end{minipage} & \begin{minipage}[b]{\linewidth}\raggedright
n
\end{minipage} & \begin{minipage}[b]{\linewidth}\raggedright
A choice
\end{minipage} & \begin{minipage}[b]{\linewidth}\raggedright
Missed deadlines
\end{minipage} \\
\midrule\noalign{}
\endhead
\bottomrule\noalign{}
\endlastfoot
sharing & Gemini & K0 & L1 & counts hidden & 2 & S0 & 12 & 7.41\% &
39.58\% \\
sharing & Gemini & K0 & L1 & counts hidden & 2 & S1 & 12 & 31.48\% &
27.60\% \\
sharing & Gemini & KT & L1 & counts hidden & 2 & S0 & 12 & 33.33\% &
25.00\% \\
sharing & Gemini & KT & L1 & counts hidden & 2 & S1 & 12 & 21.30\% &
31.77\% \\
sharing & Haiku & K0 & L0 & counts hidden & 1 & S0 & 12 & 100.00\% &
18.75\% \\
sharing & Haiku & K0 & L0 & counts hidden & 1 & S1 & 12 & 100.00\% &
18.75\% \\
sharing & Haiku & K0 & L0 & counts hidden & 2 & S0 & 12 & 0.00\% &
43.75\% \\
sharing & Haiku & K0 & L0 & counts hidden & 2 & S1 & 12 & 1.85\% &
42.71\% \\
sharing & Haiku & K0 & L1 & counts hidden & 1 & S0 & 12 & 100.00\% &
18.75\% \\
sharing & Haiku & K0 & L1 & counts hidden & 1 & S1 & 12 & 100.00\% &
18.75\% \\
sharing & Haiku & K0 & L1 & counts hidden & 2 & S0 & 12 & 0.00\% &
43.75\% \\
sharing & Haiku & K0 & L1 & counts hidden & 2 & S1 & 12 & 46.30\% &
18.23\% \\
sharing & Haiku & KT & L0 & counts hidden & 1 & S0 & 12 & 100.00\% &
18.75\% \\
sharing & Haiku & KT & L0 & counts hidden & 1 & S1 & 12 & 100.00\% &
18.75\% \\
sharing & Haiku & KT & L0 & counts hidden & 2 & S0 & 12 & 22.22\% &
31.25\% \\
sharing & Haiku & KT & L0 & counts hidden & 2 & S1 & 12 & 100.00\% &
18.75\% \\
sharing & Haiku & KT & L1 & counts hidden & 1 & S0 & 12 & 100.00\% &
18.75\% \\
sharing & Haiku & KT & L1 & counts hidden & 1 & S1 & 12 & 100.00\% &
18.75\% \\
sharing & Haiku & KT & L1 & counts hidden & 2 & S0 & 12 & 35.19\% &
23.96\% \\
sharing & Haiku & KT & L1 & counts hidden & 2 & S1 & 12 & 100.00\% &
18.75\% \\
sharing & Luna & K0 & L0 & counts hidden & 1 & S0 & 12 & 100.00\% &
18.75\% \\
sharing & Luna & K0 & L0 & counts hidden & 1 & S1 & 12 & 100.00\% &
18.75\% \\
sharing & Luna & K0 & L0 & counts hidden & 2 & S0 & 12 & 0.00\% &
43.75\% \\
sharing & Luna & K0 & L0 & counts hidden & 2 & S1 & 12 & 53.70\% &
16.67\% \\
sharing & Luna & K0 & L1 & counts hidden & 1 & S0 & 12 & 100.00\% &
18.75\% \\
sharing & Luna & K0 & L1 & counts hidden & 1 & S1 & 12 & 100.00\% &
18.75\% \\
sharing & Luna & K0 & L1 & counts hidden & 2 & S0 & 12 & 12.04\% &
36.98\% \\
sharing & Luna & K0 & L1 & counts hidden & 2 & S1 & 12 & 94.44\% &
16.15\% \\
sharing & Luna & KT & L0 & counts hidden & 1 & S0 & 12 & 93.52\% &
15.62\% \\
sharing & Luna & KT & L0 & counts hidden & 1 & S1 & 12 & 85.19\% &
11.46\% \\
sharing & Luna & KT & L0 & counts hidden & 2 & S0 & 12 & 8.33\% &
39.06\% \\
sharing & Luna & KT & L0 & counts hidden & 2 & S1 & 12 & 54.63\% &
13.02\% \\
sharing & Luna & KT & L1 & counts hidden & 1 & S0 & 12 & 87.96\% &
13.02\% \\
sharing & Luna & KT & L1 & counts hidden & 1 & S1 & 12 & 100.00\% &
18.75\% \\
sharing & Luna & KT & L1 & counts hidden & 2 & S0 & 12 & 50.93\% &
17.71\% \\
sharing & Luna & KT & L1 & counts hidden & 2 & S1 & 12 & 100.00\% &
18.75\% \\
exploration & Haiku & K0 & L0 & counts hidden & 1 & --- & 12 & 0.00\% &
--- \\
exploration & Haiku & K0 & L0 & counts hidden & 2 & --- & 12 & 0.00\% &
--- \\
exploration & Haiku & K0 & L1 & counts hidden & 1 & --- & 12 & 0.00\% &
--- \\
exploration & Haiku & K0 & L1 & counts hidden & 2 & --- & 12 & 0.00\% &
--- \\
exploration & Haiku & KT & L0 & counts hidden & 1 & --- & 12 & 9.03\% &
--- \\
exploration & Haiku & KT & L0 & counts hidden & 2 & --- & 12 & 1.39\% &
--- \\
exploration & Haiku & KT & L1 & counts hidden & 1 & --- & 12 & 89.58\%
& --- \\
exploration & Haiku & KT & L1 & counts hidden & 2 & --- & 12 & 97.92\%
& --- \\
exploration & Luna & K0 & L0 & counts hidden & 1 & --- & 12 & 0.00\% &
--- \\
exploration & Luna & K0 & L0 & counts hidden & 2 & --- & 12 & 0.00\% &
--- \\
exploration & Luna & K0 & L1 & counts hidden & 1 & --- & 12 & 0.00\% &
--- \\
exploration & Luna & K0 & L1 & counts hidden & 2 & --- & 12 & 1.39\% &
--- \\
exploration & Luna & KT & L0 & counts hidden & 1 & --- & 12 & 7.64\% &
--- \\
exploration & Luna & KT & L0 & counts hidden & 2 & --- & 12 & 2.78\% &
--- \\
exploration & Luna & KT & L1 & counts hidden & 1 & --- & 12 & 6.25\% &
--- \\
exploration & Luna & KT & L1 & counts hidden & 2 & --- & 12 & 59.03\% &
--- \\
natural & Gemini & K0 & L1 & counts hidden & 2 & --- & 12 & --- &
75.00\% \\
natural & Gemini & K0 & L1 & counts shown & 2 & --- & 12 & --- & 62.50\% \\
natural & Gemini & KT & L1 & counts hidden & 2 & --- & 12 & --- &
56.25\% \\
natural & Gemini & KT & L1 & counts shown & 2 & --- & 12 & --- & 46.35\% \\
\end{longtable}

\hypertarget{s4.-primary-and-secondary-contrasts}{%
\subsection{Supplementary Note 4. Statistical analysis and planned contrasts}\label{s4.-primary-and-secondary-contrasts}}

\subsubsection*{Independent units and inferential procedures}

The experiments were acquired in stages, each with its own protocol frozen internally before acquisition , and later stages were designed in light of earlier results. The inferential procedure followed the protocol of each stage. The 12-block stages and R1--R3 specified paired bootstrap intervals, at multiplicity-adjusted nominal levels for families of primary contrasts, as their primary tests. The five-model and grouping stages specified exact sign-flip permutation tests with Holm correction, reported with descriptive bootstrap intervals. We report each analysis as frozen rather than recomputing earlier stages under the later procedure; post hoc exact sign tests for the earlier primary contrasts (Supplementary Table~9) give a common directional check.

The original comparisons used 12 matched acquisition blocks indexed by simulator seed ($n = 12$ per contrast) as replicates, rather than individual agents. The seed changed admission lotteries and internal agent identities, while model generations remained stochastic, with no generation seed supplied. In constructed histories without private draws (Supplementary Note~10), and at the first decision of each R2 condition, the complete visible prompt was identical across blocks within a condition, so these blocks measure repeated generation from a fixed input rather than replication across distinct environments, and pairing matched acquisition indices rather than seed-dependent inputs. Free trajectories could develop different visible histories, and seeds coupled private draws in F3, R1, and R3; we retained the prespecified block pairing throughout. Where specified, contrasts were averaged across layouts or information conditions within each seed before resampling. Percentile intervals came from 10,000 paired bootstrap resamples of seeds with fixed analysis seeds. The two, four, and six primary contrasts in successive original stages used multiplicity-adjusted nominal levels of 97.5\%, 98.75\%, and 99.17\% (Supplementary Table~5). With 12 seeds, extreme percentile bounds have limited resolution, and adjusted nominal levels do not guarantee exact family-wise coverage. In the post hoc sign test (Supplementary Table~9), each of the six factorial and constructed-history contrasts showed 12/12 consistent signs (two-sided $P=0.000488$; stage-wise Bonferroni-adjusted $P<0.002$), whereas Luna's delay-1 exploration contrast had six positive differences and six ties (adjusted $P=0.1875$), and Gemini's sharing contrast had one positive and eight negative differences (adjusted $P=0.234$).

\subsubsection*{Prespecified contrasts by acquisition stage}

The subsequent experiments specified 15 additional primary contrasts (three for Sol, two for competitor counts, two for the distribution rule, and eight for report components), using the same 12-seed paired bootstrap at nominal levels of 98.33\%, 97.5\%, 97.5\%, and 99.375\% (Supplementary Table~14). Sol's completion contrast compared shared-branch completion with the attainable maximum of eight. We predicted positive effects for Sol's knowledge and sharing responses and for partial-rule minus no-rule completions (F3), and negative effects for both count-revelation contrasts; no direction was predicted for Sol's gap from eight or for the report components. Secondary contrasts used descriptive 95\% intervals (Supplementary Table~15). A zero-width bootstrap interval, which arises when every seed gives the same outcome, reflects that sample rather than certainty about future generations. Supplementary Table~9 reports post hoc exact sign tests for all 15 contrasts with within-stage Bonferroni adjustment.

R1 specified four primary contrasts: complete-rule minus partial-rule completions in each model, and the effect of count revelation on Sol's A choice. Only Luna's completion contrast carried a positive directional prediction. R2 specified six primary contrasts that compared each variant with its base shared-record condition in each model. We predicted a reduction in avoidable loss for the lower-load and extra-round variants, whereas the capacity contrasts were nondirectional. These two families of tests used 24-block paired bootstrap intervals at nominal levels of 98.75\% and 99.17\%, respectively. Six R1 rule-versus-no-rule comparisons and eight R2 shared-versus-withheld completion comparisons were treated as secondary and reported with descriptive 95\% intervals. Supplementary Table~20 lists every contrast, interval, and count of positive, negative, and tied block differences. The complete-rule benchmark and the always-A capacity formula were specified before data acquisition; they condition on fixed action policies.

R3 specified one primary comparison (private number present minus absent for completed jobs, with a positive prediction) and one secondary comparison (A-choice rate, with a negative prediction). Both used the 24-block paired bootstrap with 10,000 resamples and 95\% intervals; the secondary interval is descriptive. We compared the historical R1 prompts and examined the associations between private numbers and choices post hoc (Supplementary Note~14).

The five-model comparison specified ten two-sided planned comparisons, S1 minus S0 and S1R minus S1 in each model, evaluated with exact sign-flip permutation tests over the 24 paired blocks and Holm correction across the ten, together with descriptive 95\% paired bootstrap intervals. Models were not treated as a random sample, and differences between models were not tested directly. The grouping test specified I minus C and I minus N in each model, with Holm correction within each acquisition stage (six comparisons for the first three models, two for Luna) and, for the four-model display, across all eight; the within-stage results are the frozen primary analysis. Positive effects were predicted for Sol and Gemini; Haiku and Luna were analyzed identically without a directional prediction. Because each number appeared in nine groups under I, the grouping tests are conditional on the realized set of 24 numbers and its arrangement; they do not average over other sets of numbers.

\subsubsection*{Post hoc analyses and allocation benchmarks}

In a post hoc analysis of 72 final-round conditions with identical inputs, we compared observed mean completion with predictions from independently sampled choices. Supplementary Note~10 gives the held-out calculation and its limits.

The Gemini deadline contrast was post hoc. Supplementary Note 5 gives the independent-choice benchmark and the equilibrium references.

\textbf{Supplementary Table 5. All 12 primary contrasts.} Values and designated intervals are in percentage points.

\begin{longtable}[]{@{}
  >{\raggedright\arraybackslash}p{(\columnwidth - 10\tabcolsep) * \real{0.1667}}
  >{\raggedright\arraybackslash}p{(\columnwidth - 10\tabcolsep) * \real{0.1667}}
  >{\raggedright\arraybackslash}p{(\columnwidth - 10\tabcolsep) * \real{0.1667}}
  >{\raggedright\arraybackslash}p{(\columnwidth - 10\tabcolsep) * \real{0.1667}}
  >{\raggedright\arraybackslash}p{(\columnwidth - 10\tabcolsep) * \real{0.1667}}
  >{\raggedright\arraybackslash}p{(\columnwidth - 10\tabcolsep) * \real{0.1667}}@{}}
\toprule\noalign{}
\begin{minipage}[b]{\linewidth}\raggedright
Stage
\end{minipage} & \begin{minipage}[b]{\linewidth}\raggedright
Model
\end{minipage} & \begin{minipage}[b]{\linewidth}\raggedright
Comparison
\end{minipage} & \begin{minipage}[b]{\linewidth}\raggedright
Difference
\end{minipage} & \begin{minipage}[b]{\linewidth}\raggedright
Interval key
\end{minipage} & \begin{minipage}[b]{\linewidth}\raggedright
Interval
\end{minipage} \\
\midrule\noalign{}
\endhead
\bottomrule\noalign{}
\endlastfoot
factorial & Luna & KT-KG & −29.17 & 97.5\% & {[}−32.03, −26.56{]} \\
factorial & Haiku & KT-KG & −36.46 & 97.5\% & {[}−40.89, −32.81{]} \\
constructed history & Luna & Exploration KT $-$ K0 & +29.51 & 98.75\% & {[}22.92,
35.59{]} \\
constructed history & Luna & Sharing effect on A choice, KT & +68.52 & 98.75\% & {[}59.26,
77.31{]} \\
constructed history & Haiku & Exploration KT $-$ K0 & +50.35 & 98.75\% & {[}50.00,
51.39{]} \\
constructed history & Haiku & Sharing effect on A choice, KT & +75.00 & 98.75\% & {[}69.91,
79.63{]} \\
delay transfer & Luna & Exploration KT $-$ K0, delay 1, L1 & +6.25 & 99.17\% &
{[}1.39, 12.50{]} \\
delay transfer & Luna & Sharing effect, delay 1 $-$ delay 2, KT & −45.83 & 99.17\% &
{[}−58.03, −34.26{]} \\
delay transfer & Haiku & Exploration KT $-$ K0, delay 1, L1 & +89.58 & 99.17\% &
{[}84.03, 95.14{]} \\
delay transfer & Haiku & Sharing effect, delay 1 $-$ delay 2, KT & −71.30 & 99.17\% &
{[}−77.31, −66.20{]} \\
delay transfer & Gemini & Missed deadlines KT $-$ K0 & −17.45 & 99.17\% &
{[}−20.83, −14.84{]} \\
delay transfer & Gemini & Sharing effect, KT & −12.04 & 99.17\% & {[}−25.93,
0.93{]} \\
\end{longtable}

The intervals are nominal 97.5\%, 98.75\%, and 99.17\% percentile bootstrap intervals for the two, four, and six primary contrasts in the respective stages. Each uses 10,000 matched-seed resamples after the specified within-seed averaging. 

\textbf{Supplementary Table 6. Selected prespecified secondary contrasts relevant to
the manuscript.} Intervals are descriptive 95\% intervals.

\begin{longtable}[]{@{}
  >{\raggedright\arraybackslash}p{(\columnwidth - 10\tabcolsep) * \real{0.1667}}
  >{\raggedright\arraybackslash}p{(\columnwidth - 10\tabcolsep) * \real{0.1667}}
  >{\raggedright\arraybackslash}p{(\columnwidth - 10\tabcolsep) * \real{0.1667}}
  >{\raggedright\arraybackslash}p{(\columnwidth - 10\tabcolsep) * \real{0.1667}}
  >{\raggedright\arraybackslash}p{(\columnwidth - 10\tabcolsep) * \real{0.1667}}
  >{\raggedright\arraybackslash}p{(\columnwidth - 10\tabcolsep) * \real{0.1667}}@{}}
\toprule\noalign{}
\begin{minipage}[b]{\linewidth}\raggedright
Stage
\end{minipage} & \begin{minipage}[b]{\linewidth}\raggedright
Model
\end{minipage} & \begin{minipage}[b]{\linewidth}\raggedright
Comparison
\end{minipage} & \begin{minipage}[b]{\linewidth}\raggedright
Condition
\end{minipage} & \begin{minipage}[b]{\linewidth}\raggedright
Difference
\end{minipage} & \begin{minipage}[b]{\linewidth}\raggedright
95\% interval
\end{minipage} \\
\midrule\noalign{}
\endhead
\bottomrule\noalign{}
\endlastfoot
constructed history & Luna & Exploration KG $-$ K0 & layouts averaged & +0.00 &
{[}0.00, 0.00{]} \\
constructed history & Luna & Sharing effect KT $-$ K0 & layouts averaged & −5.56 &
{[}−13.89, 3.24{]} \\
constructed history & Luna & Sharing effect KG $-$ K0 & layouts averaged & +2.31 &
{[}−6.02, 11.11{]} \\
constructed history & Haiku & Exploration KG $-$ K0 & layouts averaged & +0.00 &
{[}0.00, 0.00{]} \\
constructed history & Haiku & Sharing effect KT $-$ K0 & layouts averaged & +48.15
& {[}42.13, 54.63{]} \\
constructed history & Haiku & Sharing effect KG $-$ K0 & layouts averaged & +73.15
& {[}68.52, 77.78{]} \\
delay transfer & Luna & Exploration L1 $-$ L0, KT, delay 1 & delay 1 & −1.39 &
{[}−6.94, 4.17{]} \\
delay transfer & Haiku & Exploration L1 $-$ L0, KT, delay 1 & delay 1 & +80.56 &
{[}73.61, 87.52{]} \\
boundary & Haiku & KT $-$ K0 & counts hidden & −3.65 & {[}−7.29, 0.52{]} \\
boundary & Haiku & KT $-$ K0 & counts shown & +0.00 & {[}−1.56, 1.56{]} \\
boundary & Luna & KT $-$ K0 & counts hidden & −10.42 & {[}−14.58,
−5.21{]} \\
boundary & Luna & KT $-$ K0 & counts shown & −7.81 & {[}−10.94, −4.17{]} \\
output limit & Luna & 1,024 $-$ 256 tokens & counts shown & −4.69 & {[}−10.42,
1.04{]} \\
output limit & Luna & 1,024 $-$ 256 tokens & counts hidden & −2.08 &
{[}−3.65, −0.52{]} \\
\end{longtable}

The Gemini KT sharing loss comparison was summarized post hoc: +6.77
percentage points (95\% interval 1.56 to 12.50), with 12 paired seeds.
The prespecified third-model outcome was A choice, whose positive
prediction failed. 

\hypertarget{s5.-request-allocation-and-unused-capacity}{%
\subsection{Supplementary Note 5. Allocation benchmarks}\label{s5.-request-allocation-and-unused-capacity}}

For nine independent A/B choices with probability $p$ of choosing A and no
WAIT, the number $M$ choosing A has a binomial distribution. Expected
final-round completions are
\[
 E[C]=\sum_{m=0}^{9}\binom{9}{m}p^m(1-p)^{9-m}
       \left[\min(m,6)+\min(9-m,2)\right].
\]
A grid search in increments of 0.001 gives a maximum near $p=0.741$ and
approximately 7.400 expected completions, below the coordinated upper bound
of eight.

A symmetric mixed-strategy equilibrium gives a second reference. Let each of
the nine agents choose A with probability $p$ and let $K\sim\mathrm{Binomial}(8,p)$
be the number of the other eight agents choosing A. A request to A succeeds
with probability $\min(1,6/(K+1))$ and a request to B with probability
$\min(1,2/(9-K))$. The equilibrium $p^{*}$ solves
\[
 \sum_{k=0}^{8}\binom{8}{k}p^k(1-p)^{8-k}
 \left[\min\!\left(1,\tfrac{6}{k+1}\right)-\min\!\left(1,\tfrac{2}{9-k}\right)\right]=0,
\]
which bisection gives as $p^{*}=0.811$, with success probability $0.810$ for
each agent and $E[C]=7.29$. The pure profile with seven requests to A and two
to B is a strict equilibrium: an A requester succeeds with probability $6/7$
and would obtain $2/3$ at B; a B requester succeeds with probability 1 and
would obtain $6/8$ at A. The observed all-A profile is not an equilibrium,
since each agent succeeds with probability $6/9$ and a unilateral move to B
succeeds with probability 1. Both references assume that agents know the
capacities and that nine agents remain, which the prompts did not state.

Seven jobs have already completed in this constructed history. The deadline-loss
fraction is therefore $(9-C)/16$, whereas A choice is measured among the nine
pending agents. The attainable minimum loss is $1/16=6.25\%$. An all-A
allocation gives $C=6$ and loss $3/16=18.75\%$: one unavoidable expiry and two
additional expiries from unused capacity at B.

\hypertarget{s6.-acquisition-settings-and-reproducibility-records}{%
\subsection{Supplementary Note 6. Acquisition settings and reproducibility
records}\label{s6.-acquisition-settings-and-reproducibility-records}}

\textbf{Supplementary Table 7. Requested model settings, copied from acquisition
freezes.} The additional Luna output-limit control is reported separately.

\begin{longtable}[]{@{}
  >{\raggedright\arraybackslash}p{(\columnwidth - 8\tabcolsep) * \real{0.2000}}
  >{\raggedright\arraybackslash}p{(\columnwidth - 8\tabcolsep) * \real{0.2000}}
  >{\raggedright\arraybackslash}p{(\columnwidth - 8\tabcolsep) * \real{0.2000}}
  >{\raggedright\arraybackslash}p{(\columnwidth - 8\tabcolsep) * \real{0.2000}}
  >{\raggedright\arraybackslash}p{(\columnwidth - 8\tabcolsep) * \real{0.2000}}@{}}
\toprule\noalign{}
\begin{minipage}[b]{\linewidth}\raggedright
Stage
\end{minipage} & \begin{minipage}[b]{\linewidth}\raggedright
Alias
\end{minipage} & \begin{minipage}[b]{\linewidth}\raggedright
Requested model ID
\end{minipage} & \begin{minipage}[b]{\linewidth}\raggedright
Output limit
\end{minipage} & \begin{minipage}[b]{\linewidth}\raggedright
Requested settings
\end{minipage} \\
\midrule\noalign{}
\endhead
\bottomrule\noalign{}
\endlastfoot
factorial & Luna & gpt-6-luna & 256 & \{``reasoning'': \{``effort'':
``none''\}, ``service\_\allowbreak{}tier'': ``default'', ``temperature'': 1.0\} \\
factorial & Haiku & claude-haiku-5-5 & 1024 & \{``effort'': ``low'',
``thinking'': \{``type'': ``adaptive''\}\} \\
constructed history & Luna & gpt-6-luna & 256 & \{``reasoning'': \{``effort'':
``none''\}, ``service\_\allowbreak{}tier'': ``default'', ``temperature'': 1.0\} \\
constructed history & Haiku & claude-haiku-5-5 & 2048 & \{``effort'': ``low'',
``thinking'': \{``type'': ``adaptive''\}\} \\
delay transfer & Luna & gpt-6-luna & 256 & \{``reasoning'': \{``effort'':
``none''\}, ``service\_\allowbreak{}tier'': ``default'', ``temperature'': 1.0\} \\
delay transfer & Haiku & claude-haiku-5-5 & 2048 & \{``effort'': ``low'',
``thinking'': \{``type'': ``adaptive''\}\} \\
delay transfer & Gemini & gemini-3.5-flash-lite & 1024 & \{``temperature'': 1,
``thinkingConfig'': \{``thinkingLevel'': ``minimal''\}\} \\
\end{longtable}

Settings for the 18 October experiments appear in Supplementary Note 8. A 192-response diagnostic selected output limits of 256 tokens for Luna and 2,048 for Haiku by a
frozen nonvalid-response rule: nonvalid counts were 1/48 and 2/48 for
Luna at 256 and 1,024 tokens, and 0/48 at each Haiku limit of 1,024 and 2,048; the tie rule selected the larger Haiku limit.

Transport failures were not interpreted as WAIT decisions, whereas invalid or refused model outputs were retained as WAIT according to the acquisition-stage rules. A corrected response extraction for Luna, a network replacement during the factorial stage, and a schema and transport replacement for the Gemini client are documented in the retained protocols and addenda; none changed any accepted response.

\par\Needspace{32mm}
\subsubsection*{Stage designs, seeds, and acquisition records}

This subsection lists the records that the Methods summarize.

\textit{Seeds and sample sizes.} The factorial stage used seeds 2301--2312 (288 trajectories, with 48 longer-contract and 48 corrected-extraction auxiliary trajectories listed in Supplementary Table~1). The constructed-history stage used seeds 3101--3112: 144 exploration batches and 144 matched pairs, plus 312 free trajectories with the same seeds (96 repeat recovery episodes, 24 Luna output-limit controls, 96 with A always available, and 96 with A unavailable throughout; Supplementary Table~3). The transfer stage used seeds 3301--3312: 192 exploration batches and 192 matched pairs in Luna and Haiku, plus 48 free trajectories and 24 matched pairs in Gemini. F1--F4 reused seeds 3101--3112 with new responses: 36 exploration batches and 24 sharing pairs in Sol, 48 matched pairs for the competitor count, 24 pairs for the distribution rule, and 144 exploration batches for the report components. R1 and R2 used seeds 5001--5024: 72 matched cohorts and 240 branches in R1, and 192 matched pairs in R2. R3 used seeds 6001--6024 (432 responses). The five-model comparison used seeds 7001--7024 (3,240 decisions), and the grouping test used seeds 8001--8024 (1,944 decisions in Sol, Gemini, and Haiku, and 648 in Luna).

\textit{Design details.} In F1--F4, the exploration tests retained the active-client count of 12, and the sharing tests hid the count of nine except in the count-revelation treatment. In the report-component experiment, each of T, TO, R, and C was compared with L0, except that TO was compared with T; L1 $-$ L0 was secondary. Cohort order was shuffled with a fixed seed, and the order of paired branches was varied by environment seed. In R2, the capacity variant changed both the capacity ratio and total capacity, and the load variant changed both the registered and pending populations. The extra-round variant allowed private outcomes and the ordinary delayed report to update, so it tested the additional decision together with its feedback rather than extra time alone. Jobs beyond the feasible maximum were not counted as allocation loss.

\textit{Private draws.} Draws were generated from a fixed hash of the environment seed and internal client identifier, rounded to 12 decimal places, and matched across models and branches; the identifier was not shown. R1 used a new hash namespace. In F3, 21 of 108 agent decisions were designated for B; sending those agents to B and all others to A gives $88/12=7.33$ mean completions, a benchmark not guaranteed under the partial rule. In R1, 42 of 216 agents were designated, and the same calculation gives 7.46 completions, below eight because the number of designated agents varies across blocks.

\textit{Model requests and replacements.} Haiku and Sol requests omitted temperature and top-p, leaving sampling to the provider's defaults; saved R1 metadata for Sol reported temperature 1.0. Sol used the default service tier. Acquisition used at most 16 concurrent calls. Eligible communication failures triggered reacquisition of the entire cohort (one model, configuration, and seed), with at most two attempts. In F1--F4, five cohorts required replacement, and all 50 calls from their failed attempts, including 36 valid responses, were excluded and stored separately (Supplementary Table~10). An initial R1 acquisition and an interrupted R2 acquisition are described in Supplementary Note~13, and the Opus diagnostic and nine excluded Gemini requests in Supplementary Note~15.

\textit{Nonvalid outputs and totals.} In the factorial cells with KT and request counts hidden, Haiku had no nonvalid outputs, and Luna had 17 among 559 decisions (3.0\%) with the expanded report and none among 664 with the compact report. In the delay-transfer stage, Luna produced four nonvalid outputs among 2,880 decisions, and Haiku and Gemini none. The original three datasets contain 744 free trajectories, 336 exploration batches, and 360 matched pairs, totaling 48,554 analyzed decisions from 50,062 requests. F1--F4 added 180 exploration batches and 96 pairs, or 3,888 behavioral decisions from 3,938 requests, with 48 diagnostic calls counted separately. R1 and R2 added 5,596 accepted decisions, all valid (2,160 in R1 and 3,436 in R2). All 432 R3 responses and all 2,592 grouping-test responses were valid. Across all behavioral datasets, 64,302 decisions were accepted from 66,460 requests, and 2,158 were excluded.

\subsection{Supplementary Note 7. Response validity, free histories, and directional sensitivity}

These post hoc checks use accepted acquisitions only and verify 2,024 stored
result and API files against their recorded SHA-256 hashes. Supplementary Table 8 separates valid
WAIT choices from invalid or refused outputs executed as WAIT. In particular,
Haiku's higher deadline loss under KT/L1 with request counts hidden had no
response failures. All 1,728 primary constructed-history exploration
responses were valid. Within every primary sharing branch, the nine pending
agents received identical prompts. They saw 16 registered clients, but no
active-client count or agent identifier in their private input.

Luna's response failures concentrated in the KT/L1 cells with request counts shown. A separate check of all 432 accepted Luna free trajectories (36 cells of 12 episodes) traced each nonvalid response to its executed WAIT action and to the outcome of the affected job. In the original factorial KT/L1 cell with counts shown, 17 invalid and 21 refused decisions among 574 responses (6.62\%) affected 38 jobs, of which 28 completed and 10 expired; 20 other jobs expired without any nonvalid response, and no failure occurred at a final decision. The repeat acquisition of that cell had 25 invalid and 37 refused decisions among 578 responses, with 10 of its 28 expired jobs affected, and the 1,024-token control had three invalid and 38 refused decisions among 577 responses, with 10 of 19 expired jobs affected. Raising the output limit reduced deadline losses by 4.69 percentage points in that comparison (95\% interval $-10.42$ to 1.04; Supplementary Table 6). These observed histories do not identify the number of jobs that replacing WAIT would have rescued, because replacement actions would also change congestion and subsequent histories. Response failures in all other acquisitions are summarized with their stages (Supplementary Table 10 and Supplementary Notes 13 to 15).

\textbf{Supplementary Table 8. Response status in the recovery-description factorial
conditions and primary exploration experiment.} Each factorial row contains
12 complete episodes; exploration rows pool the six instruction/layout cells
(72 batches per model). Nonvalid combines invalid and refused responses;
its denominator is decisions. Expired jobs use 192 jobs per factorial cell.
Exploration batches stop at the free choice and have no expiry outcome.

\begin{center}
\footnotesize
\begin{tabular}{lllrrrrr}
\toprule
Model & Layout & Counts & Decisions & Valid WAIT & Nonvalid & Nonvalid (\%) & Expired\\
\midrule
Luna & L0 & shown & 594 & 0 & 0 & 0.00 & 13 / 192\\
Luna & L1 & shown & 574 & 0 & 38 & 6.62 & 30 / 192\\
Luna & L0 & hidden & 664 & 0 & 0 & 0.00 & 108 / 192\\
Luna & L1 & hidden & 559 & 0 & 17 & 3.04 & 29 / 192\\
Haiku & L0 & shown & 607 & 2 & 0 & 0.00 & 6 / 192\\
Haiku & L1 & shown & 571 & 6 & 3 & 0.53 & 5 / 192\\
Haiku & L0 & hidden & 648 & 5 & 0 & 0.00 & 46 / 192\\
Haiku & L1 & hidden & 617 & 0 & 0 & 0.00 & 77 / 192\\
\midrule
Luna & \multicolumn{2}{l}{Exploration (all cells)} & 864 & 0 & 0 & 0.00 & --\\
Haiku & \multicolumn{2}{l}{Exploration (all cells)} & 864 & 0 & 0 & 0.00 & --\\
\bottomrule
\end{tabular}
\end{center}

The 24 Luna output-limit controls tested KT/L1 only. With counts shown,
nonvalid outputs fell from 62/578 (10.73\%) at 256 tokens to 41/577 (7.11\%)
at 1,024; deadline losses were 28/192 and 19/192. With counts hidden, nonvalid
outputs were 10/554 (1.81\%) and 9/554 (1.62\%); deadline losses were 26/192
and 22/192. The controls did not cross knowledge conditions or layouts and
cannot establish that those contrasts are insensitive to the output limit.

\begin{center}
\begin{minipage}{\textwidth}
\centering
\includegraphics[width=\linewidth]{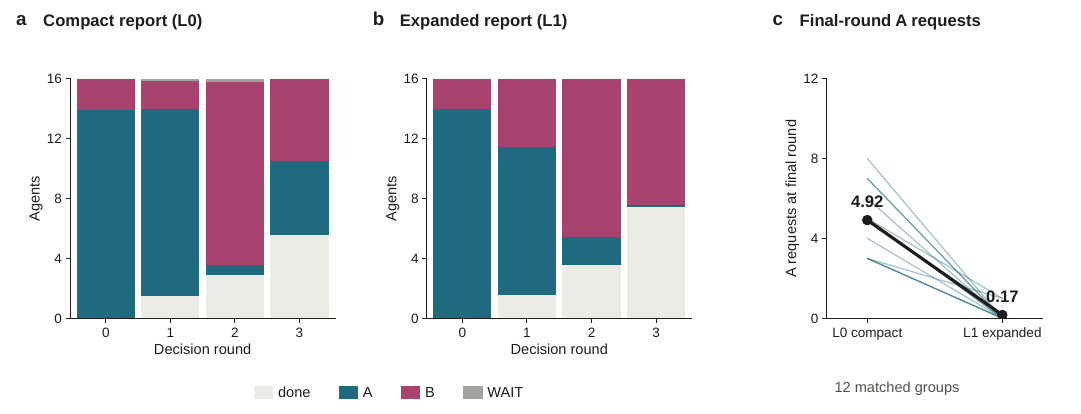}
\captionsetup{hypcap=false}
\captionof{figure}{\textbf{More early exploration need not produce continued
use in free episodes.} Haiku under the recovery description (KT), with request
counts hidden, in the original factorial acquisition (1,024-token limit).
\textbf{a, b}, Mean states of the 16 agents across 12 groups with compact
(L0) and expanded (L1) reports. WAIT represents valid choices; no response
failures occurred in either cell. \textbf{c}, Final-round A requests in the
same matched seeds (thin lines); thick line and labels show means. Totals
over groups at rounds 2 and 3 were 8 and 59 A requests with L0, and 22 and 2
with L1; at round 3 with L1, the other 101 of 103 requests went to B. These are freely evolving histories; they do not estimate the
isolated effect of a particular earlier request or population-count cue.}
\end{minipage}
\end{center}

In the round-2 delay comparison, both reports showed A at zero. A direct
prompt comparison at seed 3301 for both primary models (KT/L1) found that
only the observation timestamps differed, including their repeated raw-report
entries: round 1 under delay 1 and round 0 under delay 2. Private histories
were identical. Agents received no serialized sequence of earlier public
reports, so repeated exposure to consecutive zero reports is not an
identified mechanism in this control.

All 432 accepted Gemini sharing responses (216 K0 and 216 KT, each split
equally between S0 and S1) returned action-only JSON. Their response parts
contained opaque thought signatures, but no returned thought text or other
textual rationale. These signatures cannot establish crowd-avoidance
reasoning. Under KT, B requests averaged 6.00 per group without the record
and 7.08 with it, while A requests averaged 3.00 and 1.92; neither branch
contained WAIT.

\Needspace{72mm}
\textbf{Supplementary Table 9. Post hoc exact sign sensitivity for all primary contrasts in the original stages and in F1--F4.}
Positive, negative, and tied matched-seed differences use the contrast
orientation in Supplementary Table 5 (original stages) or Supplementary Table 14 (F1--F4). Ties are excluded. The null is equal probability of
positive and negative differences, conditional on a nonzero difference,
with signs independent across seed replicates. This tests directional
consistency, not the null of zero mean effect. For $n$ nonzero differences
and $k$ the smaller sign count, the two-sided value is
$P=\min\{1,2\sum_{j=0}^{k}\binom{n}{j}2^{-n}\}$; when all blocks tie, $P=1$ by convention.
Adjusted values multiply by the number of primary contrasts in that stage
(2, 4, and 6 in the original stages; 3, 2, 2, and 8 in F1--F4). 

\begin{center}
\footnotesize
\begin{tabular}{llp{4.4cm}rrrll}
\toprule
Stage & Model & Comparison & $+$ & $-$ & Ties & $P$ & Adjusted $P$\\
\midrule
Factorial & Luna & Deadline loss, KT $-$ KG & 0 & 12 & 0 & 0.000488 & 0.000977\\
Factorial & Haiku & Deadline loss, KT $-$ KG & 0 & 12 & 0 & 0.000488 & 0.000977\\
Controlled & Luna & Exploration, KT $-$ K0 & 12 & 0 & 0 & 0.000488 & 0.001953\\
Controlled & Luna & A choice, S1 $-$ S0 (KT) & 12 & 0 & 0 & 0.000488 & 0.001953\\
Controlled & Haiku & Exploration, KT $-$ K0 & 12 & 0 & 0 & 0.000488 & 0.001953\\
Controlled & Haiku & A choice, S1 $-$ S0 (KT) & 12 & 0 & 0 & 0.000488 & 0.001953\\
Transfer & Luna & Exploration, KT $-$ K0 (delay 1, L1) & 6 & 0 & 6 & 0.031250 & 0.187500\\
Transfer & Luna & Sharing effect, delay 1 $-$ delay 2 (KT) & 0 & 12 & 0 & 0.000488 & 0.002930\\
Transfer & Haiku & Exploration, KT $-$ K0 (delay 1, L1) & 12 & 0 & 0 & 0.000488 & 0.002930\\
Transfer & Haiku & Sharing effect, delay 1 $-$ delay 2 (KT) & 0 & 12 & 0 & 0.000488 & 0.002930\\
Transfer & Gemini & Deadline loss, KT $-$ K0 & 0 & 12 & 0 & 0.000488 & 0.002930\\
Transfer & Gemini & A choice, S1 $-$ S0 (KT) & 1 & 8 & 3 & 0.039062 & 0.234375\\
\midrule
F1 & Sol & Exploration, KT $-$ K0 (L1) & 10 & 0 & 2 & 0.001953 & 0.005859\\
F1 & Sol & A choice, shared $-$ withheld & 12 & 0 & 0 & 0.000488 & 0.001465\\
F1 & Sol & Shared completion $-$ 8 & 0 & 12 & 0 & 0.000488 & 0.001465\\
F2 & Luna & Count shown $-$ hidden & 4 & 5 & 3 & 1.000000 & 1.000000\\
F2 & Haiku & Count shown $-$ hidden & 0 & 0 & 12 & 1.000000 & 1.000000\\
F3 & Luna & Rule $-$ no rule & 1 & 9 & 2 & 0.021484 & 0.042969\\
F3 & Haiku & Rule $-$ no rule & 11 & 0 & 1 & 0.000977 & 0.001953\\
F4 & Luna & T $-$ L0 & 12 & 0 & 0 & 0.000488 & 0.003906\\
F4 & Luna & TO $-$ T & 3 & 6 & 3 & 0.507812 & 1.000000\\
F4 & Luna & R $-$ L0 & 7 & 0 & 5 & 0.015625 & 0.125000\\
F4 & Luna & C $-$ L0 & 6 & 0 & 6 & 0.031250 & 0.250000\\
F4 & Haiku & T $-$ L0 & 12 & 0 & 0 & 0.000488 & 0.003906\\
F4 & Haiku & TO $-$ T & 8 & 1 & 3 & 0.039062 & 0.312500\\
F4 & Haiku & R $-$ L0 & 12 & 0 & 0 & 0.000488 & 0.003906\\
F4 & Haiku & C $-$ L0 & 12 & 0 & 0 & 0.000488 & 0.003906\\
\bottomrule
\end{tabular}
\end{center}
\par\Needspace{32mm}
\subsection{Supplementary Note 8. Reasoning, competitor information, and distribution rules}

These four experiments were specified after the original analyses and acquired
on 18 October 2026 (UTC). The execution protocol, code, model configuration,
comparison family, and cohort order were frozen internally before each stage.
 Seeds 3101--3112 reused the physical
histories, with new API responses in every condition. Request counts were
hidden and report delay was two rounds throughout. Exploration displayed the
active-client count of 12. Sharing hid the count of nine except in the
competitor-count treatment. Every comparison used 12 matched environment seeds.

F1 repeated exploration and sharing with Sol (\texttt{gpt-6.1-sol}). F2 compared
count hidden versus \texttt{active\_clients=9}, with the same A-success record
in both branches. F3 compared a private draw alone against the same draw plus
a partial distribution rule, again with the success record in both branches.
In F2 and F3, saved branch names S0 and S1 mean control and treatment; they do
not mean withholding versus sharing the success record. F4 varied the report
components in exploration under KT. It acquired L0 and L1 contemporaneously
with T, TO, R, and C; its source-label contrast was TO minus T.

\textbf{Exact added instruction for the distribution-rule treatment.}
Both branches appended a numeric \texttt{private\_draw} to private state.
Only the treatment appended the following sentence to the instruction:
\begin{Verbatim}[breaklines=true,fontsize=\scriptsize,frame=single,framerule=0.3pt]
Distribution rule: if public_success_records shows a success at A and your private_draw is below 0.22, request B; otherwise follow your own judgment.
\end{Verbatim}

For each seed and internal client identifier, a deterministic hash generated
a uniform draw in $[0,1)$, rounded to 12 decimal places. Each agent's draw was held fixed
across branches and models; different agents received different values, and identifiers were not shown. The threshold was
exactly 0.22, an approximation to $2/9$. On average, 1.75 agents per group were designated for B; independent draws would designate 1.98 in expectation. Assigning those clients to B and all others to A would yield
7.33 mean completions. This input-based reference differs from the instruction,
which left the other agents' decisions free, and from a coordinated assignment
that fixes exactly two or three B requests and completes eight.

\textbf{Component changes in F4.}
T moved the global observation timestamp into each service's fields and used
the L1 instruction that each service's capacity and status refer to its own observation time. TO added
\texttt{capacity\_origin=delayed\_report} to T. R appended a
\texttt{raw\_capacity\_report} copy to L0; C appended
\texttt{confirmation\_records=[]} to L0. All delayed values stayed unchanged.
The T contrast therefore includes matching instruction wording. In this acquisition the complete expanded report gave Luna 41.0\% A choice, against 58.3\% and 59.0\% in two other acquisitions of the same input, so report components are compared only within an acquisition.

\textbf{Supplementary Table 10. Additional acquisition counts.}
Batches are 12-client exploration checkpoints; pairs contain two nine-client
final-round branches.
All accepted responses were valid.

\begin{longtable}{@{}lrrrrr@{}}
\toprule
Stage & Batches & Pairs & Accepted & Acquired & Excluded \\
\midrule
\endhead
\bottomrule
\endlastfoot
F1 & 36 & 24 & 864 & 890 & 26 \\
F2 & 0 & 48 & 864 & 864 & 0 \\
F3 & 0 & 24 & 432 & 432 & 0 \\
F4 & 144 & 0 & 1728 & 1752 & 24 \\
diagnostic & 0 & 0 & 48 & 48 & 0 \\
Total & 180 & 96 & 3936 & 3986 & 50 \\
\end{longtable}

Three Sol cohorts and two report-component cohorts required replacement.
The 50 excluded calls comprise 10 ambiguous communication errors, four HTTP
errors, and 36 valid responses in failed attempts. Reacquisition replaced
the entire cohort, at most once, after eligible communication failures.

Sol requested medium reasoning effort and the default service tier, with
an 8,192-token output limit in the behavioral stage. The limit was chosen by a 48-call diagnostic before behavioral acquisition, without using action choices. It compared limits of 4,096 and 8,192 tokens; both gave 24 valid responses, and the larger limit was selected by the tie rule. Luna retained no
reasoning, temperature 1, and 256 tokens; Haiku retained adaptive thinking,
low effort, and 2,048 tokens.

\subsection{Supplementary Note 9. Complete condition summaries and prespecified comparisons}

\textbf{Supplementary Table 11. Sol exploration and sharing.}
Each row contains 12 seeds. A choice is a percentage of 12 pending agents in
exploration and nine in sharing. Completion is at the tested round, not the
whole episode. All-A and full-capacity columns count seed batches or branches.
Nonvalid counts are numbers of decisions. The sharing record is withheld in
S0 and supplied in S1.

\begin{longtable}{@{}llllrrrrr@{}}
\toprule
Task & Knowledge & Layout & Record & A (\%) & Complete & All A & Full & Nonvalid \\
\midrule
\endhead
\bottomrule
\endlastfoot
Share & KT & L0 & S0 & 18.52 & 3.67 & 0 & 0 & 0 \\
Share & KT & L0 & S1 & 100.00 & 6.00 & 12 & 0 & 0 \\
Share & KT & L1 & S0 & 0.00 & 2.00 & 0 & 0 & 0 \\
Share & KT & L1 & S1 & 100.00 & 6.00 & 12 & 0 & 0 \\
Explore & K0 & L1 & -- & 0.00 & 2.00 & 0 & 0 & 0 \\
Explore & KT & L0 & -- & 9.03 & 3.08 & 0 & 0 & 0 \\
Explore & KT & L1 & -- & 20.83 & 4.50 & 0 & 0 & 0 \\
\end{longtable}

\textbf{Supplementary Table 12. Competitor counts and the private-draw rule.}
Every row contains 12 paired seeds and uses KT with a shared success record.
Completion is the final-round count. There were no nonvalid decisions.
\textit{A. Competitor-count conditions (F2).} Count hidden versus nine shown.

\begin{longtable}{@{}lllrrrr@{}}
\toprule
Model & Layout & Count & A (\%) & Complete & All A & Full \\
\midrule
\endhead
\bottomrule
\endlastfoot
Haiku & L0 & Hidden & 100.00 & 6.00 & 12 & 0 \\
Haiku & L0 & 9 shown & 100.00 & 6.00 & 12 & 0 \\
Haiku & L1 & Hidden & 100.00 & 6.00 & 12 & 0 \\
Haiku & L1 & 9 shown & 100.00 & 6.00 & 12 & 0 \\
Luna & L0 & Hidden & 71.30 & 7.67 & 0 & 9 \\
Luna & L0 & 9 shown & 91.67 & 6.75 & 5 & 2 \\
Luna & L1 & Hidden & 100.00 & 6.00 & 12 & 0 \\
Luna & L1 & 9 shown & 79.63 & 7.42 & 1 & 6 \\
\end{longtable}

\textit{B. Distribution-rule conditions (F3).} All use L1 and show the private
draw. Designated and other B choices use denominators 21 and 87, respectively,
over 108 decisions per row. WAIT denotes valid choices.

\begin{longtable}{@{}llrrrrrr@{}}
\toprule
Model & Rule & A (\%) & Complete & Designated B & Other B & WAIT & Full \\
\midrule
\endhead
\bottomrule
\endlastfoot
Haiku & Absent & 100.00 & 6.00 & 0/21 & 0/87 & 0 & 0 \\
Haiku & Present & 67.59 & 7.33 & 21/21 & 11/87 & 3 & 5 \\
Luna & Absent & 97.22 & 6.25 & 0/21 & 3/87 & 0 & 0 \\
Luna & Present & 37.04 & 5.25 & 21/21 & 47/87 & 0 & 1 \\
\end{longtable}

\textit{C. Descriptive capacity decomposition of F3.} Per-branch means;
unused capacity is six minus A completions or two minus B completions.

\begin{longtable}{@{}llrrrr@{}}
\toprule
Model & Rule & A complete & B complete & A unused & B unused \\
\midrule
\endhead
\bottomrule
\endlastfoot
Haiku & Absent & 6.00 & 0.00 & 0.00 & 2.00 \\
Haiku & Present & 5.58 & 1.75 & 0.42 & 0.25 \\
Luna & Absent & 6.00 & 0.25 & 0.00 & 1.75 \\
Luna & Present & 3.25 & 2.00 & 2.75 & 0.00 \\
\end{longtable}

\par\Needspace{32mm}
\textbf{Supplementary Table 13. Report-component exploration (F4).}
Each row contains 12 round-2 batches of 12 clients under KT.
L0 and L1 are the compact and expanded reports; T, TO, R, and C are defined
in Supplementary Note 8. Completion is at round 2. All-A and full-capacity columns
count batches; nonvalid counts decisions.

\begin{longtable}{@{}llrrrrr@{}}
\toprule
Model & Layout & A (\%) & Complete & All A & Full & Nonvalid \\
\midrule
\endhead
\bottomrule
\endlastfoot
Luna & L0 & 0.00 & 2.00 & 0 & 0 & 0 \\
Luna & T & 91.67 & 7.00 & 5 & 5 & 0 \\
Luna & TO & 87.50 & 7.17 & 3 & 5 & 0 \\
Luna & R & 5.56 & 2.67 & 0 & 0 & 0 \\
Luna & C & 5.56 & 2.67 & 0 & 0 & 0 \\
Luna & L1 & 40.97 & 6.83 & 0 & 6 & 0 \\
Haiku & L0 & 4.86 & 2.58 & 0 & 0 & 0 \\
Haiku & T & 75.69 & 7.67 & 1 & 9 & 0 \\
Haiku & TO & 88.89 & 7.08 & 3 & 4 & 0 \\
Haiku & R & 43.06 & 6.67 & 0 & 5 & 0 \\
Haiku & C & 58.33 & 7.75 & 0 & 10 & 0 \\
Haiku & L1 & 100.00 & 6.00 & 12 & 0 & 0 \\
\end{longtable}

\begin{center}
\includegraphics[width=170mm]{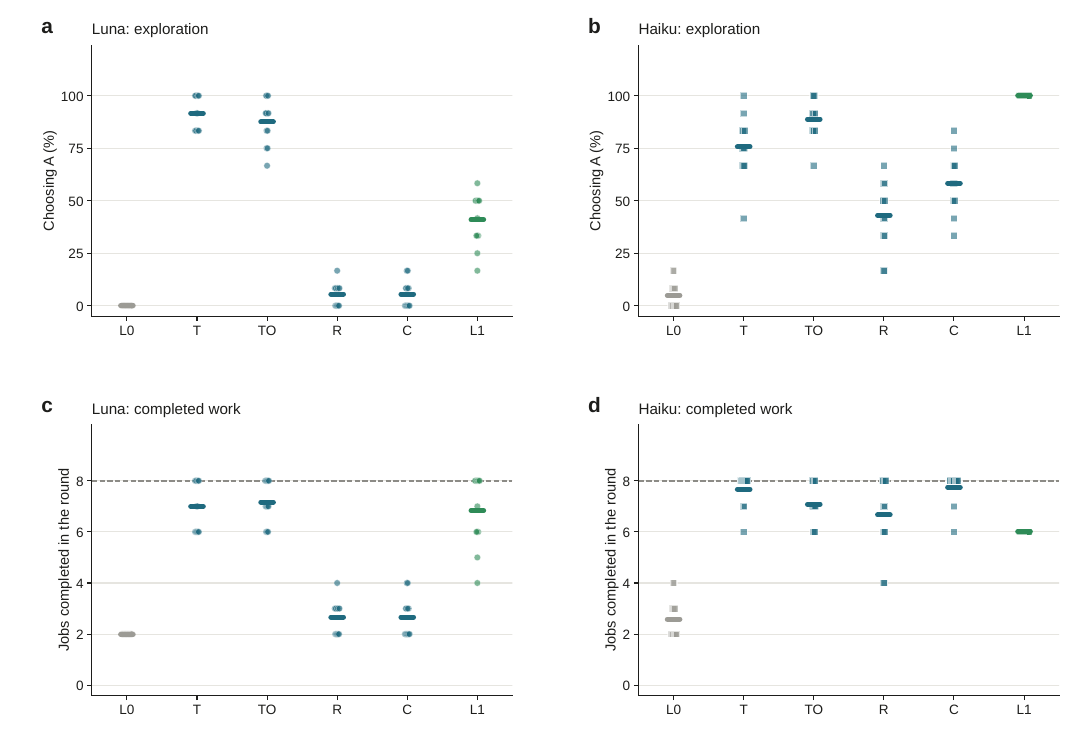}
\captionsetup{hypcap=false}
\captionof{figure}{\textbf{Report components change exploration and the work completed.}
(\textbf{a}, \textbf{b}) A-choice rates in Luna and Haiku under KT from the
same B-only history, with 12 active clients visible.
(\textbf{c}, \textbf{d}) Completions in that round. Points are the 12 seed
summaries per condition; short horizontal bars mark means. Dashed lines in c
and d mark the feasible eight completions. T gives each service its own
timestamp and states that its capacity and status refer to that time; TO adds origin labels to T;
R and C add a raw copy or empty confirmation field to L0. L1 combines all
components. All conditions were acquired anew. Higher exploration does not
always increase completed work.}
\label{fig:components}
\end{center}
\par\Needspace{32mm}
\textbf{Supplementary Table 14. All 15 primary comparisons in F1--F4.}
Each estimate uses 12 paired seed differences and 10,000 bootstrap resamples. Nominal coverage is $1-0.05/m$ for the $m$ primary
comparisons in that stage: three in F1, two each in F2 and F3, and eight in F4. pp denotes percentage
points; jobs denotes completions in the tested final round. Finite-sample limits of these intervals are discussed in Supplementary Note 4.

\begin{longtable}{@{}llp{40mm}lrrr@{}}
\toprule
Stage & Model & Contrast & Unit & Mean & Interval & Level (\%) \\
\midrule
\endhead
\bottomrule
\endlastfoot
F1 & Sol & KT $-$ K0 (L1) & pp & $+20.83$ & $+12.50$ to $+29.17$ & 98.333 \\
F1 & Sol & Shared $-$ withheld & pp & $+90.74$ & $+86.11$ to $+94.91$ & 98.333 \\
F1 & Sol & Shared completion $-$ 8 & jobs & $-2.00$ & $-2.00$ to $-2.00$ & 98.333 \\
F2 & Luna & Count shown $-$ hidden & pp & $+0.00$ & $-4.63$ to $+5.09$ & 97.500 \\
F2 & Haiku & Count shown $-$ hidden & pp & $+0.00$ & $+0.00$ to $+0.00$ & 97.500 \\
F3 & Luna & Rule $-$ no rule & jobs & $-1.00$ & $-1.75$ to $-0.17$ & 97.500 \\
F3 & Haiku & Rule $-$ no rule & jobs & $+1.33$ & $+0.92$ to $+1.75$ & 97.500 \\
F4 & Luna & T $-$ L0 & pp & $+91.67$ & $+86.11$ to $+97.22$ & 99.375 \\
F4 & Luna & TO $-$ T & pp & $-4.17$ & $-11.11$ to $+2.78$ & 99.375 \\
F4 & Luna & R $-$ L0 & pp & $+5.56$ & $+2.08$ to $+9.72$ & 99.375 \\
F4 & Luna & C $-$ L0 & pp & $+5.56$ & $+1.39$ to $+10.42$ & 99.375 \\
F4 & Haiku & T $-$ L0 & pp & $+70.83$ & $+59.72$ to $+81.94$ & 99.375 \\
F4 & Haiku & TO $-$ T & pp & $+13.19$ & $+2.78$ to $+29.00$ & 99.375 \\
F4 & Haiku & R $-$ L0 & pp & $+38.19$ & $+25.00$ to $+51.91$ & 99.375 \\
F4 & Haiku & C $-$ L0 & pp & $+53.47$ & $+40.28$ to $+63.89$ & 99.375 \\
\end{longtable}

Positive effects were predicted for Sol's knowledge and sharing contrasts
and for both distribution-rule contrasts; negative effects were predicted
for both competitor-count contrasts. No direction was assigned to Sol's
gap from eight or to component contrasts. The Luna rule contrast reversed
its positive prediction. Count-revelation effects were averaged over layouts
before resampling; their layout-specific patterns are descriptive cell summaries.
Sol's completion contrast measures distance from attainable capacity, not a
loss caused by sharing.

\textbf{Supplementary Table 15. Prespecified secondary contrasts.}
Intervals have unadjusted 95\% nominal coverage and are descriptive.

\begin{longtable}{@{}llp{45mm}lrr@{}}
\toprule
Stage & Model & Contrast & Unit & Mean & 95\% interval \\
\midrule
\endhead
\bottomrule
\endlastfoot
F2 & Luna & Count shown $-$ hidden & jobs & $+0.25$ & $-0.17$ to $+0.71$ \\
F2 & Haiku & Count shown $-$ hidden & jobs & $+0.00$ & $+0.00$ to $+0.00$ \\
F3 & Luna & Rule $-$ no rule & pp & $-60.19$ & $-70.37$ to $-49.07$ \\
F3 & Haiku & Rule $-$ no rule & pp & $-32.41$ & $-38.89$ to $-25.00$ \\
F4 & Luna & L1 $-$ L0 & pp & $+40.97$ & $+34.03$ to $+47.22$ \\
F4 & Haiku & L1 $-$ L0 & pp & $+95.14$ & $+91.67$ to $+98.61$ \\
\end{longtable}
\par\Needspace{32mm}
\subsection{Supplementary Note 10. Repeated inputs and the allocation reference}

We compared the complete saved prompts in all constructed-history exploration
and sharing conditions from the common-history stage, the delay/model-transfer
stage, and F1--F4. A prompt included the instruction, serialized shared report,
and private state. Within each branch, all actions were chosen before service admission,
and no agent observed another agent's current choice. The design excluded
direct action-to-action feedback within that branch; common inputs could
still concentrate response probabilities. Seed, internal owner identifier, and execution order were not part of
the prompt. Supplementary Table 16 distinguishes repeated generation from distinct visible
histories. In the 115 conditions without a private draw, all 12 batches within
a condition had the same complete prompt, shared by all active agents.
The four F3 conditions instead had 108 unique prompts each, because of the
private draws.

\textbf{Supplementary Table 16. Prompt identity in the existing controlled conditions.}
M denotes the common-history stage; T denotes delay/model transfer.
Unique prompts are counted within each condition. Blocks are generation and
admission-lottery replicates, not different environmental configurations.

\begin{longtable}{@{}llrrrr@{}}
\toprule
Stage & Task & Conditions & Blocks & Decisions & Unique/condition \\
\midrule
\endhead
\bottomrule
\endlastfoot
F1 & Share & 4 & 48 & 432 & 1 \\
F1 & Explore & 3 & 36 & 432 & 1 \\
F2 & Share & 8 & 96 & 864 & 1 \\
F3 & Share & 4 & 48 & 432 & 108 \\
F4 & Explore & 12 & 144 & 1728 & 1 \\
M & Share & 24 & 288 & 2592 & 1 \\
M & Explore & 12 & 144 & 1728 & 1 \\
T & Share & 36 & 432 & 3888 & 1 \\
T & Explore & 16 & 192 & 2304 & 1 \\
\end{longtable}

For the 72 final-round conditions with identical inputs (864 branches, including success-record withholding/sharing and F2 count revelation), we
estimated the probabilities $p_A$, $p_B$, and $p_W$ from the other 11 batches
in that same acquisition and condition. Under independent categorical draws,
the marginal counts are binomial, giving
\[
E[C]=\sum_{k=0}^{9}\binom{9}{k}p_A^k(1-p_A)^{9-k}\min(k,6)
 +\sum_{k=0}^{9}\binom{9}{k}p_B^k(1-p_B)^{9-k}\min(k,2).
\]
WAIT includes valid WAIT choices and executed fallback actions. The held-out
batch supplied only the observed completion count. We repeated this
leave-one-batch-out calculation 12 times and averaged within condition.
This post hoc comparison evaluates the conditional mean, not temporal
independence or the full joint distribution. 

Across these conditions, observed and predicted mean completion were 5.0324 and 5.0286; the mean absolute condition-level discrepancy was 0.0265 jobs. Residuals ranged from -0.2909 to 0.2855. All 27 conditions that were unanimous at A in every batch (324 branches) had both observed and predicted completion of six. 

\begin{center}
\begin{minipage}{\textwidth}
\includegraphics[width=170mm]{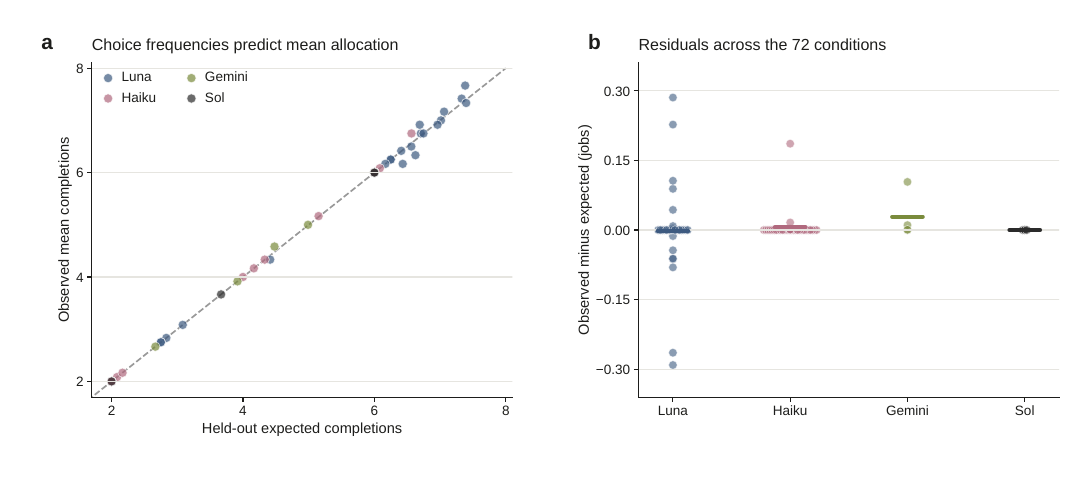}
\captionsetup{hypcap=false}
\captionof{figure}{\textbf{Observed mean allocation and a reference based on held-out choice probabilities.}
(\textbf{a}) Each point is one of 72 final-round conditions, with 12 branches per
condition and nine choices per branch. The dashed line denotes equality.
(\textbf{b}) Condition-level residuals by model; short bars mark means.
Points may overlap. Prediction uses the other 11 batches, within the same
condition and acquisition.  Supplementary Table 17 reports every point.}
\label{fig:response_distribution}
\end{minipage}
\end{center}
\textbf{Supplementary Table 17. Observed and held-out expected completion in all 72 conditions.}
Lag is the report delay; S0/S1 mean record withheld/shared in M, T, and F1,
but count hidden/shown in F2 (both with the success record). Residual is
observed minus expected. All values are jobs completed in the final round.

\begin{longtable}{@{}llllrlrrr@{}}
\toprule
Stage & Model & K & Layout & Lag & Arm & Observed & Expected & Residual \\
\midrule
\endhead
\bottomrule
\endlastfoot
F1 & Sol & KT & L0 & 2 & S0 & 3.667 & 3.666 & $+0.00$ \\
F1 & Sol & KT & L0 & 2 & S1 & 6.000 & 6.000 & $+0.00$ \\
F1 & Sol & KT & L1 & 2 & S0 & 2.000 & 2.000 & $+0.00$ \\
F1 & Sol & KT & L1 & 2 & S1 & 6.000 & 6.000 & $+0.00$ \\
F2 & Haiku & KT & L0 & 2 & S0 & 6.000 & 6.000 & $+0.00$ \\
F2 & Haiku & KT & L0 & 2 & S1 & 6.000 & 6.000 & $+0.00$ \\
F2 & Haiku & KT & L1 & 2 & S0 & 6.000 & 6.000 & $+0.00$ \\
F2 & Haiku & KT & L1 & 2 & S1 & 6.000 & 6.000 & $+0.00$ \\
F2 & Luna & KT & L0 & 2 & S0 & 7.667 & 7.381 & $+0.29$ \\
F2 & Luna & KT & L0 & 2 & S1 & 6.750 & 6.707 & $+0.04$ \\
F2 & Luna & KT & L1 & 2 & S0 & 6.000 & 6.000 & $+0.00$ \\
F2 & Luna & KT & L1 & 2 & S1 & 7.417 & 7.328 & $+0.09$ \\
M & Haiku & K0 & L0 & 2 & S0 & 2.000 & 2.000 & $+0.00$ \\
M & Haiku & K0 & L0 & 2 & S1 & 2.083 & 2.083 & $+0.00$ \\
M & Haiku & K0 & L1 & 2 & S0 & 2.000 & 2.000 & $+0.00$ \\
M & Haiku & K0 & L1 & 2 & S1 & 6.750 & 6.564 & $+0.19$ \\
M & Haiku & KG & L0 & 2 & S0 & 2.000 & 2.000 & $+0.00$ \\
M & Haiku & KG & L0 & 2 & S1 & 6.000 & 6.000 & $+0.00$ \\
M & Haiku & KG & L1 & 2 & S0 & 2.000 & 2.000 & $+0.00$ \\
M & Haiku & KG & L1 & 2 & S1 & 6.000 & 6.000 & $+0.00$ \\
M & Haiku & KT & L0 & 2 & S0 & 4.167 & 4.165 & $+0.00$ \\
M & Haiku & KT & L0 & 2 & S1 & 6.000 & 6.000 & $+0.00$ \\
M & Haiku & KT & L1 & 2 & S0 & 4.333 & 4.331 & $+0.00$ \\
M & Haiku & KT & L1 & 2 & S1 & 6.000 & 6.000 & $+0.00$ \\
M & Luna & K0 & L0 & 2 & S0 & 2.000 & 2.000 & $+0.00$ \\
M & Luna & K0 & L0 & 2 & S1 & 7.000 & 7.013 & $-0.01$ \\
M & Luna & K0 & L1 & 2 & S0 & 2.833 & 2.833 & $+0.00$ \\
M & Luna & K0 & L1 & 2 & S1 & 6.250 & 6.248 & $+0.00$ \\
M & Luna & KG & L0 & 2 & S0 & 2.000 & 2.000 & $+0.00$ \\
M & Luna & KG & L0 & 2 & S1 & 6.750 & 6.746 & $+0.00$ \\
M & Luna & KG & L1 & 2 & S0 & 2.000 & 2.000 & $+0.00$ \\
M & Luna & KG & L1 & 2 & S1 & 6.250 & 6.248 & $+0.00$ \\
M & Luna & KT & L0 & 2 & S0 & 2.750 & 2.750 & $+0.00$ \\
M & Luna & KT & L0 & 2 & S1 & 7.333 & 7.396 & $-0.06$ \\
M & Luna & KT & L1 & 2 & S0 & 4.333 & 4.414 & $-0.08$ \\
M & Luna & KT & L1 & 2 & S1 & 6.167 & 6.166 & $+0.00$ \\
T & Gemini & K0 & L1 & 2 & S0 & 2.667 & 2.667 & $+0.00$ \\
T & Gemini & K0 & L1 & 2 & S1 & 4.583 & 4.480 & $+0.10$ \\
T & Gemini & KT & L1 & 2 & S0 & 5.000 & 4.989 & $+0.01$ \\
T & Gemini & KT & L1 & 2 & S1 & 3.917 & 3.916 & $+0.00$ \\
T & Haiku & K0 & L0 & 1 & S0 & 6.000 & 6.000 & $+0.00$ \\
T & Haiku & K0 & L0 & 1 & S1 & 6.000 & 6.000 & $+0.00$ \\
T & Haiku & K0 & L0 & 2 & S0 & 2.000 & 2.000 & $+0.00$ \\
T & Haiku & K0 & L0 & 2 & S1 & 2.167 & 2.167 & $+0.00$ \\
T & Haiku & K0 & L1 & 1 & S0 & 6.000 & 6.000 & $+0.00$ \\
T & Haiku & K0 & L1 & 1 & S1 & 6.000 & 6.000 & $+0.00$ \\
T & Haiku & K0 & L1 & 2 & S0 & 2.000 & 2.000 & $+0.00$ \\
T & Haiku & K0 & L1 & 2 & S1 & 6.083 & 6.082 & $+0.00$ \\
T & Haiku & KT & L0 & 1 & S0 & 6.000 & 6.000 & $+0.00$ \\
T & Haiku & KT & L0 & 1 & S1 & 6.000 & 6.000 & $+0.00$ \\
T & Haiku & KT & L0 & 2 & S0 & 4.000 & 3.999 & $+0.00$ \\
T & Haiku & KT & L0 & 2 & S1 & 6.000 & 6.000 & $+0.00$ \\
T & Haiku & KT & L1 & 1 & S0 & 6.000 & 6.000 & $+0.00$ \\
T & Haiku & KT & L1 & 1 & S1 & 6.000 & 6.000 & $+0.00$ \\
T & Haiku & KT & L1 & 2 & S0 & 5.167 & 5.150 & $+0.02$ \\
T & Haiku & KT & L1 & 2 & S1 & 6.000 & 6.000 & $+0.00$ \\
T & Luna & K0 & L0 & 1 & S0 & 6.000 & 6.000 & $+0.00$ \\
T & Luna & K0 & L0 & 1 & S1 & 6.000 & 6.000 & $+0.00$ \\
T & Luna & K0 & L0 & 2 & S0 & 2.000 & 2.000 & $+0.00$ \\
T & Luna & K0 & L0 & 2 & S1 & 6.333 & 6.624 & $-0.29$ \\
T & Luna & K0 & L1 & 1 & S0 & 6.000 & 6.000 & $+0.00$ \\
T & Luna & K0 & L1 & 1 & S1 & 6.000 & 6.000 & $+0.00$ \\
T & Luna & K0 & L1 & 2 & S0 & 3.083 & 3.083 & $+0.00$ \\
T & Luna & K0 & L1 & 2 & S1 & 6.417 & 6.408 & $+0.01$ \\
T & Luna & KT & L0 & 1 & S0 & 6.500 & 6.562 & $-0.06$ \\
T & Luna & KT & L0 & 1 & S1 & 7.167 & 7.060 & $+0.11$ \\
T & Luna & KT & L0 & 2 & S0 & 2.750 & 2.750 & $+0.00$ \\
T & Luna & KT & L0 & 2 & S1 & 6.917 & 6.689 & $+0.23$ \\
T & Luna & KT & L1 & 1 & S0 & 6.917 & 6.960 & $-0.04$ \\
T & Luna & KT & L1 & 1 & S1 & 6.000 & 6.000 & $+0.00$ \\
T & Luna & KT & L1 & 2 & S0 & 6.167 & 6.431 & $-0.26$ \\
T & Luna & KT & L1 & 2 & S1 & 6.000 & 6.000 & $+0.00$ \\
\end{longtable}

\par\Needspace{32mm}
\subsection{Supplementary Note 11. Information and complete assignment rules}

R1 used KT/L1, capacities 6/2, nine pending agents, and one remaining decision.
All branches displayed the success record. Each agent's private draw was
held fixed across branches; different agents received different draws. Sol had four branches: count hidden with no rule; count of nine shown
with no rule; count shown with the partial rule; count shown with the complete
rule. Luna and Haiku had the latter three branches. Seeds 5001--5024 supplied
24 matched acquisition blocks. Draws used a new fixed hash namespace and were
matched across models and branches. The comparison therefore tests the
complete rule against the partial rule with competitor count and draw held
fixed; it is distinct from F3, which hid the count.

\textbf{Exact partial rule.}
\begin{Verbatim}[breaklines=true,fontsize=\scriptsize,frame=single,framerule=0.3pt]
Distribution rule: if public_success_records shows a success at A and your private_draw is below 0.22, request B; otherwise follow your own judgment.
\end{Verbatim}
\textbf{Exact complete rule.}
\begin{Verbatim}[breaklines=true,fontsize=\scriptsize,frame=single,framerule=0.3pt]
Distribution rule: if public_success_records shows a success at A and your private_draw is below 0.22, request B; otherwise request A.
\end{Verbatim}
If $k$ of the nine draws are below 0.22, complete adherence yields
$\min(9-k,6)+\min(k,2)$ completions. This need not equal eight, because the
number assigned to B varies across blocks. We compare actual completion
with this draw-specific reference, as well as measuring adherence separately
in the designated and undesignated agents.

\Needspace{40mm}
\textbf{Supplementary Table 18. All information and assignment conditions in R1.}
Each row contains 24 blocks and 216 initial choices. Count denotes whether
the total of nine active agents, including the focal agent, was shown. Designated means a private draw below 0.22; other means a
draw of at least 0.22. B designated and A other are the choices matching the
complete assignment, with their respective denominators. Completion and
reference are block means; full counts branches attaining eight.

\begin{longtable}{@{}lllrrrrrrr@{}}
\toprule
Model & Count & Rule & A (\%) & Complete & Reference & B designated & A other & Full & Nonvalid \\
\midrule
\endhead
\bottomrule
\endlastfoot
Haiku & Shown & None & 100.0 & 6.00 & 7.46 & 0/42 & 174/174 & 0 & 0 \\
Haiku & Shown & Complete & 80.6 & 7.46 & 7.46 & 42/42 & 174/174 & 13 & 0 \\
Haiku & Shown & Partial & 69.9 & 7.38 & 7.46 & 41/42 & 150/174 & 12 & 0 \\
Luna & Shown & None & 84.3 & 7.12 & 7.46 & 6/42 & 146/174 & 7 & 0 \\
Luna & Shown & Complete & 81.5 & 7.42 & 7.46 & 40/42 & 174/174 & 12 & 0 \\
Luna & Shown & Partial & 74.5 & 7.67 & 7.46 & 41/42 & 160/174 & 17 & 0 \\
Sol & Shown & None & 83.3 & 7.38 & 7.46 & 0/42 & 138/174 & 13 & 0 \\
Sol & Shown & Complete & 80.6 & 7.46 & 7.46 & 42/42 & 174/174 & 13 & 0 \\
Sol & Hidden & None & 77.8 & 7.21 & 7.46 & 0/42 & 126/174 & 10 & 0 \\
Sol & Shown & Partial & 80.6 & 7.46 & 7.46 & 42/42 & 174/174 & 13 & 0 \\
\end{longtable}

The complete-rule threshold corresponds to $p=0.78$ for choosing A over fresh independent draws, with an expected completion of 7.365745; the reference of 7.458333 in Supplementary Table 18 is instead the exact mean conditional on the 24 blocks of draws actually realized in R1. 

\par\Needspace{32mm}
\subsection{Supplementary Note 12. Capacity, load, and time to use a recovery}

R2 used Luna and Haiku, KT/L1, and 24 blocks per environment. Within each
block the same constructed round-3 state was branched into record withheld
and shared conditions. Request counts and active-client counts were hidden;
the registered population remained visible. There was no private draw or
distribution rule. All conditions used new responses, including the base.

The base had 16 initial agents, nine pending at round 3, capacities 6/2, and
deadline 4. The capacity condition changed A to three throughout the contract
and recovered state, retaining B at two. The load condition used 13 initial
agents, leaving six pending at round 3. The time condition extended the
deadline to five, allowing pending agents to act at rounds 3 and 4. At round 4, their private histories included the round-3 outcome, and the
ordinary report described the recovered capacity at round 2 in both branches. The
disclosed success remained the record of round 2. Thus the time comparison
includes renewed action and the arrival of the updated ordinary report.
The capacity manipulation changes both the capacity ratio and its sum;
the load manipulation changes both registered and remaining populations.

For $n$ pending jobs, $h$ available decisions, and constant capacities $a,b$,
the attainable completion count is $C^*=\min[n,h(a+b)]$. The avoidable
completion gap is $C^*-C$. It excludes $n-C^*$ jobs that no allocation can
finish. Always choosing A gives $\min(n,ha)$ completions, predicting a gap of two jobs
in the base and capacity conditions, and zero in the load and time conditions.
Completed jobs in the coded prefix are excluded from all these counts.

\textbf{Supplementary Table 19. All R2 conditions.}
Each row contains 24 blocks. First completion and A choice refer to round 3;
terminal completion includes round 4 only in the time condition. Gap is
attainable minus terminal completion. Full counts branches attaining their
environment's maximum, which need not be eight.

\begin{longtable}{@{}lllrrrrrrr@{}}
\toprule
Model & Env. & Record & A (\%) & First & Terminal & Max. & Gap & Full & Nonvalid \\
\midrule
\endhead
\bottomrule
\endlastfoot
Haiku & Base & Shared & 100.0 & 6.00 & 6.00 & 8 & 2.00 & 0 & 0 \\
Haiku & Base & Withheld & 32.9 & 4.96 & 4.96 & 8 & 3.04 & 0 & 0 \\
Haiku & A=3 & Shared & 100.0 & 3.00 & 3.00 & 5 & 2.00 & 0 & 0 \\
Haiku & A=3 & Withheld & 1.9 & 2.17 & 2.17 & 5 & 2.83 & 0 & 0 \\
Haiku & n=6 & Shared & 100.0 & 6.00 & 6.00 & 6 & 0.00 & 24 & 0 \\
Haiku & n=6 & Withheld & 39.6 & 4.38 & 4.38 & 6 & 1.62 & 3 & 0 \\
Haiku & h=2 & Shared & 100.0 & 6.00 & 9.00 & 9 & 0.00 & 24 & 0 \\
Haiku & h=2 & Withheld & 79.6 & 7.33 & 9.00 & 9 & 0.00 & 24 & 0 \\
Luna & Base & Shared & 99.5 & 6.04 & 6.04 & 8 & 1.96 & 0 & 0 \\
Luna & Base & Withheld & 40.7 & 5.58 & 5.58 & 8 & 2.42 & 4 & 0 \\
Luna & A=3 & Shared & 90.7 & 3.75 & 3.75 & 5 & 1.25 & 3 & 0 \\
Luna & A=3 & Withheld & 36.1 & 4.71 & 4.71 & 5 & 0.29 & 18 & 0 \\
Luna & n=6 & Shared & 99.3 & 6.00 & 6.00 & 6 & 0.00 & 24 & 0 \\
Luna & n=6 & Withheld & 9.0 & 2.54 & 2.54 & 6 & 3.46 & 0 & 0 \\
Luna & h=2 & Shared & 100.0 & 6.00 & 9.00 & 9 & 0.00 & 24 & 0 \\
Luna & h=2 & Withheld & 39.4 & 5.50 & 9.00 & 9 & 0.00 & 24 & 0 \\
\end{longtable}

\par\Needspace{32mm}
\subsection{Supplementary Note 13. Planned contrasts and acquisition records}

\textbf{Supplementary Table 20. Planned comparisons in R1 and R2.}
Each comparison uses 24 matched blocks, with 10,000 bootstrap resamples. The four R1 primary comparisons use 98.75\% nominal
coverage; the six R2 comparisons use 99.1667\%. Secondary intervals are
unadjusted 95\% and descriptive. pp denotes percentage points. 

The positive prediction for R1 concerned Luna's complete-minus-partial rule
effect. The other R1 primary comparisons were nondirectional. R2 predicted
smaller shared-branch gaps with six pending agents and with two remaining
decisions in each model; capacity-three contrasts were nondirectional.

\textit{A. Primary comparisons.}

\begin{longtable}{@{}llp{46mm}lrrr@{}}
\toprule
Stage & Model & Contrast & Unit & Mean & Interval & +/--/0 \\
\midrule
\endhead
\bottomrule
\endlastfoot
R1 & Sol & Complete $-$ partial & jobs & $+0.00$ & $+0.00$ to $+0.00$ & 0/0/24 \\
R1 & Luna & Complete $-$ partial & jobs & $-0.25$ & $-0.50$ to $+0.00$ & 1/7/16 \\
R1 & Haiku & Complete $-$ partial & jobs & $+0.08$ & $-0.38$ to $+0.58$ & 8/8/8 \\
R1 & Sol & Count shown $-$ hidden & pp & $+5.56$ & $+0.46$ to $+10.65$ & 11/3/10 \\
R2 & Luna & A=3 $-$ base gap & jobs & $-0.71$ & $-1.12$ to $-0.33$ & 1/15/8 \\
R2 & Luna & n=6 $-$ base gap & jobs & $-1.96$ & $-2.00$ to $-1.83$ & 0/24/0 \\
R2 & Luna & h=2 $-$ base gap & jobs & $-1.96$ & $-2.00$ to $-1.83$ & 0/24/0 \\
R2 & Haiku & A=3 $-$ base gap & jobs & $+0.00$ & $+0.00$ to $+0.00$ & 0/0/24 \\
R2 & Haiku & n=6 $-$ base gap & jobs & $-2.00$ & $-2.00$ to $-2.00$ & 0/24/0 \\
R2 & Haiku & h=2 $-$ base gap & jobs & $-2.00$ & $-2.00$ to $-2.00$ & 0/24/0 \\
\end{longtable}

\textit{B. Secondary comparisons.}

\begin{longtable}{@{}llp{46mm}lrrr@{}}
\toprule
Stage & Model & Contrast & Unit & Mean & Interval & +/--/0 \\
\midrule
\endhead
\bottomrule
\endlastfoot
R1 & Sol & Partial $-$ no rule & jobs & $+0.08$ & $-0.33$ to $+0.50$ & 8/7/9 \\
R1 & Sol & Complete $-$ no rule & jobs & $+0.08$ & $-0.33$ to $+0.50$ & 8/7/9 \\
R1 & Luna & Partial $-$ no rule & jobs & $+0.54$ & $+0.17$ to $+0.92$ & 12/3/9 \\
R1 & Luna & Complete $-$ no rule & jobs & $+0.29$ & $-0.12$ to $+0.71$ & 9/6/9 \\
R1 & Haiku & Partial $-$ no rule & jobs & $+1.38$ & $+1.08$ to $+1.62$ & 21/0/3 \\
R1 & Haiku & Complete $-$ no rule & jobs & $+1.46$ & $+1.21$ to $+1.71$ & 22/0/2 \\
R2 & Luna & Shared $-$ withheld (base) & jobs & $+0.46$ & $-0.21$ to $+1.12$ & 13/8/3 \\
R2 & Luna & Shared $-$ withheld (A=3) & jobs & $-0.96$ & $-1.33$ to $-0.54$ & 2/17/5 \\
R2 & Luna & Shared $-$ withheld (n=6) & jobs & $+3.46$ & $+3.04$ to $+3.79$ & 24/0/0 \\
R2 & Luna & Shared $-$ withheld (h=2) & jobs & $+0.00$ & $+0.00$ to $+0.00$ & 0/0/24 \\
R2 & Haiku & Shared $-$ withheld (base) & jobs & $+1.04$ & $+0.50$ to $+1.54$ & 16/4/4 \\
R2 & Haiku & Shared $-$ withheld (A=3) & jobs & $+0.83$ & $+0.67$ to $+0.96$ & 20/0/4 \\
R2 & Haiku & Shared $-$ withheld (n=6) & jobs & $+1.62$ & $+1.25$ to $+2.00$ & 21/0/3 \\
R2 & Haiku & Shared $-$ withheld (h=2) & jobs & $+0.00$ & $+0.00$ to $+0.00$ & 0/0/24 \\
\end{longtable}

The original R1 acquisition over HTTP/2 stopped after repeated communication failures at the two-attempt bound; all 591 calls from that incomplete series, including 561 valid responses, were excluded before any condition-level analysis, and a frozen addendum that changed only the transport to HTTP/1.1 reacquired the complete design (2,160 accepted responses, all valid). R2 used the same transport and later stopped on a local database commit error after 44 completed cohorts; a frozen recovery record replayed the five saved valid responses of the affected cohort at their original contexts and acquired only the unsent decisions, so that no recorded response was replaced and no cohort attempt was added (3,436 accepted responses, all valid). No response was selected for its action or outcome. Model settings remained those in Supplementary Note 8.
\par\Needspace{32mm}
\subsection{Supplementary Note 14. A private random number without a distribution rule}

The active-count-hidden, no-rule Sol branch in R1 motivated a matched test, R3. A post hoc comparison of complete prompts verified that F1 shared KT/L1 inputs and R1 count-hidden no-rule inputs
differed only in the private number. F1 had 108/108 A choices and six completions
per branch; R1 had 168/216 A choices, 7.21 mean completions, and four all-A
branches out of 24. All 48 R1 B choices had numbers of at least 0.7582, and
all 42 numbers below 0.22 produced A. However, A choices extended to 0.9906,
overlapping B choices. These associations do not identify a single deterministic
threshold or prove that Sol computed a mixed-strategy equilibrium.

R3 was internally specified after inspecting R1 and acquired on 18 October
2026 (UTC). It used 24 new coded-history seeds (6001--6024), KT/L1, capacities
6/2, nine pending agents, the success record shared, and no competitor count
or distribution rule. Both branches used newly acquired responses; the sole
visible difference was the presence of the numeric \texttt{private\_draw}
field. Numbers used R1's hash namespace, the new seed and internal owner,
rounded to 12 decimal places. No owner identifier was displayed. The intervention combines the
numeric field and its name; arbitrary private identifiers were not tested.

\textbf{Supplementary Table 21. Contemporaneous private-number comparison in Sol (R3).}
Each condition has 24 blocks of nine choices. Counts are first-decision actions;
complete denotes mean jobs finished, and full counts blocks completing eight.

\begin{longtable}{@{}lrrrrrrrr@{}}
\toprule
Number & A & B & WAIT & A (\%) & Complete & All A & Full & Nonvalid \\
\midrule
\endhead
\bottomrule
\endlastfoot
Absent & 216 & 0 & 0 & 100.0 & 6.00 & 24 & 0 & 0 \\
Present & 172 & 44 & 0 & 79.6 & 7.50 & 2 & 14 & 0 \\
\end{longtable}

The single primary comparison was number present minus absent in completed
jobs, with a positive prediction. The secondary comparison was the change in
A-choice rate, with a negative prediction. Intervals use 10,000 paired bootstrap samples of 24 blocks. Each has nominal 95\%
coverage; the secondary interval is descriptive.

\begin{longtable}{@{}llrrr@{}}
\toprule
Outcome & Unit & Difference & 95\% interval & +/--/0 \\
\midrule
\endhead
\bottomrule
\endlastfoot
Completion (primary) & jobs & $+1.50$ & $+1.25$ to $+1.75$ & 22/0/2 \\
A choice (secondary) & pp & $-20.37$ & $-24.54$ to $-16.20$ & 0/22/2 \\
\end{longtable}

All 44 B choices in R3 had numbers of at least 0.7724; A choices extended to 0.9914. No B choice occurred among the 150 decisions with numbers below 0.7, compared with 44 of 66 decisions at or above 0.7.  The completion comparison appears in main-text Fig.~6f.

\begin{center}
\begin{minipage}{\textwidth}
\includegraphics[width=170mm]{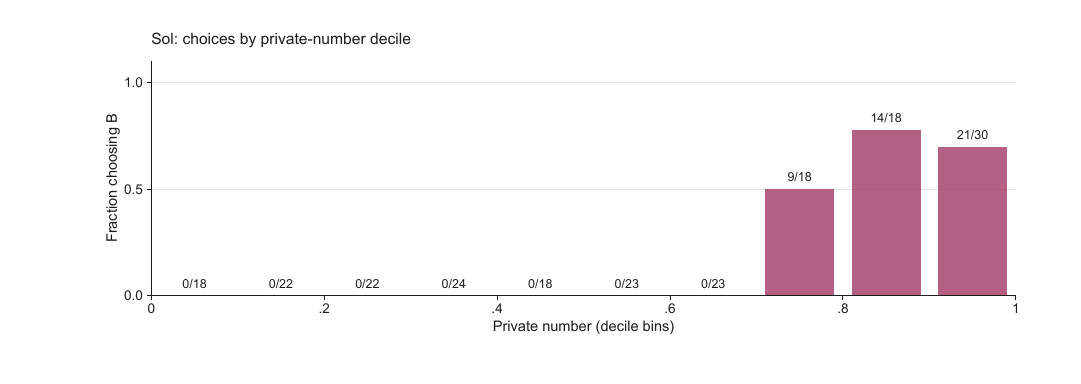}
\captionsetup{hypcap=false}
\captionof{figure}{\textbf{Sol's B choices concentrate at high private numbers.}
Fraction choosing B within each number decile in the number-present branch of
R3, with KT/L1 and the success record shared but no competitor count or
distribution rule. Labels give B choices/total decisions. All 216 decisions
from 24 blocks are included.  The
completion effect of displaying a number is shown in main-text Fig.~6f.}
\label{fig:private_number}
\end{minipage}
\end{center}

\Needspace{35mm}
\textbf{Acquisition and provenance.} Sol retained medium reasoning, the default tier, and an 8,192-token output limit. All 432 acquired responses were accepted and valid, and no replacement was required.

\par\medskip
\Needspace{65mm}
\subsection{Supplementary Note 15. Allocation loss across five models and the grouping of numerical inputs}

Both experiments used KT/L1, a two-round reporting delay and capacities of six (A) and two (B). Nine jobs remained at round 3. Request and active-client counts were hidden. Each condition comprised 24 blocks (216 decisions). Both protocols were frozen internally before acquisition on 25 October 2026 (UTC).

\textbf{Five-model comparison.} Luna, Haiku, Sol, Opus and Gemini received no success record (S0), a shared record (S1), or the record and a private number (S1R). Within S0 and S1, all blocks and models received identical prompts; in S1R, only the private numbers varied, with the same draws across models. Of 3,249 acquired requests, 3,240 were included; nine Gemini requests failed the initial JSON specification. A separate nine-response Opus diagnostic yielded seven valid choices and two refusals. Before acquisition, the start criterion was revised to accept explicit refusals; the nine diagnostic responses were excluded. Refusals in the experiment counted as WAIT and contributed to mean completion. The Methods give the model settings, and Supplementary Note~6 lists the seeds. Supplementary Table~23 reports the ten prespecified paired comparisons with Holm correction; bootstrap intervals are descriptive (Supplementary Note~4).

Sharing raised A choice to nearly 100\% in Luna and Haiku, as in every earlier acquisition, but its benefit to completion depended on how many agents chose A without the record. In the same KT/L1, delay-2, withheld-record condition, the share of requests to A was 26.85\%, 50.93\%, 40.7\%, and 52.3\% for Luna in the original constructed-history stage, the delay-transfer stage, the R2 base condition, and this comparison, and 25.93\%, 35.19\%, 32.9\%, and 44.9\% for Haiku (Supplementary Tables~2, 4, and 19 and Supplementary Table~22). The corresponding changes in final-round completions with sharing were $+1.83$, $-0.17$, $+0.46$, and $-0.25$ jobs for Luna and $+1.67$, $+0.83$, $+1.04$, and $+0.25$ for Haiku, so the completion benefit was smaller when more agents already chose A. The shared-record behavior was unanimous or nearly unanimous for A in every acquisition (215 of 216 in Luna here). We therefore report the sharing benefit per acquisition rather than as a pooled estimate.

\Needspace{95mm}
\textbf{Supplementary Table 22. Five-model comparison: cell summaries.} Each cell has 24 blocks of nine decisions. A, B, and WAIT count executed first decisions; WAIT includes refusals, listed separately. Complete is the mean number of jobs finished (attainable eight); All A and Full count blocks in which every request went to A and blocks completing eight jobs.

\begin{longtable}{@{}llrrrrrrrr@{}}
\toprule
Model & Condition & A & B & WAIT & Refused & Complete & 95\% interval & All A & Full \\
\midrule
\endhead
\bottomrule
\endlastfoot
Luna & S0 & 113 & 103 & 0 & 0 & 6.29 & 5.88 to 6.71 & 0 & 3 \\
Luna & S1 & 215 & 1 & 0 & 0 & 6.04 & 6.00 to 6.12 & 23 & 0 \\
Luna & S1R & 214 & 2 & 0 & 0 & 6.08 & 6.00 to 6.21 & 22 & 0 \\
Haiku & S0 & 97 & 112 & 7 & 0 & 5.75 & 5.12 to 6.38 & 0 & 3 \\
Haiku & S1 & 216 & 0 & 0 & 0 & 6.00 & 6.00 to 6.00 & 24 & 0 \\
Haiku & S1R & 216 & 0 & 0 & 0 & 6.00 & 6.00 to 6.00 & 24 & 0 \\
Sol & S0 & 3 & 213 & 0 & 0 & 2.12 & 2.00 to 2.25 & 0 & 0 \\
Sol & S1 & 216 & 0 & 0 & 0 & 6.00 & 6.00 to 6.00 & 24 & 0 \\
Sol & S1R & 165 & 51 & 0 & 0 & 7.38 & 7.12 to 7.62 & 2 & 11 \\
Opus & S0 & 201 & 5 & 10 & 10 & 6.21 & 6.04 to 6.42 & 11 & 1 \\
Opus & S1 & 212 & 0 & 4 & 4 & 6.00 & 6.00 to 6.00 & 20 & 0 \\
Opus & S1R & 111 & 33 & 72 & 72 & 5.71 & 5.17 to 6.25 & 0 & 3 \\
Gemini & S0 & 96 & 120 & 0 & 0 & 5.88 & 5.29 to 6.38 & 0 & 1 \\
Gemini & S1 & 69 & 147 & 0 & 0 & 4.88 & 4.29 to 5.42 & 0 & 1 \\
Gemini & S1R & 172 & 44 & 0 & 0 & 7.42 & 7.17 to 7.67 & 2 & 12 \\
\end{longtable}

\Needspace{78mm}
\textbf{Supplementary Table 23. Five-model comparison: planned contrasts.} Paired differences in completed jobs over 24 blocks; exact sign-flip $P$ values with Holm correction across the ten comparisons.

\begin{longtable}{@{}llrrrr@{}}
\toprule
Model & Comparison & Difference & 95\% interval & $P$ (exact) & $P$ (Holm) \\
\midrule
\endhead
\bottomrule
\endlastfoot
Luna & S1 $-$ S0 & -0.25 & -0.71 to 0.21 & 0.383 & 1.000 \\
Luna & S1R $-$ S1 & 0.04 & -0.08 to 0.17 & 1.000 & 1.000 \\
Haiku & S1 $-$ S0 & 0.25 & -0.38 to 0.88 & 0.527 & 1.000 \\
Haiku & S1R $-$ S1 & 0.00 & 0.00 to 0.00 & 1.000 & 1.000 \\
Sol & S1 $-$ S0 & 3.88 & 3.75 to 4.00 & $1.2\times10^{-7}$ & $1.2\times10^{-6}$ \\
Sol & S1R $-$ S1 & 1.38 & 1.12 to 1.62 & $4.8\times10^{-7}$ & $4.3\times10^{-6}$ \\
Opus & S1 $-$ S0 & -0.21 & -0.42 to -0.04 & 0.125 & 0.750 \\
Opus & S1R $-$ S1 & -0.29 & -0.83 to 0.25 & 0.368 & 1.000 \\
Gemini & S1 $-$ S0 & -1.00 & -1.83 to -0.17 & 0.044 & 0.311 \\
Gemini & S1R $-$ S1 & 2.54 & 1.96 to 3.17 & $1.4\times10^{-6}$ & $1.1\times10^{-5}$ \\
\end{longtable}

\textbf{Grouping test (N/C/I).} Seeds 8001--8024 defined the histories, with the success record shared in all conditions. Condition N added no number. Twenty-four uniform numbers were generated once (12 decimal places, as in R3). Condition C gave each group one of these numbers, shown identically to all nine agents. Condition I arranged the same 24 numbers so that each group received nine distinct numbers and each number appeared in exactly nine groups; the multisets of numbers and of complete prompts are therefore identical between C and I, and only their assignment to groups differs. I is a balanced arrangement of a finite set, not an independent draw, and the field name and format were those of R3. No condition stated whether other agents' numbers matched. Sol, Gemini, and Haiku were acquired first (1,944 decisions); Luna was added in a separate stage with the same inputs (648 decisions); Opus was not run. All 2,592 responses were valid. Models were chosen from the five-model comparison: positive effects of I were predicted for Sol and Gemini, which had responded to the private number, and Haiku and Luna were analyzed identically without a directional prediction. The planned comparisons were I minus C and I minus N per model, with Holm correction within each stage (six comparisons in the first stage, two in the Luna stage) as the frozen primary analysis and, for the combined four-model display in the main text, across all eight. All comparisons used the planned exact sign-flip permutation test. Because each number appeared in nine groups under I, outcomes of different blocks share inputs; the tests are therefore conditional on this set of 24 numbers and its arrangement and do not average over other sets of numbers.

\Needspace{85mm}
\textbf{Supplementary Table 24. Grouping test: cell summaries.} Columns as in Supplementary Table 22; no responses were refused or invalid.

\begin{longtable}{@{}llrrrrrrr@{}}
\toprule
Model & Condition & A & B & WAIT & Complete & 95\% interval & All A & Full \\
\midrule
\endhead
\bottomrule
\endlastfoot
Luna & N & 216 & 0 & 0 & 6.00 & 6.00 to 6.00 & 24 & 0 \\
Luna & C & 216 & 0 & 0 & 6.00 & 6.00 to 6.00 & 24 & 0 \\
Luna & I & 212 & 4 & 0 & 6.17 & 6.04 to 6.33 & 20 & 0 \\
Haiku & N & 216 & 0 & 0 & 6.00 & 6.00 to 6.00 & 24 & 0 \\
Haiku & C & 216 & 0 & 0 & 6.00 & 6.00 to 6.00 & 24 & 0 \\
Haiku & I & 216 & 0 & 0 & 6.00 & 6.00 to 6.00 & 24 & 0 \\
Sol & N & 216 & 0 & 0 & 6.00 & 6.00 to 6.00 & 24 & 0 \\
Sol & C & 160 & 56 & 0 & 5.12 & 4.54 to 5.67 & 17 & 0 \\
Sol & I & 158 & 58 & 0 & 7.25 & 7.00 to 7.50 & 1 & 9 \\
Gemini & N & 74 & 142 & 0 & 5.04 & 4.54 to 5.62 & 0 & 2 \\
Gemini & C & 142 & 73 & 1 & 5.50 & 4.83 to 6.12 & 8 & 1 \\
Gemini & I & 144 & 69 & 3 & 7.25 & 6.79 to 7.62 & 1 & 13 \\
\end{longtable}

\Needspace{100mm}
\textbf{Supplementary Table 25. Grouping test: planned contrasts.} Paired differences in completed jobs over 24 blocks. $P$ (stage) applies Holm correction within the acquisition stage (six comparisons for Sol, Gemini, and Haiku; two for Luna); $P$ (eight) applies it across all eight comparisons.

\begin{longtable}{@{}llrrrrr@{}}
\toprule
Model & Comparison & Difference & 95\% interval & $P$ (exact) & $P$ (stage) & $P$ (eight) \\
\midrule
\endhead
\bottomrule
\endlastfoot
Sol & I $-$ C & 2.12 & 1.50 to 2.79 & $9.5\times10^{-7}$ & $5.7\times10^{-6}$ & $7.6\times10^{-6}$ \\
Sol & I $-$ N & 1.25 & 1.00 to 1.50 & $9.5\times10^{-7}$ & $5.7\times10^{-6}$ & $7.6\times10^{-6}$ \\
Gemini & I $-$ C & 1.75 & 1.08 to 2.46 & $7.3\times10^{-5}$ & $2.2\times10^{-4}$ & $3.7\times10^{-4}$ \\
Gemini & I $-$ N & 2.21 & 1.46 to 2.88 & $5.0\times10^{-5}$ & $2.0\times10^{-4}$ & $3.0\times10^{-4}$ \\
Haiku & I $-$ C & 0.00 & 0.00 to 0.00 & 1.000 & 1.000 & 1.000 \\
Haiku & I $-$ N & 0.00 & 0.00 to 0.00 & 1.000 & 1.000 & 1.000 \\
Luna & I $-$ C & 0.17 & 0.04 to 0.33 & 0.125 & 0.250 & 0.500 \\
Luna & I $-$ N & 0.17 & 0.04 to 0.33 & 0.125 & 0.250 & 0.500 \\
\end{longtable}

In Sol, the same number for all nine agents lowered completions below the no-number condition because whole groups responded to their number alike: 17 of 24 C groups sent every request to A and completed six jobs, as without a number, whereas the other seven sent zero to two requests to A and completed two to four jobs. Under I the aggregate A share was nearly unchanged (158 against 160 of 216) but the within-group mixture let nine groups complete all eight jobs (three with seven requests to A and six with six). A group completes eight jobs when at least six requests go to A and at least two to B; with WAIT choices possible, the count of requests to A alone does not fix completion. Gemini showed the same pattern with a lower A share.

\begin{thebibliography}{99}

\bibitem{pan2026} Pan, M. et al. Measuring agents in production. In \textit{Proceedings of the 43rd International Conference on Machine Learning}, \textit{Proc. Mach. Learn. Res.} \textbf{306}, 95649--95695 (2026).

\bibitem{kleinberg2021} Kleinberg, J. \& Raghavan, M. Algorithmic monoculture and social welfare. \textit{Proc. Natl. Acad. Sci. U.S.A.} \textbf{118}, e2018340118 (2021).

\bibitem{acemoglu2018} Acemoglu, D., Makhdoumi, A., Malekian, A. \& Ozdaglar, A. Informational Braess' paradox: The effect of information on traffic congestion. \textit{Oper. Res.} \textbf{66}, 893--917 (2018).

\bibitem{mitzenmacher2000} Mitzenmacher, M. How useful is old information? \textit{IEEE Trans. Parallel Distrib. Syst.} \textbf{11}, 6--20 (2000).

\bibitem{march1991} March, J. G. Exploration and exploitation in organizational learning. \textit{Organ. Sci.} \textbf{2}, 71--87 (1991).

\bibitem{laland2004} Laland, K. N. Social learning strategies. \textit{Learn. Behav.} \textbf{32}, 4--14 (2004).

\bibitem{boyd1995} Boyd, R. \& Richerson, P. J. Why does culture increase human adaptability? \textit{Ethol. Sociobiol.} \textbf{16}, 125--143 (1995).

\bibitem{bolton1999} Bolton, P. \& Harris, C. Strategic experimentation. \textit{Econometrica} \textbf{67}, 349--374 (1999).

\bibitem{rendell2010} Rendell, L. et al. Why copy others? Insights from the social learning strategies tournament. \textit{Science} \textbf{328}, 208--213 (2010).

\bibitem{lazer2007} Lazer, D. \& Friedman, A. The network structure of exploration and exploitation. \textit{Adm. Sci. Q.} \textbf{52}, 667--694 (2007).

\bibitem{mason2012} Mason, W. \& Watts, D. J. Collaborative learning in networks. \textit{Proc. Natl. Acad. Sci. U.S.A.} \textbf{109}, 764--769 (2012).

\bibitem{barkoczi2016} Barkoczi, D. \& Galesic, M. Social learning strategies modify the effect of network structure on group performance. \textit{Nat. Commun.} \textbf{7}, 13109 (2016).

\bibitem{toyokawa2019} Toyokawa, W., Whalen, A. \& Laland, K. N. Social learning strategies regulate the wisdom and madness of interactive crowds. \textit{Nat. Hum. Behav.} \textbf{3}, 183--193 (2019).

\bibitem{bernstein2018} Bernstein, E., Shore, J. \& Lazer, D. How intermittent breaks in interaction improve collective intelligence. \textit{Proc. Natl. Acad. Sci. U.S.A.} \textbf{115}, 8734--8739 (2018).

\bibitem{couzin2005} Couzin, I. D., Krause, J., Franks, N. R. \& Levin, S. A. Effective leadership and decision-making in animal groups on the move. \textit{Nature} \textbf{433}, 513--516 (2005).

\bibitem{shaffer2013} Shaffer, Z., Sasaki, T. \& Pratt, S. C. Linear recruitment leads to allocation and flexibility in collective foraging by ants. \textit{Anim. Behav.} \textbf{86}, 967--975 (2013).

\bibitem{arthur1994} Arthur, W. B. Inductive reasoning and bounded rationality. \textit{Am. Econ. Rev.} \textbf{84}, 406--411 (1994).

\bibitem{roughgarden2002} Roughgarden, T. \& Tardos, \'{E}. How bad is selfish routing? \textit{J. ACM} \textbf{49}, 236--259 (2002).

\bibitem{morris2002} Morris, S. \& Shin, H. S. Social value of public information. \textit{Am. Econ. Rev.} \textbf{92}, 1521--1534 (2002).

\bibitem{pender2017} Pender, J., Rand, R. H. \& Wesson, E. Queues with choice via delay differential equations. \textit{Int. J. Bifurcation Chaos} \textbf{27}, 1730016 (2017).

\bibitem{ashery2025} Ashery, A. F., Aiello, L. M. \& Baronchelli, A. Emergent social conventions and collective bias in LLM populations. \textit{Sci. Adv.} \textbf{11}, eadu9368 (2025).

\bibitem{flint2026} Flint, A., Aiello, L. M., Pastor-Satorras, R. \& Baronchelli, A. Group size effects and collective misalignment in LLM multi-agent systems. \textit{Proc. Natl. Acad. Sci. U.S.A.} \textbf{123}, e2531697123 (2026).

\bibitem{piatti2024} Piatti, G. et al. Cooperate or collapse: Emergence of sustainable cooperation in a society of LLM agents. In \textit{Advances in Neural Information Processing Systems 37} (2024), pp. 111715--111759.

\bibitem{park2023} Park, J. S. et al. Generative agents: Interactive simulacra of human behavior. In \textit{Proceedings of the 36th Annual ACM Symposium on User Interface Software and Technology} (ACM, 2023), article 2, pp. 1--22.

\bibitem{zhang2024} Zhang, C. et al. ProAgent: Building proactive cooperative agents with large language models. \textit{Proc. AAAI Conf. Artif. Intell.} \textbf{38}, 17591--17599 (2024).

\bibitem{agashe2025} Agashe, S., Fan, Y., Reyna, A. \& Wang, X. E. LLM-Coordination: Evaluating and analyzing multi-agent coordination abilities in large language models. In \textit{Findings of the Association for Computational Linguistics: NAACL 2025} (2025), pp. 8053--8072.

\bibitem{cemri2025} Cemri, M. et al. Why do multi-agent LLM systems fail? In \textit{Advances in Neural Information Processing Systems 38} (2025), pp. 135676--135729.

\bibitem{ezaki2026a} Ezaki, T., Imura, N. \& Nishinari, K. How a shared state is described determines whether AI agents synchronize. Preprint at \url{https://arxiv.org/abs/2608.06968} (2026).

\bibitem{ezaki2026b} Ezaki, T., Imura, N. \& Nishinari, K. Warned alike, AI agents avoid the less-crowded road while people take it. Preprint at \url{https://arxiv.org/abs/2609.30883} (2026).

\bibitem{ezaki2026c} Ezaki, T., Imura, N. \& Nishinari, K. Shared and structured inputs undermine collective random choice by reasoning AI agents. Preprint at \url{https://arxiv.org/abs/2610.09667} (2026).

\bibitem{bikhchandani1992} Bikhchandani, S., Hirshleifer, D. \& Welch, I. A theory of fads, fashion, custom, and cultural change as informational cascades. \textit{J. Polit. Econ.} \textbf{100}, 992--1026 (1992).

\bibitem{lorenz2011} Lorenz, J., Rauhut, H., Schweitzer, F. \& Helbing, D. How social influence can undermine the wisdom of crowd effect. \textit{Proc. Natl. Acad. Sci. U.S.A.} \textbf{108}, 9020--9025 (2011).

\bibitem{kaul2012} Kaul, S., Yates, R. \& Gruteser, M. Real-time status: How often should one update? In \textit{2012 Proceedings IEEE INFOCOM} (2012), pp. 2731--2735.

\bibitem{hong2004} Hong, L. \& Page, S. E. Groups of diverse problem solvers can outperform groups of high-ability problem solvers. \textit{Proc. Natl. Acad. Sci. U.S.A.} \textbf{101}, 16385--16389 (2004).

\bibitem{microsoft_cb} Microsoft. Circuit Breaker pattern. Azure Architecture Center; \url{https://learn.microsoft.com/en-us/azure/architecture/patterns/circuit-breaker} [accessed 15 October 2026].

\bibitem{brooker2019} Brooker, M. Timeouts, retries, and backoff with jitter. Amazon Builders' Library (2019); \url{https://builder.aws.com/content/3EumjoZascWd1oZiEgL8ORlv3qE/timeouts-retries-and-backoff-with-jitter} [accessed 15 October 2026].

\end{thebibliography}
\end{document}